\newcommand{\IcPINNs}{\textit{IcPINNs}}
\newcommand{\VcPINNs}{\textit{VcPINNs}}
\newcommand{\VoFicPINNs}{\textit{VoF-IcPINNs}}
\newcommand{\VoFvcPINNs}{\textit{VoF-VcPINNs}}
\newcommand{\PFicPINNs}{\textit{PF-IcPINNs}}
\newcommand{\PFvcPINNs}{\textit{PF-VcPINNs}}
\newcommand{\RefvcPINNs}{\textit{VcNet}}
\newcommand{\ReficPINNs}{\textit{IcNet}}
\newcommand{\DFSC}{\textit{IcNet}}
\newcommand{\VoFPINNs}{\textit{VoF-PINNs}}
\newcommand{\PFPINNs}{\textit{PF-PINNs}}
\newcommand{\DFSVOF}{\textit{VoF-IcPINNs}}
\newcommand{\DFSPFa}{\textit{PF-IcPINNs (no MIM)}}
\newcommand{\DFSPFb}{\textit{PF-IcPINNs}}
\newcommand{\loss}{\mathcal{L}}
\newcommand\vf[1]{\mathbf{#1}} % this is how vectors are specified in equations
\newcommand\rm[1]{\mathrm{#1}}
\DeclareUnicodeCharacter{200E}{\hspace{0pt}} % for \mathcal{L} symbol

\documentclass[final,5p,times,twocolumn,compress]{elsarticle}

\usepackage{amssymb}
\usepackage{amsmath}
\usepackage{textgreek}
\usepackage{hyperref}
\usepackage{subcaption}
\usepackage{natbib}
\usepackage{float}
\usepackage{xcolor}
\usepackage{relsize}
\usepackage{standalone}
\usepackage{tikz}
\usepackage{pgfplots}
\usetikzlibrary{shapes,arrows,spy}
\pgfplotsset{compat=newest}
\usepackage{placeins}
\usepackage{comment}
\usepackage{array}

\journal{-}

\begin{document}

% KIT colors
\definecolor{KITred}{RGB}{162, 34, 35}
\definecolor{KITorange}{RGB}{223 155 27}
\definecolor{KITyellow}{RGB}{252 229 0}
\definecolor{KITmaygreen}{RGB}{140 182 60}
\definecolor{KITgreen}{RGB}{0 150 130}
\definecolor{KITcyan}{RGB}{35 161 224}
\definecolor{KITblue}{RGB}{70 100 170}
\definecolor{KITpurple}{RGB}{163 16 124}

\begin{frontmatter}



\title{PINNs4Drops: Video-conditioned physics-informed neural networks \\for two-phase flow reconstruction}
%\title{PINNs4Drops: Image-conditioned physics-informed neural networks \\for reconstructing two-phase flows}
%\tnotetext[label1]{\textbf{Short title:} Video-conditioned PINNs for two-phase flows}
%\tnotetext[label2]{\textbf{Teaser:} Video-conditioned PINNs reconstruct hidden 3D velocity and pressure fields in two-phase flows from 2D planar videos}


\author[inst1]{Maximilian Dreisbach \corref{cor1}}
%\ead{maximilian.dreisbach@kit.edu}
\cortext[cor1]{Corresponding author. Email: maximilian.dreisbach@kit.edu}
\author[inst2]{Elham Kiyani}
\author[inst1]{Jochen Kriegseis}
\author[inst2,inst4]{George Em Karniadakis}
\author[inst1]{Alexander Stroh}

\affiliation[inst1]{organization={Institute of Fluid Mechanics, Karlsruhe Institute of Technology},
            addressline={Kaiserstraße 10}, 
            city={Karlsruhe},
            postcode={76131},
            country={Germany}}
            
\affiliation[inst2]{organization={Division of Applied Mathematics, Brown University},
            city={Providence},
            postcode={02912}, 
            state={RI},
            country={USA}}

%\affiliation[inst3]{organization={Division of Applied Mathematics and School of Engineering, Brown University},
%            city={Providence},
%            postcode={02912}, 
%            state={RI},
%            country={USA}}

\affiliation[inst4]{organization={Pacific Northwest National Laboratory},
            city={Richland},
            postcode={99354}, 
            state={WA},
            country={USA}}

\begin{abstract}
Two-phase flow phenomena underpin critical technologies such as hydrogen fuel cells, spray cooling, and combustion, where droplet dynamics govern performance and efficiency. Conventional optical diagnostics, including shadowgraphy and particle image velocimetry, provide valuable insights but are limited to two-dimensional projections of inherently three-dimensional flows. We employ a specialized optical technique that encodes droplet surface information through color-coded glare points, enabling enhanced reconstruction of gas-liquid interfaces. To interpret these measurements, we introduce video-conditioned physics-informed neural networks (\VcPINNs{}), which integrate experimental observations with governing fluid dynamics equations. This hybrid framework leverages the strengths of both data-driven learning and physical constraints, allowing accurate volumetric flow reconstruction from limited input images. Applied to droplet impingement experiments, our method yields highly resolved and physically consistent 3D interface and flow dynamics. The combined imaging and PINN reconstruction strategy provides a powerful platform for advancing multiphase-flow analysis, with broad potential impact across energy, cooling, and propulsion applications.

%Two-phase flow phenomena underpin critical technologies such as hydrogen fuel cells, spray cooling, and combustion, where the shape and dynamics of droplets govern performance and efficiency.
%Conventional optical diagnostics, including shadowgraphy and particle image velocimetry, provide valuable insights but are limited to two-dimensional projections of inherently three-dimensional (3D) flows.
%Here, we employ a purposefully developed optical measurement technique that encodes droplet surface information through color-coded glare points from lateral illumination, enabling enhanced reconstruction of gas-liquid interfaces.
%To interpret these measurements, we introduce video-conditioned physics-informed neural networks (\VcPINNs{}), which integrate experimental observations with the governing fluid dynamics equations.
%This hybrid framework leverages the strengths of both data-driven learning and physical constraints, allowing accurate volumetric flow reconstruction from limited input images.
%Applied to droplet impingement experiments, our method yields highly resolved and physically consistent 3D interface and flow dynamics.
%The combined optical imaging and PINN reconstruction strategy provides a powerful platform for advancing the quantitative study of multiphase flows, with broad potential impact across energy, cooling, and propulsion applications.

%\vspace{1em}
%\noindent
%\textbf{Teaser} \\
%\vspace{1em}
%\noindent
%Video-conditioned PINNs reconstruct hidden 3D velocity and pressure fields in two-phase flows from 2D planar videos
\end{abstract}

\begin{keyword}
two-phase flow \sep interfacial dynamics \sep physics-informed neural networks \sep volumetric reconstruction \sep deep learning
\end{keyword}

\end{frontmatter}

\section*{Introduction}

Two-phase flows, involving liquid droplets or gaseous bubbles, are ubiquitous in nature and critical in numerous technical applications.
These include the impingement of liquid droplets on wet or dry surfaces, such as in spray cooling~\cite{Moreira2011}, spray coating~\cite{Andrade2013, Dalili2016}, inkjet printing \cite{Lohse2022}, and adhering droplets in external flows for applications like cleaning and drying~\cite{Thoreau2006, Seevaratnam2010}, oil recovery~\cite{Thompson1994, Schleizer1999, Gupta2008, Madani2014}, heat exchangers~\cite{Korte2001, Kandlikar2002}, airfoil icing prevention~\cite{Karlsson2019}, and fuel cells, where efficient removal of water droplets is essential for optimal performance~\cite{Theodorakakos2006, Kumbur2006, Esposito2010, Burgmann2013}.
These problems are governed by moving gas-liquid interfaces, which induce complex interactions between the external and internal flow of droplets~\cite{Minor2009}.
Despite extensive experimental and numerical investigations, the internal flow topology remains insufficiently understood.
Numerical simulation has advanced the study of droplet dynamics, including impingement on structured surfaces~\cite{fink2018, worner2021, Samkhaniani2021, Toprak2024} and shear-driven deformation~\cite{Maurer2016, Kramer2021, Burgmann2022}.
While simulations can accurately predict three-dimensional (3D) droplet dynamics, they are computationally expensive, limited to idealized cases, and require experimental validation for surface tension and contact angle modeling~\cite{kistler1993, cox1986}.
Experimental acquisition of 3D data is therefore crucial for gaining deeper insight into two-phase flow dynamics.
Shadowgraphy and Particle Image Velocimetry (PIV) are commonly employed to measure interface location and velocity fields, respectively.
Shadowgraphy is favored for 3D interface reconstruction due to its simplicity and spatial accuracy, though capturing complex deformations typically requires symmetry assumptions~\cite{Tomiyama2002,Higashine2008} or multiple camera angles~\cite{Fu2018, Masuk2019}.
However, non-convex shapes causing self-occlusion remain challenging even with a large number of viewpoints, while limited optical access in the experiments makes single-view techniques highly desirable.
PIV can reveal complex internal flow patterns in droplets related to Marangoni effects~\cite{Erbil2002,Rowan1997} and interface oscillation~\cite{Minor2009, Burgmann2021}.
However, the refraction of light at the gas-liquid interface distorts the measured velocity field and necessitates complicated correction procedures~\cite{Kang2004, Minor2007} that require precise knowledge of the instantaneous 3D interface shape.

Recent advances in deep learning, particularly implicit neural representations~\cite{Chen2019, Mescheder2019, Park2019}, offer a promising solution to this challenge.
These methods train neural networks to approximate a continuous implicit representation of 3D shapes through a level-set or occupancy function.
To enable reconstruction from images, the network is typically conditioned on features extracted from the input images through convolutional neural networks~\cite{Saito2019}.
This approach has been successfully applied to reconstruct the interface dynamics of impinging~\cite{Dreisbach2024a} and adhering droplets~\cite{Dreisbach2025a} from monocular optical experiments.
However, these models learn the underlying laws of physics governing the two-phase flow only implicitly from numerical data and do not enforce temporal coherence, which limits reconstruction accuracy.
Here, physics-informed neural networks (PINNs~\cite{Raissi2019}) are anticipated to enhance the learning of spatio-temporal droplet dynamics by incorporating both these untapped sources of prior knowledge into the network training.
By minimizing the residuals of governing equations, PINNs encode the underlying laws of physics during the optimization process.
PINNs have emerged as a powerful framework for solving complex problems due to their expressivity in approximating nonlinear functions and their unified approach for solving both forward and inverse problems. 
This framework enables the flexible integration of measurement data, making PINNs particularly well-suited for addressing fluid mechanical challenges~\cite{Cai2021a, Sharma2023, Mao2020, Kiyani2024, Eivazi2022}.
The foundational work by \citet{Raissi2019} demonstrated the potential of PINNs for solving inverse problems in fluid mechanics by encoding the continuity and non-dimensional incompressible Navier–Stokes equations (NSE).
PINNs can infer flow fields in regions without measurement data~\cite{Xu2021}, including continuous 3D velocity and pressure fields from sparse 2D two-component (2D2C) velocity measurements~\cite{Cai2021a} by solving the governing equations in the entire domain.
This capability extends to the prediction of quantitatively accurate 3D velocity and pressure fields from auxiliary measurements of tracer concentration, without requiring any data for the quantities of interest~\cite{Raissi2020}.
This \emph{hidden fluid mechanics} approach~\cite{Raissi2020} was applied to infer continuous 3D velocity and pressure fields in buoyancy-driven flows from tomographic temperature measurements~\cite{Cai2021b}.
Similarly, hard-to-access temperature fields in Rayleigh--B\'enard convection can be inferred from velocity measurements~\cite{Toscano2025}.
%The continuous and differentiable nature of the approximated solution allows for the reconstruction of the flow at higher spatio-temporal resolution than the experimental data \cite{Cai2021b}, and enables inference of derived quantities such as shear forces and vorticity \cite{Raissi2020}, as well as viscous and thermal dissipation rates \cite{Toscano2025}.

Applying PINNs to multiphase flows introduces additional challenges due to discontinuities of the flow fields at interfaces, which lead to steep gradients in the solution and locally high errors, hindering the network optimization.
Adaptive sampling strategies that dynamically track discontinuities and increase sampling density in critical regions have been developed to mitigate this issue~\cite{Mao2020, Lu2021}.
PINNs have been used to investigate forward and inverse problems in multiphase flows, including bubble and droplet dynamics~\cite{Buhendwa2021, Qiu2022, Zhai2022}, and flows involving heat transfer~\cite{Jalili2024} or electrochemical corrosion~\cite{Chen2024}.
For two-phase flows, the single-field Navier-Stokes and continuity equations are typically encoded alongside an interface evolution equation adopted from Volume of Fluid (VoF)~\cite{Buhendwa2021, Jalili2024}, phase-field~\cite{Qiu2022, Chen2024}, or level-set methods~\cite{Zhai2022}. 
High density and viscosity ratios (\emph{e.g.} water and air) remain challenging due to the large disparity in material properties, resulting in sharp changes in the magnitude of the physics-informed losses across the interface~\cite{Zhai2022, Qiu2022, Zhu2023}.
Therefore, accurate prediction of the phase distribution is critical, as it directly governs the quality of the physics-informed regularization.
Moreover, appropriate loss weighting is required to prevent divergence of the optimization caused by abrupt changes in the magnitude of the surface tension force at the interface~\cite{Buhendwa2021}.
% V: Scope: motivation, objective, strategy
While PINNs show promise for modeling two-phase flows, most studies to date are limited to synthetic data and simplified 2D benchmark problems.
Addressing realistic 3D problems and incorporating experimental data remains an open challenge, primarily for two reasons:

1) The complexity of the multi-objective PINNs optimization in the context of two-phase flows. 
The discontinuity of the flow field associated with sharp gradients and localized surface tension forces leads to a stiff problem that demands the accurate solution of the phase distribution.

2) The lack of reliable volumetric measurement data for training and validation purposes.
The commonly used shadowgraphy technique provides only sparse planar measurements of the phase distribution, while PIV measurements of the velocity within the liquid phase are hindered by optical distortion due to refraction at the interface.

Moreover, no direct comparison has been conducted between PINNs based on the VoF and phase-field methods, leaving unclear which formulation is best suited for inverse two-phase flow problems, such as 3D flow field reconstruction.
To address these challenges, we introduce \emph{PINNs4Drops}, a novel PINNs framework to infer continuous 3D phase, velocity, and pressure fields in two-phase flows from sequences of experimental images.
First, to address the complexity of the PINNs optimization, we propose image- and video-conditioned PINNs (\IcPINNs{} and \VcPINNs{}).
Conditioning the neural network with spatio-temporal features extracted from images provides localized priors that guide the reconstruction of sharp interfaces and thereby improve the stability of the optimization.
Second, to address the sparsity of volumetric experimental data, we employ a purposefully developed optical measurement technique based on shadowgraphy and color-coded glare points~\cite{Dreisbach2023} that encodes additional 3D information in the images.
We validate \emph{PINNs4Drops} against direct numerical simulations, demonstrating accurate 3D reconstruction of the gas-liquid interface, as well as the velocity and pressure fields.
Subsequently, we apply the proposed approach to reconstruct an impinging droplet using planar experimental data, highlighting its practical applicability and significant potential for real-world fluid dynamics analysis.
We integrate both the VoF and phase-field formulations in the \emph{PINNs4Drops} framework and assess their performance comparatively.
The insights in this work suggest that video-conditioned PINNs can directly infer continuous 3D velocity, pressure, and phase distribution fields in two-phase flows from image sequences obtained in experiments, effectively bypassing complicated measurements through curved interfaces.
In consequence, this opens a new pathway for processing data in experimental fluid mechanics and enables the study of complex internal flow structures using a simple single-camera experimental setup.

\section*{Problem setup}

\begin{figure*}[h]
    \centering
	\includegraphics[width=1.0\textwidth]{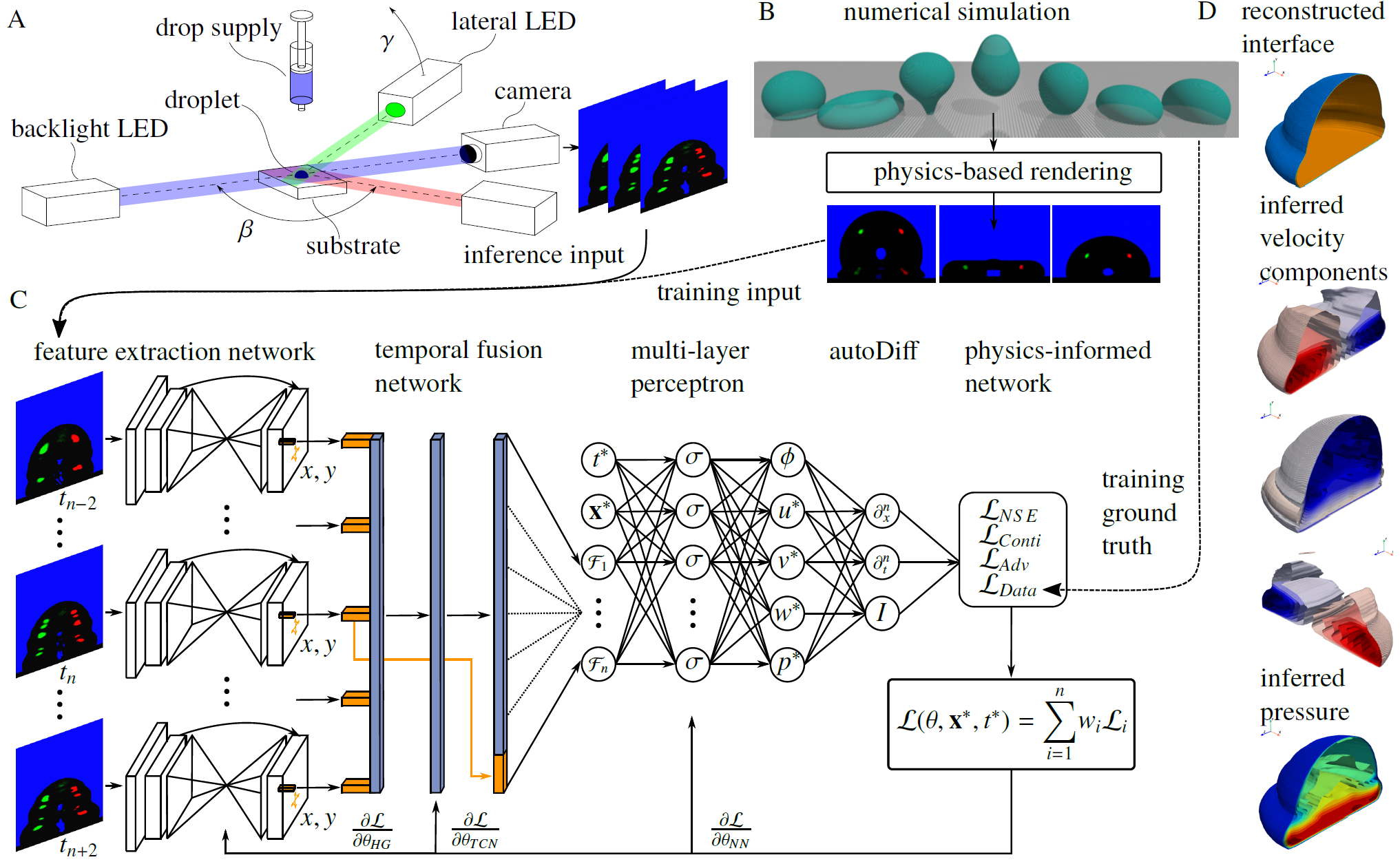}
    \caption{\textbf{Componential overview of the \emph{PINNs4Drops} framework for prediction of the three-dimensional gas-liquid interface, as well as the velocity and pressure distributions.} (\textbf{A}) Experimental setup of the glare-point shadowgraphy technique consisting of a blue backlight, green and red lateral light sources, and a high-speed RGB camera. The impingement of liquid droplets on solid substrates is recorded as an image sequence. (\textbf{B}) A sequence visualizing the droplet dynamics during impingement on a hydrophobic substrate obtained by direct numerical simulation. Physics-based rendering is employed to generate synthetic glare-point shadowgraphy images from the simulated gas-liquid interface geometries. (\textbf{C}) Schematics of the proposed video-conditioned PINNs (\VcPINNs{}). Initially, glare-point shadowgraphy images are processed using a convolutional hourglass network, which extracts pixel-aligned features from the input image at the pixel location $(x, y)$ on the image plane. Subsequently, temporal correspondences are extracted from the sequence of spatial features. The resulting spatio-temporal features, along with the temporal coordinate $t^*$ and the spatial coordinates $\vf{x}^*$, are provided as inputs to an MLP. The MLP predicts the phase distribution $\phi$, the three components of the dimensionless velocity vector $\vf{u}^*=(u^*,v^*,w^*)^T$, and the dimensionless pressure $p^*$ at the spatio-temporal coordinates $(x^*,y^*,z^*,t^*)$. The loss function $\loss$ comprises data loss terms for $\phi$, $\vf{u}^*$, and $p^*$, as well as physics-informed loss terms enforcing the Navier-Stokes equations, the continuity equation, and the advection equation for the phase distribution.}
    \label{fig:problem_setup}
\end{figure*}

To obtain experimental data suitable for the reconstruction of the 3D droplet dynamics from monocular recordings, we employ an optical measurement technique that embeds additional information on the 3D gas-liquid interface shape in the images.
To achieve this, for droplet impingement experiments, we apply the glare-point shadowgraphy technique~\cite{Dreisbach2023}, which extends the canonical shadowgraphy technique by color-coded glare points from additional lateral light sources.
As illustrated in Figure~\ref{fig:problem_setup} (A), a blue LED is used as the backlight for the shadowgraphy setup, which produces an accurate projection of the gas-liquid interface in the image.
Additionally, two lateral red and green LED light sources are positioned at specific scattering ($\beta$) and elevation angles ($\gamma$) relative to the droplet to produce colored glare points on the gas-liquid interface.
Given the known geometric configuration of the light propagation, additional 3D information of the gas-liquid is encoded in the position and the shape of the glare points, as previously reported in~\cite{Dreisbach2023} and demonstrated by the successful data-driven reconstruction of the spatio-temporal interface dynamics~\cite{Dreisbach2024a}.
To evaluate the generalization capability of the developed framework, we conduct droplet impingement experiments involving different impact velocities and surfaces.
Specifically, we investigate the impingement of $D_0=2.2-2.3$\,mm water droplets on structured hydrophobic Polydimethylsiloxane (PDMS) and hydrophilic polylactide (PLA) substrates, at velocities ranging from $U_0 = 0.43-0.88$\,m/s.
% Further details on experimental setup in supplementary material S1

We obtain training data for the PINNs by synthetic data generation on the basis of direct numerical simulation (DNS)~\cite{fink2018}.
Numerical simulation provides suitable ground truth data of the phase distribution, as well as the velocity and pressure fields in both phases, for the supervision of the network optimization through the data loss terms.
These simulations involved water droplets with an equivalent diameter of $D_0 = 2.1$\,mm impacting at a velocity of $U_0 = 0.62$\,m/s on flat and structured hydrophobic Polydimethylsiloxane (PDMS) substrates.
As indicated in Figure~\ref{fig:problem_setup} (B), synthetic images are generated based on the gas-liquid interface geometries extracted from the numerical simulation results through physics-based rendering~\cite{Dreisbach2025b}.
This enables the generation of synthetic images that visually match the experimental recordings (\emph{cp.} Figures~\ref{fig:problem_setup} (A) and~\ref{fig:problem_setup}(B)) and also correspond exactly to the numerical ground truth.
It is important to note that the numerical simulation and the experiments were deliberately conducted under different kinematic conditions to assess the generalization capability of the framework.
In particular, the experiments with the PLA substrate differed substantially from the simulation due to the substrate's hydrophilic nature and lower kinetic energy of the droplet upon impact, leading to notably different dynamics of the gas-liquid interface.
% Further details on synthetic training data generation in supplementary material S2

For the reconstruction of the 3D two-phase flow dynamics, we introduce image- and video-conditioned PINNs that leverage a tailored spatio-temporal feature extraction approach for the direct parameterization of PINNs with image data from optical measurements.
Figure~\ref{fig:problem_setup} (C) illustrates the architecture of the proposed video-conditioned PINNs, comprising four main components: a convolutional feature extraction network (CNN), a temporal convolutional network (TCN, here referred to as the temporal fusion network), a multi-layer perceptron (MLP), and the physics-informed network.
First, the glare-point shadowgraphy images are processed using a convolutional hourglass network~\cite{Newell2016}, which extracts pixel-aligned features $\mathcal{I}_i$ from the input images at the location $x,y$ on the image plane (orange blocks).
This operation is performed simultaneously for a sequence of $N=5$ images, with the snapshot considered for reconstruction in the middle of the sequence.
In the temporal fusion network, temporal correlations between the spatial pixel-aligned features $\mathcal{I}_i$ are extracted through 1D-convolutional operations.
The extracted spatio-temporal features $\mathcal{F}_i$ (last blue block), along with their corresponding spatial coordinates $x,y$, are forwarded to the MLP.
To preserve sharp spatial information of the considered time step, the pixel-aligned features from the central image are directly forwarded to the MLP as well.
Additionally, the temporal coordinate $t$ and the spatial coordinate $z$ are given as inputs to the MLP.
On this basis, the MLP predicts the phase distribution $\phi$, the three components of the dimensionless velocity vector $\vf{u}^*=(u^*,v^*,w^*)^T$, as well as the dimensionless pressure $p^*$ at $x,y,z,t$.
The point-wise residuals of the single-field two-phase Navier-Stokes, continuity, and interface evolution equations are computed using automatic differentiation applied to the predicted output with respect to the spatio-temporal input coordinates.
These residuals form the basis for physics-informed loss terms, \emph{i.e.}, the continuity equation ($\loss_\text{Conti}$), the advection equation governing the interface ($\loss_\text{Adv}$) and the Navier-Stokes momentum equations ($\loss_{\text{NSE},j}$), with $j=(x,y,z)$ (see equations~\ref{eq:dimensionless_NSE} --~\ref{eq:CahnHilliard} in the Supplementary Materials).
Each loss term is defined as the mean squared error (MSE) of the respective residuals.
In addition to these physics-informed losses, ground truth labels for the phase distribution, velocity, and pressure fields are extracted from numerical simulations to define data loss terms ($\loss_\text{Data}$) for the predicted quantities $\phi,u^*,v^*,w^*,p^*$, computed as the MSE between the predictions and ground truth.
The weights $\theta$ of the joint neural networks are updated by minimizing the composite loss representing the weighted sum of the physics-informed and data loss terms
\begin{equation}
    \begin{aligned}
    \mathcal{L}(\theta,\vf{x},t)=\sum_{i=1}^n w_i \mathcal{L}_i~,
    \end{aligned}
    \label{eq:loss_composite}
\end{equation}
with weighting coefficients $w_i$. %Lagrange multipliers
The formulation of governing equations, dedicated sampling schemes, and particular design choices for the neural network architecture play a critical role in the successful application of PINNs for two-phase flow problems.
The methods employed for the proposed \IcPINNs{} and \VcPINNs{} are detailed in the Materials and Methods section and the Supplementary Materials.
We train the developed PINNs on limited amounts of high-fidelity DNS data and apply them to the reconstruction of real data from experiments.
Specifically, the \VcPINNs{} are trained on a single simulation case featuring droplet impingement on a structured PDMS substrate, while \IcPINNs{} are trained additionally on the case of droplet impingement on the flat substrate.
For validation purposes, we train a purely data-driven version of the neural network exclusively with data loss terms.
Table~\ref{tab:model_overview} provides an overview of the PINNs configurations developed and evaluated in this work.

\begin{table}[ht]
    \centering
    \caption{\textbf{Overview and terminology of the developed PINNs configurations.} We integrate PINNs based on the Volume of Fluid method (\VoFPINNs{}) and the phase-field method (\PFPINNs{}) with two conditioning strategies, specifically, image-conditioned PINNs (\IcPINNs{}) and video-conditioned PINNs (\VcPINNs{}). For each conditioning approach, a data-driven reference model is trained solely on data losses.}
    \setlength{\tabcolsep}{6pt} % horizontal padding
    \renewcommand{\arraystretch}{1.5} % spacing for main table
    \begin{tabular}{l|ll}
        \textbf{equations} 
        & \begin{tabular}[c]{@{}l@{}}
            {\renewcommand{\arraystretch}{0.65}%
            \begin{tabular}[t]{@{}l@{}}
                image-conditioned \\ (\IcPINNs{})
            \end{tabular}}
          \end{tabular}
        & \begin{tabular}[c]{@{}l@{}}
            {\renewcommand{\arraystretch}{0.65}%
            \begin{tabular}[t]{@{}l@{}}
                video-conditioned \\ (\VcPINNs{})
            \end{tabular}}
          \end{tabular} \\
        \hline
        \begin{tabular}[c]{@{}l@{}}
            {\renewcommand{\arraystretch}{0.65}%
            \begin{tabular}[t]{@{}l@{}}
                Volume of fluid \\ (\VoFPINNs{})
            \end{tabular}}
          \end{tabular}
        & \VoFicPINNs{} 
        & \VoFvcPINNs{} \\
        \begin{tabular}[c]{@{}l@{}}
            {\renewcommand{\arraystretch}{0.65}%
            \begin{tabular}[t]{@{}l@{}}
                Phase field \\ (\PFPINNs{})
            \end{tabular}}
          \end{tabular}
        & \PFicPINNs{} 
        & \PFvcPINNs{} \\
        \begin{tabular}[c]{@{}l@{}}
            {\renewcommand{\arraystretch}{0.65}%
            \begin{tabular}[t]{@{}l@{}}
                data only \\ (reference)
            \end{tabular}}
          \end{tabular}
        & \ReficPINNs{} 
        & \RefvcPINNs{} \\
    \end{tabular}
    \label{tab:model_overview}
\end{table}

\section*{Results}
\label{sec:results}

\begin{figure}[h]
    \centering
    \includegraphics[width=1.0\linewidth]{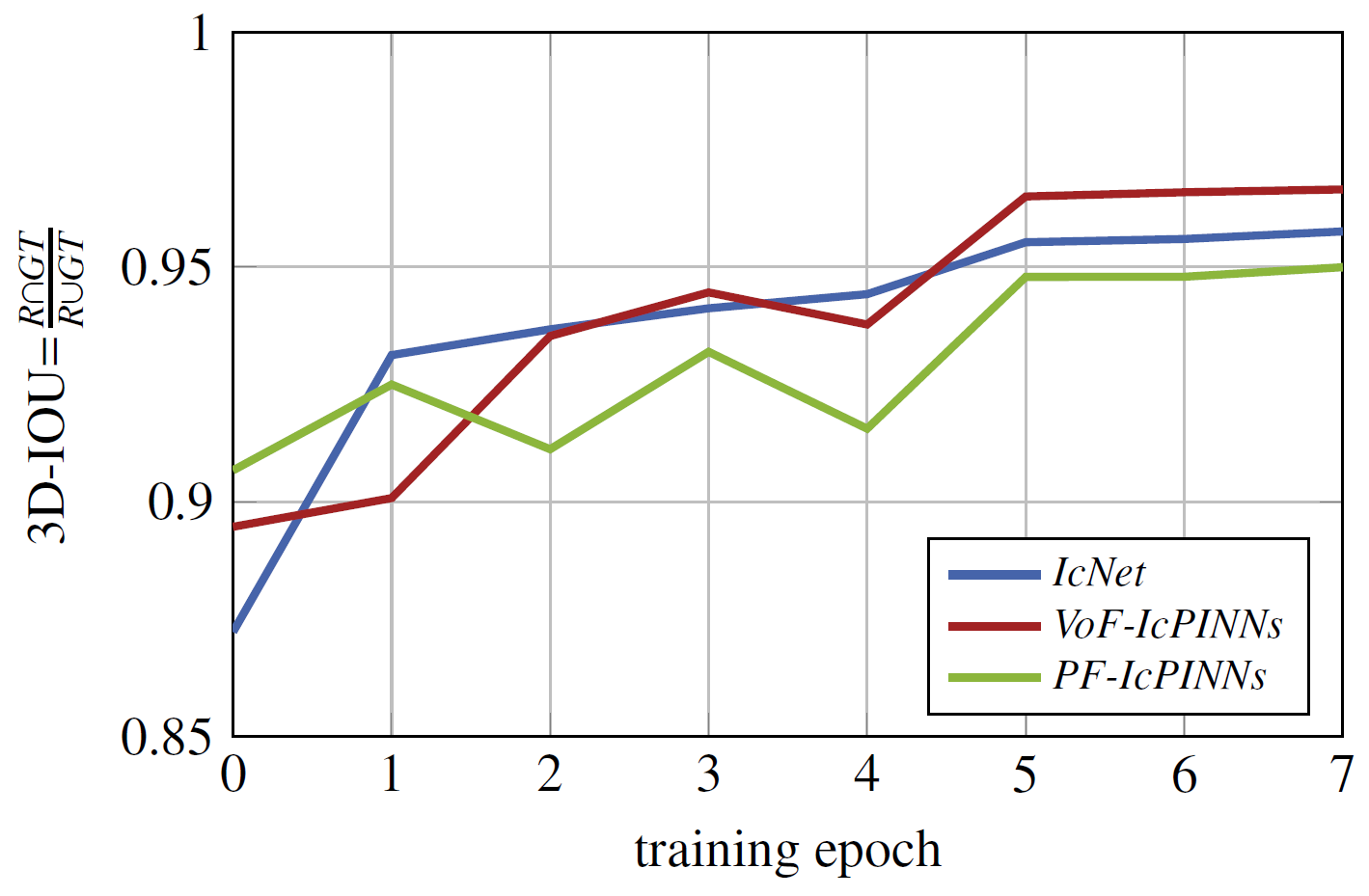}
    \caption{\textbf{Volumetric interface reconstruction accuracy of \IcPINNs{} during training.} We measure the volumetric accuracy by the average 3D-IOU on the validation dataset at the end of each training epoch and compare the \IcPINNs{} variants with the data-driven baseline \DFSC{}. \DFSVOF{} reach a higher reconstruction accuracy in comparison to \DFSC{}, while the best-performing variant of \DFSPFb{} does not reach the accuracy of \DFSC{}.}
    \label{fig:PINNs:3D-IOU}
\end{figure}

\begin{figure*}[h]
    \centering
    \includegraphics[width=1.0\linewidth]{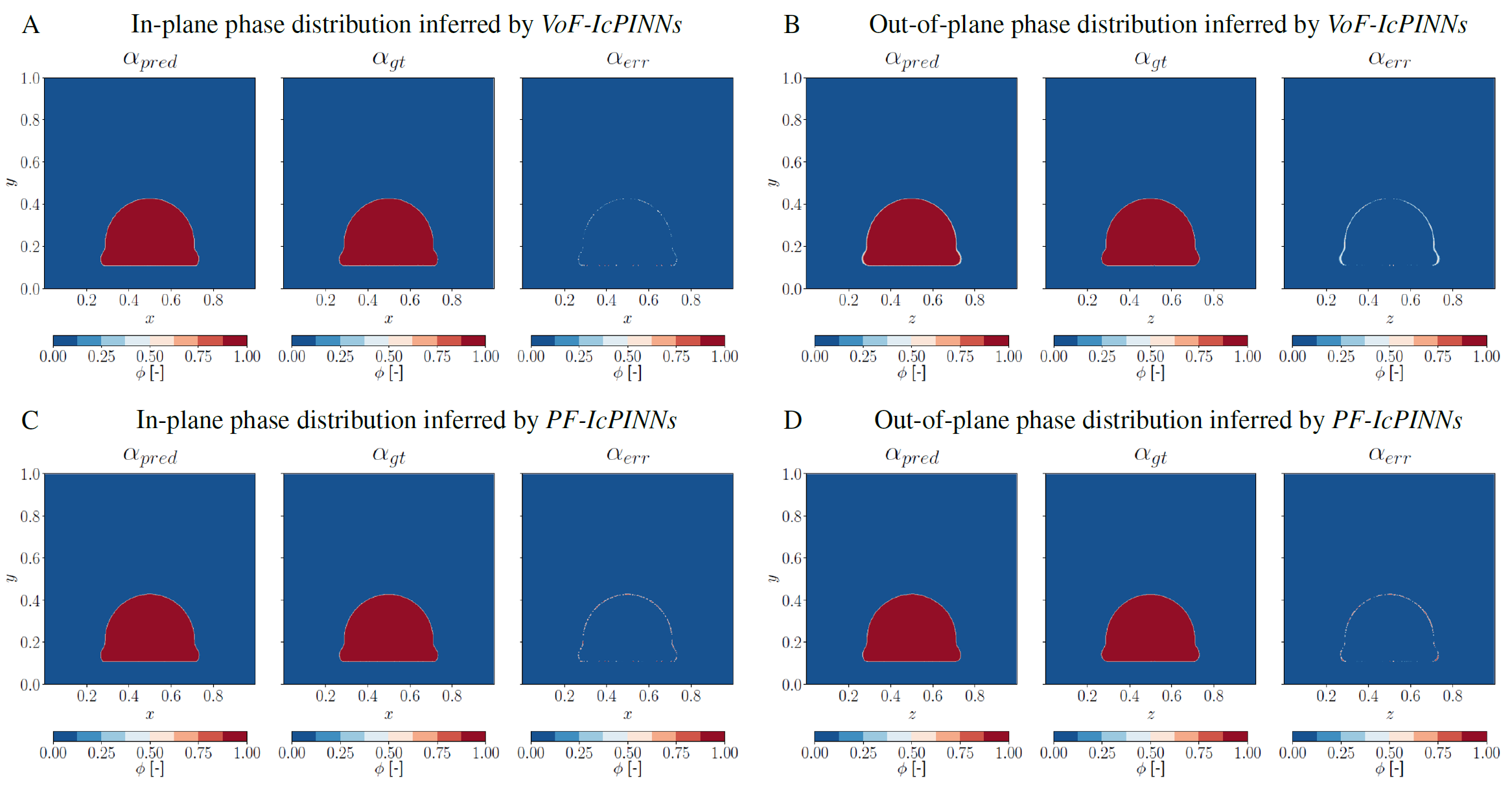}
    \caption{\textbf{Phase distribution predicted by \DFSVOF{} and \DFSPFb{} along the center planes of the droplet.} Shown are the predictions by \DFSVOF{} (top) and \DFSPFb{} with an initial diffuse-interface width of $\epsilon_0 = 0.01$ (bottom) in the in-plane and out-of-plane directions. The predictions (left) are shown in comparison to the ground truth simulation data (middle) and the spatial distribution of the absolute error between the prediction and the ground truth (right). Both \DFSVOF{} and \DFSPFb{} predict a sharp interface in the in-plane direction as indicated by the narrow error distribution (\emph{cp.} (A) and (C)), while in the out-of-plane direction \DFSPFb{} predicts a sharper interface than \DFSVOF{} (\emph{cp.} (B) and (C)).}
    \label{fig:PINNs:phase_dist}
\end{figure*}

In the following, we first validate the proposed \IcPINNs{} and \VcPINNs{} by means of synthetic data obtained through simulation, considering the predictive accuracy for the reconstruction of the 3D gas-liquid interface and the inference of the latent 3D velocity and pressure fields.
Afterward, we discuss the results obtained by the application of the developed PINNs for the reconstruction of images recorded in the experiments.
Here, we show the results of \IcPINNs{} optimized for interface reconstruction and \VcPINNs{} optimized for velocity and pressure inference, with a more detailed comparison in the Supplementary Materials.

%\subsection*{Gas-liquid interface reconstruction}
\subsection*{\textbf{Interface reconstruction from synthetic data}}
\label{sec:results:interface}

We quantitatively evaluate the reconstruction accuracy of the developed \IcPINNs{} by the results for the reconstructed gas-liquid interfaces on the synthetic validation dataset.
Figure~\ref{fig:PINNs:3D-IOU} shows the evolution of the three-dimensional intersection over union (3D-IOU) on the synthetic validation data during the training of the \IcPINNs{} in comparison to the purely data-driven \DFSC{}. 
As can be seen, \DFSVOF{} reach a higher reconstruction accuracy in comparison to \DFSC{}, while \DFSPFb{} do not yield an improvement over \DFSC{}.
More specifically, \DFSVOF{} reach an accuracy of 3D-IOU=$0.967$, which translates to a $0.9\%$ improvement over the data-driven \DFSC{} with 3D-IOU=$0.958$, and is significant considering the proximity to optimal reconstruction results at $\mathrm{3D{\texttt{-}}IOU}_\mathrm{ideal}=1$.
Moreover, the temporal consistency of the interface reconstruction is significantly increased by both variants of \IcPINNs{} in comparison to \DFSC{}.
A more detailed comparison of the different training dynamics between \VoFPINNs{} and \PFPINNs{}, as well as their influence on the reconstruction accuracy of the gas-liquid interface, can be found in the Supplementary Materials.

We further investigate the impact of the different formulations of the governing equations in \VoFPINNs{} and \PFPINNs{} on the prediction of the gas-liquid interface by comparing the spatial distribution of the phases in the predictions returned by \DFSVOF{} and \DFSPFb{}, with a particular focus on how varying the initial diffuse-interface width $\epsilon_0$ in the phase-field model affects the results.
Figure~\ref{fig:PINNs:phase_dist} shows the predicted phase distribution in the in-plane and out-of-plane direction for one sample of the validation dataset in comparison to the ground truth and the respective error distributions.
The comparison of the error distributions reveals that the prediction of \DFSPFb{} with $\epsilon_0=0.01$ features a less diffuse interface in the out-of-plane direction compared to \DFSVOF{}, while both models predict a rather sharp in-plane phase distribution.
These results indicate that the in-plane reconstruction adheres closely to the shadowgraph contour, while the out-of-plane reconstruction heavily relies on the trained model of droplet dynamics.
For \DFSPFb{} with $\epsilon_0=0.05$, the out-of-plane prediction becomes more diffuse, resembling the results of \DFSVOF{}.
The different sharpness of the out-of-plane phase distribution highlights the influence of the physics-informed loss based on the interface evolution equation on the learned droplet model.
A sufficiently small value for $\epsilon$ in the phase-field approach encourages the learning of a more accurately localized gas-liquid interface in comparison to the algebraic VoF approach, in which the interface thickness is not explicitly considered, but instead, a sharp interface is assumed, while a certain degree of numerical diffusion is accepted.
The less restrictive formulation of VoF apparently enables a better optimization of the neural network, which aims at approximating a sharp interface through a continuous function.

\begin{figure*}[h]
    \centering
    \includegraphics[width=1.0\linewidth]{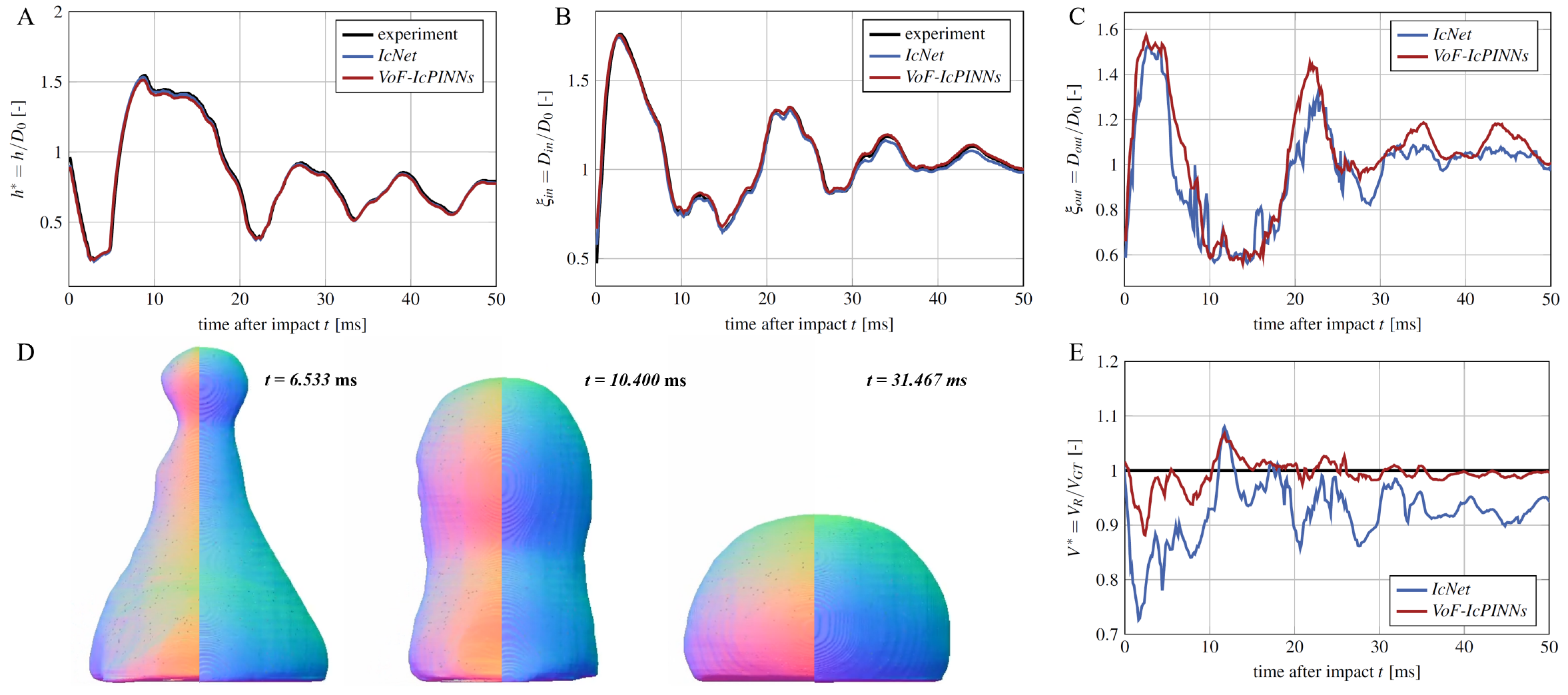}
    \caption{\textbf{Gas-liquid interface reconstruction by \DFSVOF{} and \DFSC{} for droplet impingement experiments on the structured PDMS substrate.} Temporal evolution of (\textbf{A}) the dimensionless droplet height $h^*$ and (\textbf{B}) the in-plane spreading factor $\xi_{in}$, obtained from the reconstructed gas-liquid interface geometries and compared with experimental measurements. We obtained the experimental data of $h^*$ and $\xi_{in}$ from the shadowgraphy contours of the droplet in the input images. (\textbf{C}) Development of the out-of-plane spreading factor $\xi_{out}$ over time. (\textbf{D}) Out-of-plane reconstruction by \DFSC{} (left half) and \DFSVOF{} (right half) for three different recordings from the experiments. The droplet geometry reconstructed by \DFSVOF{} is mirrored vertically to allow for a direct comparison of the reconstructed contours. (\textbf{E}) Temporal evolution of the normalized integral volume of the droplet $V^*$; the black reference line represents mass-conservative reconstruction results.}
    \label{fig:vol_PDMS_90_PINNs}
\end{figure*}

\subsection*{\textbf{Interface reconstruction from experimental data}}

We now employ the \IcPINNs{} that were trained and validated on synthetic data for the reconstruction of images recorded in experiments using the glare-point shadowgraphy technique, to evaluate their effectiveness in real-world applications.
Overall, we found a substantial improvement in the reconstruction accuracy by both \VoFicPINNs{} and \PFicPINNs{} in comparison to the data-driven \DFSC{} for the prediction of the gas-liquid interface on experimental data.
The visual inspection of the reconstruction results for experiments with both substrates at different observation angles reveals that the gas-liquid interfaces reconstructed by the \IcPINNs{} are smoother and more consistent over time in comparison to the reconstruction of \DFSC{}.
The in-plane reconstruction of both \IcPINNs{} and \DFSC{} aligns very accurately with the shadowgraphy contour, which is reflected in the temporal evolution of the droplet height and in-plane spreading factor closely following the measurements from the experiment, as illustrated in Figure~\ref{fig:vol_PDMS_90_PINNs} (A) and (B), respectively.
The almost identical agreement of the in-plane contour with the input images demonstrates that image-conditioning enforces direct consistency with the experimental data, providing local support for the 3D reconstruction of the interface.
Furthermore, the temporal evolution of the gas-liquid interface appears to be smoother for the reconstruction of the \IcPINNs{} in comparison to the data-driven baseline, as indicated by the significantly reduced fluctuations in the out-of-plane spreading factor presented in Figure~\ref{fig:vol_PDMS_90_PINNs} (C).
A comparison of the reconstruction results obtained by \DFSC{} and \DFSVOF{} is shown in Figure~\ref{fig:vol_PDMS_90_PINNs} (D).
The curvature of the interface reconstruction by \DFSVOF{} appears to be more physical, as small-scale features with a high curvature that can be observed for the purely data-driven reconstruction are not present in the reconstruction by the PINNs.
At the considered $\mathit{We}$-number, such features are unphysical, as the surface tension of the gas-liquid interface counteracts the formation of high surface curvatures.
These results indicate that the consideration of surface tension in the momentum equation of the \IcPINNs{} had a positive regularizing effect on the optimization of the neural network that led to a more physical reconstruction of the gas-liquid interface.
As the in-plane reconstruction is highly accurate, volumetric errors are predominantly caused by fluctuations in the depth estimation.
Consequently, the smoother temporal evolution of the gas-liquid interface, particularly in the out-of-plane direction, results in a more volume-conservative reconstruction by \IcPINNs{} in comparison to the purely data-driven model.
These results are illustrated in Figure~\ref{fig:vol_PDMS_90_PINNs} (E), which shows the temporal evolution of the normalized integral droplet volume for the reconstruction of \DFSVOF{} and \DFSC{} for experiments that featured droplet impingement on the structured PDMS substrate.
Throughout the entire period of time, the reconstruction of \DFSVOF{} lies significantly closer to the ground truth volume, indicated by the black line, and hence remains more conservative.
Both reconstruction techniques exhibit larger errors immediately after the impingement of the droplet, which correlates with the strongest droplet deformation.
However, the maximum error is significantly reduced by \DFSVOF{} compared to \DFSC{}.
Furthermore, the prediction of \DFSC{} rather underestimates the volume of the droplet, while the prediction of \DFSVOF{} remains closer to the ground truth volume.
Consequently, \DFSVOF{} achieve a significantly lower uncertainty and bias error of the reconstructed volume of the droplet in comparison to \DFSC{}.
Specifically, \DFSVOF{} reach an average uncertainty of $\sigma_{\rm{V}} = 1.6\%$ and bias error of $\delta_{\rm{V}} = 1.6\%$ in comparison to $\sigma_{\rm{V}} = 6.5\%$ and bias error of $\delta_{\rm{V}} = 5.1\%$ for \DFSC{}, effectively reducing the errors by factors of four and three, respectively.
The substantial improvement of the reconstruction accuracy of the gas-liquid interface by the \IcPINNs{} suggests that the introduction of prior knowledge through the physical constraints during training is an effective measure to enhance the trained model of the droplet dynamics.
The lower errors by the \IcPINNs{} in comparison to the data-driven baseline are consistently achieved for the reconstruction of all considered experiments, involving the impingement of droplets on structured PLA and PDMS substrates at different observation angles and velocities, as detailed in the Supplementary Materials, Table~\ref{tab:uncertainty_bias_IcPINNs}.
This includes varying degrees of gas-liquid interface deformation, as the impingement on the hydrophobic PDMS substrate leads to noticeably greater deformation of the droplet, demonstrating that the developed \IcPINNs{} generalize to different fluid mechanical conditions.
Furthermore, the enhancement of the predictive accuracy for the gas-liquid interface that was observed for the reconstruction of the synthetic validation data carried over well to the prediction of experiments, which indicates that the approach of training on synthetic data is suitable for the proposed \IcPINNs{}.

%\subsection*{Velocity and pressure inference}
\subsection*{\textbf{Velocity and pressure inference from synthetic data}}
\label{sec:results:velocity}

In the following, we present results for the velocity and pressure inference obtained by \VcPINNs{}, which demonstrated significantly improved accuracy compared to \IcPINNs{}, including instances of \IcPINNs{} with equivalent sampling and sequential training configurations.

\begin{figure}[h]
    \centering
    \includegraphics[width=1.0\columnwidth]{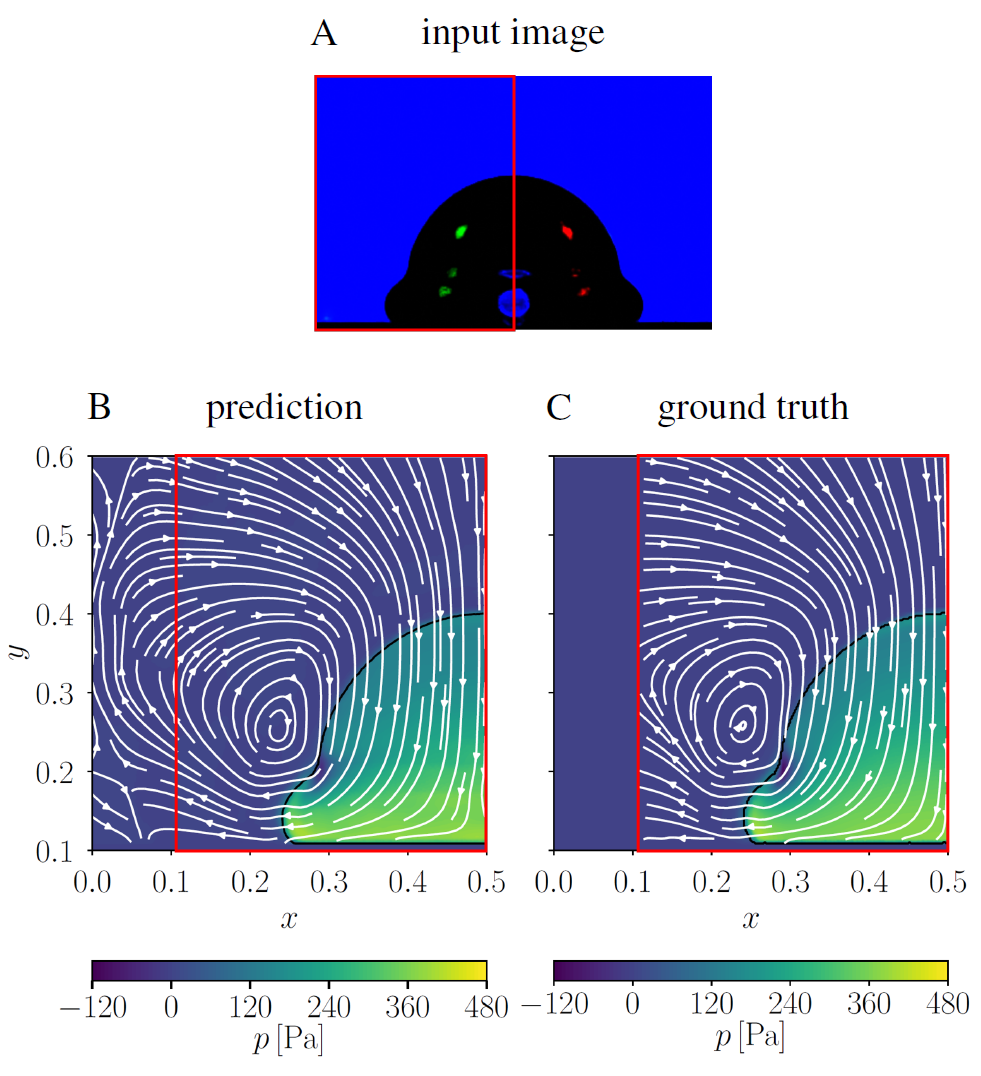}
    \caption{\textbf{Velocity and pressure inference of \VoFvcPINNs{} for one snapshot of the validation dataset.} (\textbf{A}) Synthetic input image rendered from a simulated droplet at $t=1.05$\,ms after impingement. (\textbf{B}) Streamline visualization of the inferred velocity field and pressure contours along the center plane of the droplet in the in-plane direction in comparison to (\textbf{C}) the ground truth simulation data. The contour of the gas-liquid interface is indicated by the black solid line. The PINNs complete the flow field in a physically consistent way beyond the boundary of the training data domain, which is indicated by the red box.}
    \label{fig:PINNs:compound_z}
\end{figure}

We evaluate the accuracy of the velocity and pressure prediction of the proposed \VcPINNs{} on the validation dataset by the comparison to the ground truth velocity and pressure data obtained by direct numerical simulation.
Figure~\ref{fig:PINNs:compound_z} shows the in-plane pressure and velocity fields in the center plane of the droplet (left) predicted by \VoFvcPINNs{} in comparison to the ground truth (right) for one snapshot of the synthetic validation data (top).
As can be seen, there is a good topological agreement of the predicted pressure and velocity fields with the ground truth.
Similar results were obtained for the inferred fields in the out-of-plane direction, as illustrated by figure~\ref{fig:PINNs:compound_x} in the Supplementary Materials.
Furthermore, the prediction of the pressure reaches a good quantitative agreement in both the in-plane and out-of-plane directions. 
The velocity field is accurately reconstructed in both the liquid and gaseous phases and remains physically consistent in regions beyond the simulation domain where no training data were available.

\begin{figure*}[h!]
    \centering
    \includegraphics[width=1.0\linewidth]{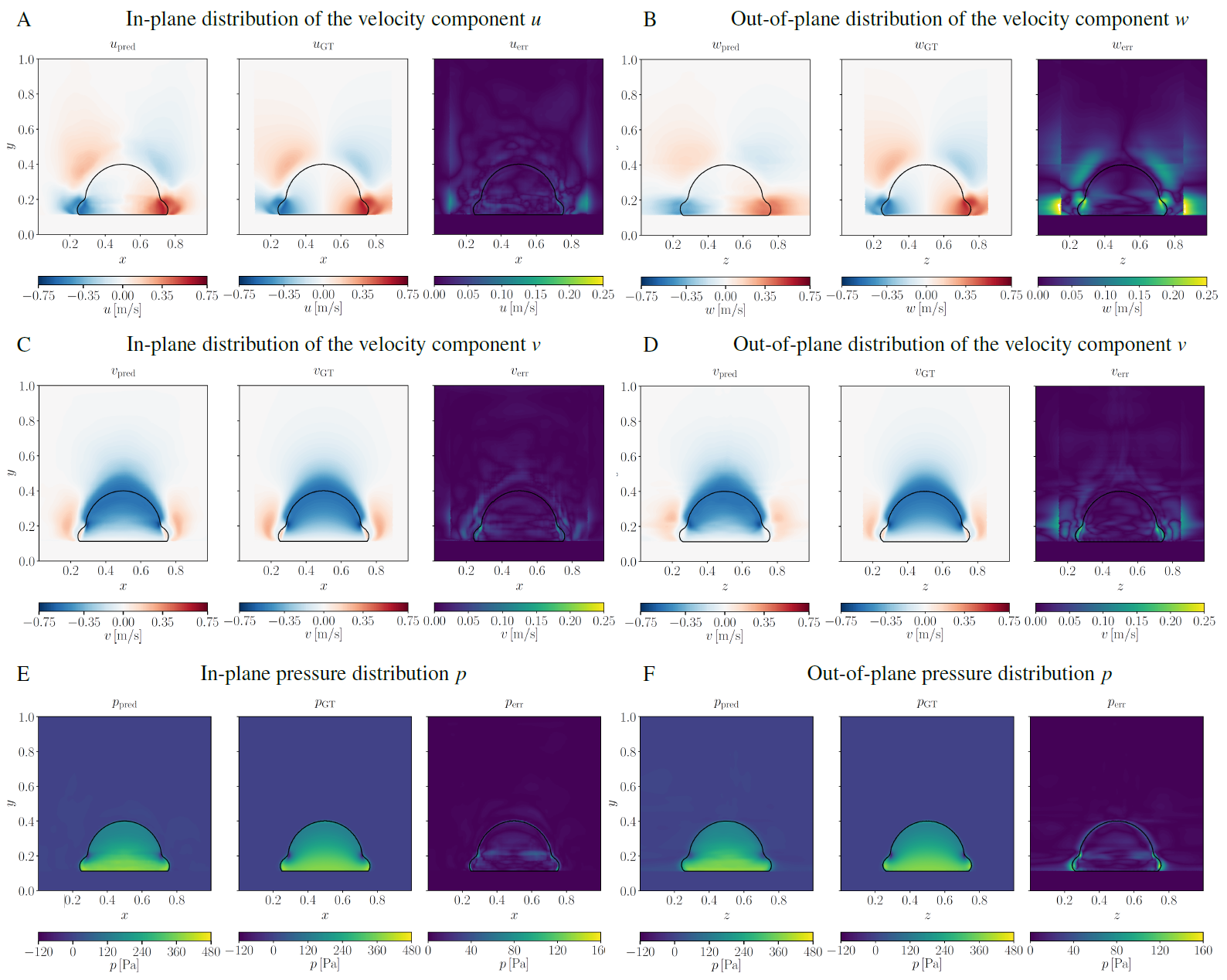}
    \caption{\textbf{Velocity and pressure fields predicted by \VoFvcPINNs{} for one snapshot from the validation dataset.} Velocity components $u$, $v$, and $w$, and pressure distribution $p$ at $t=1.05$\,ms after impingement along the center planes of the droplet in the in-plane and out-of-plane directions (left) in comparison to the ground truth simulation data (middle) and the absolute error between the prediction and the ground truth (right). The predicted contour of the droplet's gas-liquid interface is indicated by the black solid line overlayed on the prediction, and analogously, the ground truth contour is overlayed on the ground truth velocity and pressure distributions.}
    \label{fig:PINNs:val_vel_pres}
\end{figure*}

The results for the velocity and pressure prediction are illustrated in further detail in Figure~\ref{fig:PINNs:val_vel_pres}, showing the prediction for the horizontal velocity components $u$ and $w$ in Figures~\ref{fig:PINNs:val_vel_pres} (A) and~\ref{fig:PINNs:val_vel_pres} (B), respectively, the vertical velocity component $v$ in Figures~\ref{fig:PINNs:val_vel_pres} (C) and~\ref{fig:PINNs:val_vel_pres} (D), as well as the pressure in Figures~\ref{fig:PINNs:val_vel_pres} (E) and~\ref{fig:PINNs:val_vel_pres} (F), each in comparison to the ground truth data and the absolute error distribution for the same sample of the validation data as shown in Figure~\ref{fig:PINNs:compound_z} (top).
As can be seen, all three components of the predicted velocity field show an overall good topological agreement with the ground truth data.
%The symmetry of the flow was learned accurately by the network.
%Moreover, due to this symmetry, the out-of-plane prediction for $u$ and the in-plane prediction for $w$ are zero, which was precisely inferred as well.
Furthermore, small-scale details, as well as regions with high gradients in the inferred velocity fields, were accurately predicted, as demonstrated, for instance, by the reconstruction of the complicated flow field above the contact line.
As indicated by the plots of the absolute errors, the prediction of the vertical velocity component $v$ is more accurate in comparison to the horizontal velocity components. The in-plane velocity $u$ is, however, predicted more accurately than the out-of-plane velocity $w$.
This observation is consistent across the entire validation dataset and is, furthermore, reflected in the relative errors of the predicted quantities.
The averaged relative $L^1$ and $L^2$ errors over all predicted quantities inside the droplet amount to $RL^1_{u,v,w,p}=26.0\%$ and $RL^2_{u,v,w,p}=33.6\%$ for \VoFvcPINNs{}, with marginally lower errors obtained for \PFvcPINNs{} and \RefvcPINNs{}.
The lowest errors are obtained for the inferred pressure field ($RL^2_p = 10.7\%$), followed by the vertical velocity component $v$ ($RL^2_v = 20.7\%$), the horizontal velocity component $u$ ($RL^2_u = 48.4\%$), while the highest errors are observed for the horizontal velocity component $w$ ($RL^2_w = 54.4\%$).
%The lowest errors were obtained for the inferred pressure field with $RL^1_p = 6.7\%$ and $RL^2_p = 10.7\%$, followed by the vertical velocity component with $RL^1_v = 16.6\%$ and $L^2_v = 20.7\%$, the horizontal velocity component $u$ with $RL^1_u = 36.5\%$ and $RL^2_u = 48.4\%$, and the highest errors were obtained for the horizontal velocity component $w$ with $RL^1_w = 44.1\%$ and $L^2_w = 54.4\%$.

The dynamic deformation of the droplet follows a dampened oscillation of the interface, leading to a decaying velocity magnitude that covers a wide dynamic range between $U_0=0.62$\,m/s upon impact and $U \approx 0$\,m/s at the end of the oscillation.
Moreover, the flow direction completely reverses at each maximum and minimum of the oscillation, leading to several zero crossings of the velocity components.
Due to these fluctuations in the velocity magnitude, the relative errors vary significantly with time and increase towards later time steps.
The absolute errors of the inferred fields are distributed more uniformly across the validation dataset, but are locally higher around the extrema of the droplet oscillation.
This is likely caused by the rapid change of the flow topology and velocity magnitude, occurring simultaneously with a low interface motion in the input images, which diminishes the cues for the velocity inference in the spatio-temporal features and leads to a higher uncertainty of the prediction.
However, both relative and absolute errors of the velocity components remain limited, even for very low velocity magnitudes.
Similarly, \citet{Qiu2022} found that low velocity magnitudes correlated with high errors, due to the focus of the PINNs optimization on training samples with large velocities and, consequently, relatively high losses.
Due to the aforementioned damped nature of droplet impingement, the majority of the training samples had low velocities, which in turn could balance out the influence of fewer high-velocity samples.
Additionally, the conditioning of the PINNs by spatio-temporal features provides direct velocity information that guides the inference and apparently facilitates more consistent results.
A detailed overview of the velocity and pressure errors, as well as their temporal evolution, can be found in tables~\ref{tab:VoFVcPINNs_err} and~\ref{tab:PFVcPINNs_err} and figures~\ref{fig:PINNs:err_rel_complete} to~\ref{fig:PINNs:err_abs_drop} in the Supplementary Materials, respectively.

The comparison of the in-plane and the out-of-plane predictions reveals a higher in-plane accuracy for all predicted quantities, which demonstrates the beneficial effect of the spatio-temporal features.
However, in comparison to the prediction of the gas-liquid interface, the available information for the inference of the latent velocity and pressure fields is even more limited, as only the temporal evolution of the shadowgraph contour and the glare points provide cues for the reconstruction, while the bulk of the shadowgraph does not carry any information for the velocity and pressure.
Consequently, only the in-plane velocity at the contour and the glare points can be predicted with the support of spatio-temporal features, while the rest of the flow topology and pressure distribution have to be inferred from the trained model of droplet dynamics.
%While comprehensive spatio-temporal features from the shadowgraphy contour are available for the in-plane prediction, the out-of-plane prediction depends on the trained model of the droplet dynamics and the 3D information encoded in the glare points.
The ability of the proposed \VcPINNs{} to infer the three-dimensional distribution of these latent quantities from limited image information highlights the effectiveness of the physics-informed learning approach.

The accuracy of the velocity prediction inside the liquid phase is higher compared to the gas phase, as indicated by around $31\%$ lower $RL^1$ errors and $20\%$ lower $RL^2$ errors inside the droplet compared to the rest of the domain.
A likely cause for this difference is the significantly higher density of the sampling points for both the data and physics-informed loss calculation in the liquid domain, and especially at the interface.
Moreover, the high density ratio of water to air results in greater residuals in the liquid phase, while the residual-based weighting further reinforces the physics-informed losses in the liquid domain.
Another source of prediction discrepancies in the outer region, farther from the interface, might arise from the outer boundary condition applied in the simulation (zero-gradient for velocity), which was intentionally omitted in the PINNs implementation to facilitate the inference of arbitrary, realistic distributions at the boundaries.
As can be seen most saliently in Figure~\ref{fig:PINNs:val_vel_pres} (D), the prediction by the PINNs completes the velocity field in a physically reasonable manner, which indicates a certain capability for the extrapolation of the prediction outside of the training data domain.

\subsection*{\textbf{Velocity and pressure inference from experimental data}}

\begin{figure}[h]
    \centering
    \includegraphics[width=1.0\columnwidth]{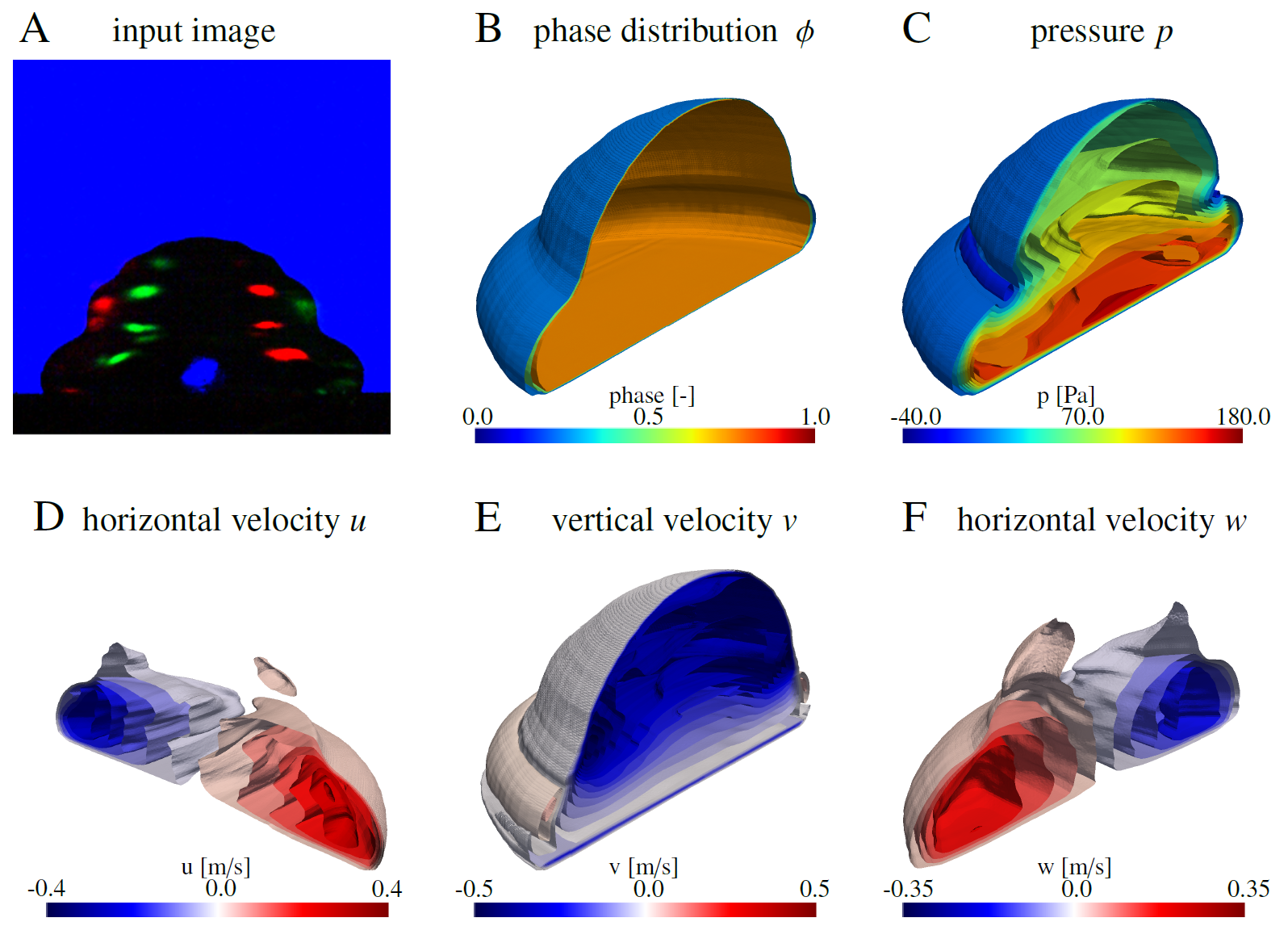}
    \caption{\textbf{Snapshot from droplet impingement experiment and corresponding predictions by \VoFvcPINNs{}.} (\textbf{A}) Snapshot at $t=1.73$\,ms from the experiment involving the structured PLA substrate and iso-contours of the \VoFvcPINNs{} predictions for (\textbf{B}) the phase distribution $\phi$ indicating the gas-liquid interface, (\textbf{C}) the pressure $p$, (\textbf{D}) the horizontal velocity component $u$, (\textbf{E}) the vertical velocity component $v$, and (\textbf{F}) the horizontal velocity component $w$. \VoFvcPINNs{} successfully infer continuous 3D flow fields, capturing local details of the flow topology such as the positive vertical velocity component above the contact line (see (E)) and the low-pressure region related to the concave interface (see (C)), as well as the overall symmetry of the flow.}
    \label{fig:PINNs:exp_rec}
\end{figure}

%%% Setup
The promising results obtained for velocity and pressure inference on synthetic validation data demonstrate that the developed \VcPINNs{} successfully capture the two-phase flow dynamics of the simulation case used for training and validation.
In the following, we investigate the extent to which these results translate to real experimental data, especially under varying initial conditions.
For that purpose, we conduct droplet impingement experiments involving two different substrates with distinct wetting behaviors and varying impact velocities, resulting in different magnitudes of initial kinetic energy.
These variations induce significantly different flow dynamics among the experimental cases compared to the training case.
%- simulation: rebound, experiments deposition without rebound of droplet, just liquid column formation for PDMS, while PLA experiments deposited without significant recoil -> overall recoil less pronounced in experiments

%%% Evaluation of velocity and pressure inference
Figure~\ref{fig:PINNs:exp_rec} shows the inferred velocity and pressure distribution inside the droplet as predicted by \VoFvcPINNs{} for a representative snapshot of the experiments involving droplet impingement on the structured PLA substrate.
As can be seen, \VcPINNs{} successfully infer continuous 3D flow fields within the droplet, capturing local details such as the positive vertical velocity component above the contact line (see Figure~\ref{fig:PINNs:exp_rec}, (E)) and the low-pressure region related to the concave interface and locally high velocities (\emph{cp.} Figure~\ref{fig:PINNs:exp_rec}, (C) and (E)).
Furthermore, the symmetry of the flow is well preserved across all predicted fields, indicating physically consistent results.
%The comparison with the numerical simulation at similar time instances reveals that the velocity and pressure are consistently inferred in a physically plausible manner.
The reconstructed fields exhibit a continuous temporal development, with flow topologies qualitatively matching those observed in the simulations throughout the dynamic deformation of the droplet. 
In particular, \VoFvcPINNs{} exhibit a more consistent temporal evolution of velocity and pressure compared to \RefvcPINNs{}, which occasionally produced rapid, nonphysical fluctuations during the time sequence (not shown here).

\begin{figure*}[h]
\centering
  \includegraphics[width=0.85\textwidth]{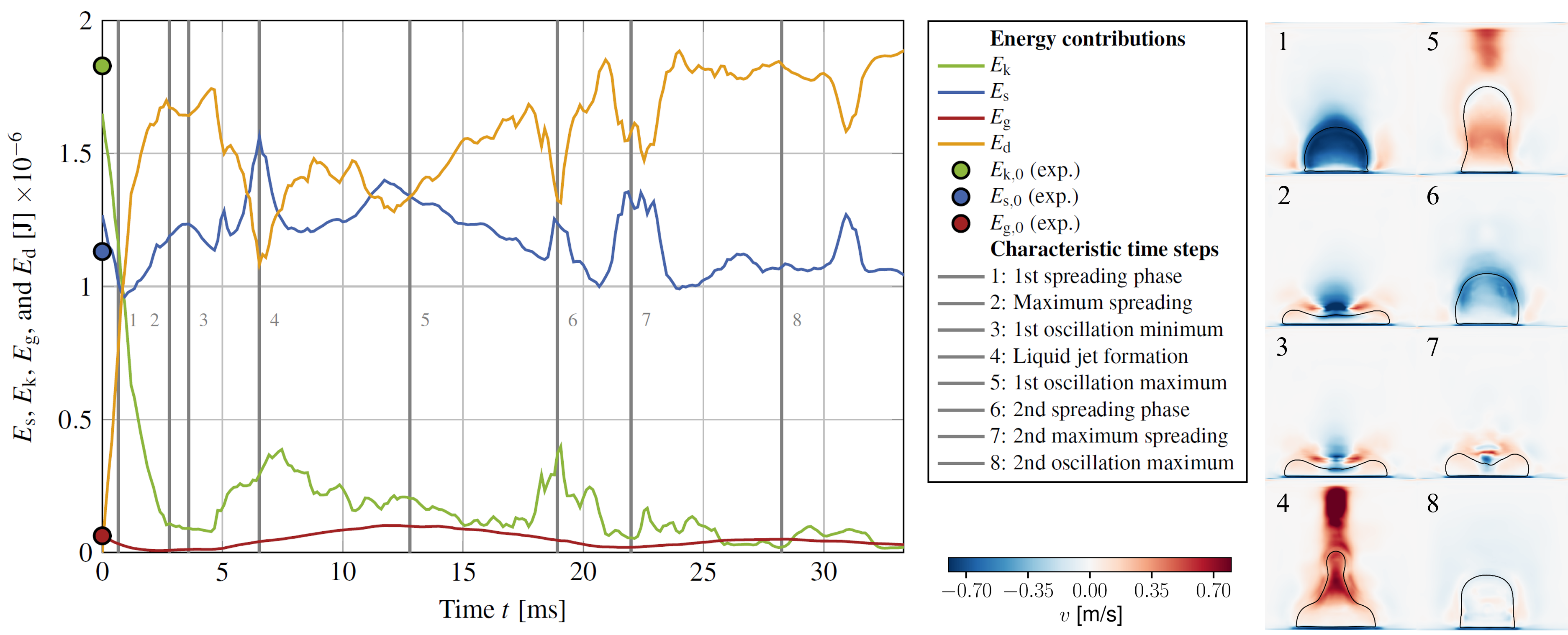}
  \caption{\textbf{Temporal development of the energy contributions predicted by \VoFvcPINNs{} for a droplet impingement experiment on the structured PDMS substrate.} The colored lines represent the kinetic energy of the droplet $E_{\rm{k}}$, the cumulative surface energy $E_{\rm{s}}$ of the gas-liquid and the liquid-solid interfaces, and the potential energy $E_{\rm{g}}$. The colored circles indicate measured energy contributions from the experiment just before impingement of the droplet. The details (1) to (8) display the in-plane prediction of the vertical velocity component $v$ at characteristic time steps, which are indicated by the gray vertical lines in the main plot. The alternating sign of the vertical velocity reflects the oscillatory nature of the droplet dynamics after impingement, which is also captured in a physically consistent manner in the energy contributions reconstructed by \VoFvcPINNs{}.}
  \label{fig:energy_PDMS_90_VoF-VcPINNs}
\end{figure*}

This advantageous impact of the physics-informed regularization is most evident during critical phases of the droplet impingement. 
At the oscillation extrema, where the velocity of the gas-liquid interface approaches zero and the acceleration changes sign, both the \RefvcPINNs{} and the \VcPINNs{} exhibit increased uncertainty, reflected in localized errors within the predicted velocity fields.
The first such extremum occurs during the transition from the spreading phase to the receding phase (see \citet{Rioboo2001, Rioboo2002} for further details).
In these frames, the interface motion is minimal while the velocity field undergoes rapid changes, which poses a particularly difficult reconstruction task, as the spatio-temporal features are less informative.
Moreover, self-occlusion of the gas-liquid interface further reduces the information content (\emph{cp.} Figure~\ref{fig:energy_PDMS_90_VoF-VcPINNs}, detail 3). 
Under these conditions, \VcPINNs{} yielded fewer erroneous frames and consistently lower error magnitudes than the baseline model.
These results indicate that the encoded two-phase flow dynamics from training with physics-informed losses support the velocity and pressure inference when the spatio-temporal features are less reliable, thereby enhancing the robustness and adaptability of \VcPINNs{} for varying droplet dynamics.

\VcPINNs{} successfully reconstructed complete sequences of experimental data, even for durations exceeding those present in the training data (simulation: $117$\,ms; PLA experiments: $133$\,ms).
Notably, the simulation covers three oscillation periods, after which the oscillation was almost entirely damped, while the droplet oscillation in the experiments persisted for the entire captured sequence ($7-16$ periods), highlighting the difference between the training and the test data.
Even over these extended times, the flow topology inferred by \VcPINNs{}, and particularly \VoFvcPINNs{}, remained physically consistent.
%aligning with the interface motion and matching corresponding phases from the simulation.
These results demonstrate the enhanced capability of \VcPINNs{} for long-term prediction compared to the data-driven baseline.

The accurate inference of velocity and pressure observed on the synthetic validation data was successfully reproduced on real experimental data, which featured varying kinematic conditions and, consequently, different flow dynamics compared to the training data. 
Furthermore, the differences between synthetic and real images demanded \VcPINNs{} to adapt to different spatio-temporal features, which they accomplished successfully, highlighting the practicality of training on synthetic image data.
Overall, the reliable prediction across a wide range of experimental conditions outside the training regime demonstrates a good generalization of \VcPINNs{}.

\subsection*{\textbf{Residuals of the governing equations}}

We further analyze the physical consistency of the inferred velocity, pressure, and phase distribution fields by evaluating the residuals of the governing equations across the entire domain.
Both versions of \VcPINNs{} reached significantly lower residuals for all governing equations compared to the data-driven \RefvcPINNs{}.
A detailed overview of the residuals obtained by all models can be found in the Supplementary Materials.
\PFvcPINNs{} consistently reached the lowest residuals on the training, validation, and test data, closely followed by \VoFvcPINNs{}, for all governing equations, except for the interface evolution equation, where \VoFvcPINNs{} obtained the lowest residuals.
Specifically, \PFvcPINNs{} achieved mean absolute errors of $\text{MAE}_{\text{Conti}} = 1.21 \times 10^{-2}$ for the dimensionless continuity equation (eq.~\ref{eq:fund:conti}), $\text{MAE}_{\text{Adv}} = 6.68 \times 10^{-4}$ for the dimensionless interface evolution equation (eq.~\ref{eq:CahnHilliard}), and $\text{MAE}_{\text{NSE},x} = 9.14 \times 10^{-4}$, $\text{MAE}_{\text{NSE},y} = 2.66 \times 10^{-3}$, and $\text{MAE}_{\text{NSE},z} = 1.10 \times 10^{-2}$ for the dimensionless momentum conservation equations in the $x$, $y$, and $z$ directions (eq.~\ref{eq:dimensionless_NSE}), respectively, averaged across all experiments in the test dataset.

These results further demonstrate that encoding the governing equations by training on physics-informed losses leads to a more physically consistent prediction of the two-phase flow dynamics.
%This falls in line with the earlier observation of more physically reasonable velocity and pressure predictions by \VcPINNs{}.
Notably, the residuals for the predictions from real experimental data reached a similar magnitude compared to the results for the synthetic training and validation data.
%, which likewise had a similar magnitude.
Moreover, the residuals are uniform across the experiments with significantly different droplet dynamics, which indicates a good generalization of the learned two-phase flow dynamics by \VcPINNs{} from one simulation to various experimental conditions.

The largest improvement by \VcPINNs{} occurs for the residuals on the continuity equation, which is also the equation with the overall highest residuals.
%This is followed by a significant improvement for residuals of the momentum equations in $X$, $Y$, and $Z$, and finally, only a marginal improvement for the interface evolution equation, which has, however, several orders of magnitude lower residuals than the continuity equation for all models.
While the optimization of \VcPINNs{} focused on the highest residuals, all physics-informed losses are converging, resulting in a balanced level of satisfaction for all governing equations at the end of the network training.
These results indicate an appropriate weighting of the physics-informed losses by the employed loss weighting schemes.

While the developed \VcPINNs{} were optimized towards a high accuracy of the velocity and pressure prediction, the accuracy of the interface reconstruction was maintained at the same level as \DFSC{}.
The successful prediction of the 3D flow topology in combination with an accurate gas-liquid interface reconstruction on the real data validates the proposed \VcPINNs{} approach.
The optimization of the phase distribution and velocity field is mutually dependent, as they are coupled by the physics-informed loss derived from the residuals of the interface evolution equation and the surface tension term in the momentum equations.
Thereby, the learning of an accurate velocity distribution at the interface promotes the learning of the interface location and vice versa.
The accurate prediction for both the phase distribution and the velocity field, in combination with low residuals for the interface evolution equation at the end of the training, demonstrates that the underlying physics were successfully encoded in the neural network during the optimization of the PINNs.

\subsection*{\textbf{Energy contributions evaluated for validation data}}
\label{sec:results:energy}

To further evaluate how well the global dynamics of the impinging droplet are learned, we calculate the kinetic energy $E_{\rm{k}}$, the surface energy $E_{\rm{s}}$, and the potential energy $E_{\rm{g}}$ from the inferred velocity and phase distributions and compare them with DNS reference data \cite{fink2018}.
Using the principle of energy conservation, we additionally calculate the viscous energy dissipation $E_{\rm{d}}$ as $E_{\rm{d}} = E_{\rm{t},0} - E_{\rm{k}} - E_{\rm{s}} - E_{\rm{g}} $, where $E_{\rm{t},0}$ is the total energy calculated immediately before impingement.

The comparison between the energy contributions inferred by \VcPINNs{} and those obtained by DNS reveals a good agreement over the entire temporal development.
Particularly, the potential energy $E_{\rm{g}}$ is predicted consistently well, with relative $L^2$ errors of $RL^2_{E_{\rm{g}}}=1.7\%$ for \PFvcPINNs{} and $RL^2_{E_{\rm{g}}}=3.0\%$ for \VoFvcPINNs{}.
Likewise, low relative errors are achieved for the surface energy $E_{\rm{s}}$, with $RL^2_{E_{\rm{s}}}=2.0\%$ by \PFvcPINNs{} and $RL^2_{E_{\rm{s}}}=2.3\%$ by \VoFvcPINNs{}.
These results fall in line with the findings of an accurate interface prediction, as both $E_{\rm{g}}$ and $E_{\rm{s}}$ are determined by the interface location.
A detailed overview of the energy contributions and a discussion of their temporal evolution are presented in the Supplementary Materials.

Among all energy terms, $E_{\rm{k}}$ has the highest relative deviations, reflecting its sensitivity to the velocity prediction.
Nevertheless, the overall agreement remains high, with $RL^2_{E_{\rm{k}}}=6.6\%$ reached by \VoFvcPINNs{} and $RL^2_{E_{\rm{k}}}=9.5\%$ by \PFvcPINNs{}.
Note that $E_{\rm{k}}$ inherits the broad dynamic range of the velocity field, leading to high relative velocity errors towards later time steps where the droplet oscillation has almost subsided.
This is further amplified by the quadratic relation of $E_{\rm{k}}$ to the velocity field, making $E_{\rm{k}}$ a sensitive global indicator for the accuracy of the predicted flow dynamics.
Despite this, $E_{\rm{k}}$ is predicted with significantly lower relative errors in comparison to the inferred velocity fields.
As the relative error of $E_{\rm{k}}$ reflects a global measure, whereas velocity errors are evaluated point-wise, the comparatively low errors for $E_{\rm{k}}$ suggest that the global flow topology and overall magnitude are learned well, while spatial inaccuracies of the velocity field arise on a local level.

Both \VcPINNs{} variants accurately predict the temporal evolution of $E_{\rm{d}}$ and maintain a better long-term agreement with the reference data compared to \RefvcPINNs{}, which is reflected in significantly lower $RL^2$ errors of $RL^2_{E_{\rm{d}}}=3.0\%$ for \PFvcPINNs{} and $RL^2_{E_{\rm{d}}}=5.0\%$ for \VoFvcPINNs{}, compared to $6.0\%$ for \RefvcPINNs{}.
Overall, these results indicate that \VcPINNs{}, and especially \PFvcPINNs{}, reliably capture the dissipative dynamics during droplet impingement while ensuring energy conservation over the entire temporal evolution.

\subsection*{\textbf{Temporal evolution of the energy contributions for real data}}

%%% Method
In the following, we evaluate the energy contributions calculated from the inferred flow fields to quantify the prediction accuracy for the experiments.
Figure~\ref{fig:energy_PDMS_90_VoF-VcPINNs} shows the evolution of the energy contributions for one representative experiment involving droplet impingement on the structured PDMS substrate predicted by \VoFvcPINNs{}.
In this experiment, a liquid jet was formed during the retraction phase, but no rebound of the droplet occurred in contrast to the simulation (\emph{cp.} Figure~\ref{fig:energy_validation}, details (4) and (5)).
In the experiments involving the PLA substrate, the droplet dynamics differed even more, with an overall reduced droplet motion resulting from the lower initial kinetic energy.

%%% Observations:
First, we compare the energy contributions upon impact with experimental measurements. 
Approximating the droplet just before impingement as a sphere allows us to accurately estimate $E_{\rm{k},0}$, $E_{\rm{s},0}$, $E_{\rm{g},0}$, and $E_{\rm{t},0}$.
As can be seen in Figure~\ref{fig:energy_PDMS_90_VoF-VcPINNs}, by the comparison of the measurements (colored circles) with the reconstruction at $t=0$ (colored lines), a good agreement with the experimentally determined energy estimates was achieved by \VoFvcPINNs{}.
Similar results were obtained across all experiments.
Particularly, \VoFvcPINNs{} achieved the most accurate prediction of $E_{\rm{t},0}$, with a relative error of $|\delta_{E_{\rm{t},0}}|=5.0\%$, followed by \RefvcPINNs{} at $|\delta_{E_{\rm{t},0}}|=5.8\%$ and \PFvcPINNs{} at $|\delta_{E_{\rm{t},0}}|=9.4\%$.

To further validate the reconstructed dynamics, we assess the temporal development of the energy contributions. 
As can be seen in Figure~\ref{fig:energy_PDMS_90_VoF-VcPINNs}, $E_{\rm{k}}$ decreases rapidly during the spreading phase (1) until the droplet reaches its maximum spreading diameter (2).
Simultaneously, $E_{\rm{g}}$ decreases and reaches a minimum around the first minimum of the droplet oscillation (3).
The surface energy $E_{\rm{s}}$ decreases during the initial wetting of the substrate, but subsequently increases again as the droplet spreads, reaching a local maximum near the oscillation minimum.
Consequently, $E_{\rm{d}}$ rises rapidly, reflecting energy dissipation due to viscous effects.
Overall, the evolution of the energy contributions during the spreading phase is physically consistent and agrees phenomenologically well with the DNS.
In particular, the continuous and monotonic decrease of $E_{\rm{k}}$ is in good agreement with both the simulation and the expected physical behavior.
During the subsequent retraction phase (4), $E_{\rm{k}}$ increases while a liquid jet is formed.
This process expands the gas-liquid interface, leading to a corresponding rise in $E_{\rm{s}}$.
While the overall development of $E_{\rm{k}}$, $E_{\rm{g}}$, and $E_{\rm{s}}$ during jet formation is physically reasonable, the simultaneous decrease in $E_{\rm{d}}$ suggests that one or more energy contributions are overestimated. 
The development of $E_{\rm{k}}$ and $E_{\rm{g}}$ remains consistent with the simulation, whereas $E_{\rm{s}}$ increases rapidly towards a high local maximum in the retraction stage.
Comparison with the simulation reveals a similar decrease in $E_{\rm{d}}$, though to a significantly smaller extent, while \VcPINNs{} overestimate $E_{\rm{s}}$ at comparable times in the validation data, further amplifying the drop in $E_{\rm{d}}$.
Consequently, an overestimation of $E_{\rm{s}}$ is the most probable cause for the observed decrease in $E_{\rm{d}}$.
As the retraction progresses, $E_{\rm{k}}$ reaches a local maximum before decreasing again as the liquid column grows, accompanied by rising $E_{\rm{g}}$ and $E_{\rm{s}}$, which is physically consistent with the continuous deceleration of the flow during its upward motion.
Subsequently, $E_{\rm{g}}$ reaches a local maximum at the maximum of the droplet oscillation (5) and decreases again as the liquid column collapses.
At the end of the retraction phase, $E_{\rm{k}}$ reaches a second local minimum, after which the droplet enters a second spreading phase in which $E_{\rm{k}}$ rises again to reach a third local maximum (6).
Simultaneously, $E_{\rm{g}}$ decreases further and reaches a local minimum corresponding to the oscillation minimum (7).
This energy exchange continues throughout the damped oscillations of the droplet and is in good qualitative agreement with the simulation.
Overall, \VcPINNs{} capture the droplet dynamics in a physically consistent manner, further indicated by the long-term subsiding of $E_{\rm{k}}$ and growth of $E_{\rm{d}}$.
As a cumulative quantity, $E_{\rm{d}}$ should ideally increase monotonically and thus serves as an indicator for energy conservation of the predictions.
Further local drops in $E_{\rm{d}}$ correlate with rises in $E_{\rm{s}}$, underlining that an overestimation of the gas-liquid interface area leads to fluctuations in $E_{\rm{d}}$.
Conversely, the temporal evolution of $E_{\rm{k}}$ is continuous and appears physically consistent throughout, reflecting the earlier observation that the evolution of the inferred velocity fields is physically consistent.
Accordingly, the velocity magnitude is reliably inferred across all experiments, which cover a wide dynamic range of initial kinetic energies, ranging from $0.6$ to $2.5 \times E_{\rm{k},0}$ of the DNS training data. % corresponding to impact velocities between $U_0=0.43$ and $U_0=0.88$\,m/s.
Consequently, \VcPINNs{} successfully predict physically plausible droplet dynamics for experiments with significantly different energy content, demonstrating a reliable reconstruction across different experimental conditions.
Overall, the good phenomenological agreement of the energy contributions with the DNS highlights that both the interface and velocity fields are inferred in a physically consistent manner.
These results demonstrate that \VcPINNs{} are capable of generalizing, indicating that the model has effectively learned universal two-phase flow dynamics.

\section*{Discussion}
\label{sec:discussion}

The presented results for the accurate three-dimensional reconstruction of the two-phase flow dynamics from planar measurement data, exemplified for an impinging droplet, demonstrate the success of the proposed image- and video-conditioned physics-informed neural networks (\IcPINNs{} and \VcPINNs{}).
The integration of localized spatio-temporally aligned features extracted by convolutional neural networks allows the developed PINNs to effectively leverage the volumetric information encoded in the images recorded through the purposefully developed glare-point shadowgraphy technique.
%%% Relevance of image-/video-conditioning -> Accuracy of gas-liquid interface prediction & generalizability
The convolutional features provide a robust and information-rich foundation for the reconstruction, which facilitates significant generalization capabilities, demonstrated by the accurate reconstruction of experiments with varying fluid mechanical conditions in comparison to the synthetic training data.
In particular, the in-plane reconstruction generalizes well to unfamiliar interface shapes and varying flow topologies because the shadowgraphy contours provide a reliable and spatially accurate information on the development of the planar gas-liquid interface location.
For out-of-plane reconstruction, the model relies more on the encoded droplet dynamics within the PINNs, supplemented by the additional information of the 3D gas-liquid interface encoded in the glare points, which allows for an accurate inference in the entire 3D domain.
While purely spatial features in \IcPINNs{} are optimal for accurate interface reconstruction, the additional temporal dependencies in the spatio-temporal features leveraged by \VcPINNs{} proved essential for predicting velocity and pressure.
%%% Synthetic training data generation
The successful reconstruction of 3D flow dynamics from real measurement data indicates that synthetic training data can effectively prepare PINNs for accurate experimental applications.
To the best of the author's knowledge, the present work marks the first application of PINNs to real measurement data of interface-resolved two-phase flows, demonstrating the feasibility of the approach.
Currently, the accurate measurement of velocity and pressure inside liquid droplets by optical techniques is generally inhibited due to refraction of light at the curved interface.
The developed PINNs overcome this limitation by providing a new pathway for internal two-phase flow measurements that allows for the quantitative recovery of hidden velocity and pressure distributions from flow visualization techniques.
Remarkably, even a limited training dataset, consisting of as few as one numerical simulation case, proved sufficient for training the PINNs.
This also offers the prospect for further enhancing the predictive accuracy by expanding the training dataset to a greater size and variance.

%%% Successful encoding of NSE
The incorporation of the governing equations of two-phase flows through physics-informed losses derived from the Navier-Stokes, continuity, and interface evolution equations results in a smoother, more physically accurate flow reconstruction compared to the data-driven baseline, as evidenced by lower residuals and more coherent velocity and pressure fields.
Furthermore, the temporal consistency of the interface reconstruction was significantly improved, resulting in better conservation of the droplet's volume over time and a notable reduction in uncertainty.
Importantly, the curvature of the reconstructed interface reflected the action of surface tension, demonstrating that the inclusion of surface tension models in the physics-informed losses regularized the learning process.
These findings demonstrate the successful learning of two-phase flow dynamics by the PINNs.
%%% learning of interfaces
The challenge of learning sharp liquid/gas interfaces by means of PINNs was successfully addressed by the local refinement of sampling points around the interface.
Prioritizing the learning of the phase distribution was found to be crucial, as the high density ratio of the considered water-air flows necessitates precise local phase information for accurate physics-informed losses of the momentum equation.
This was achieved by first pre-training the network on data to accurately capture the phase distribution, followed by the learning of velocity and pressure, and finally incorporating the physics-informed losses.
The physics-informed loss of the interface evolution equation couples the learning of the gas-liquid interface location and the velocity distribution at the interface.
The accurate reconstruction of the interface location and the velocity field around the gas-liquid interface, in combination with low residuals of the interface evolution equation, reveals that this relationship was learned well by the PINNs.
%%% VoF vs phase-field PINNs
The simplicity of the VoF approach, in combination with a superior accuracy of the predicted phase distribution and interface location in comparison to the phase-field approach, renders it more suitable for the considered two-phase droplet flows.
The optimization of the neural network, which is tasked with approximating a sharp interface via a continuous function, apparently is facilitated by the less restrictive formulation of algebraic VoF, which permits numerical diffusion.
%The less restrictive formulation of algebraic VoF -- permitting numerical diffusion -- apparently enables a better optimization of the neural network, which aims at approximating a sharp interface through a continuous function.
In contrast, the explicit modeling of interface thickness in the phase-field approach enables the learning of thin interfaces, which may be valuable for certain applications. 
The control of interface thickness was identified as the critical parameter for optimizing phase-field PINNs. 
Furthermore, the application of the Deep Mixed Residual Method (MIM) \cite{Lyu2022} significantly improved the optimization of the phase-field PINNs, resulting in a higher reconstruction accuracy, while reducing computational demands.
%by avoiding the calculation of fourth-order derivatives in the Cahn-Hilliard equations, leading to faster training with fewer computational resources.

%%% Prediction of hidden quantities for inference of velocity and pressure distribution -> VcPINNs
The developed \VcPINNs{} accurately infer continuous 3D velocity and pressure field for both phases, while maintaining a comparable interface reconstruction accuracy to the data-driven baseline.
The physically consistent prediction for the flow topology on real measurement data, including sharp gradients in the predicted velocity fields, demonstrates that \VcPINNs{} can reliably and accurately infer these hidden flow quantities from optical experiments.
The conditioning on features that capture the spatio-temporal evolution of the gas-liquid interface establishes a robust and accurate basis for velocity and pressure inference, even from very limited data in the images.
This approach enables generalization to experiments with substantially different droplet dynamics from the training case, particularly due to variations in initial kinetic energy upon impact and varying wettability of the substrates.
Under these challenging conditions, the residuals of the governing equation remain low, and the temporal evolution of energy contributions, including kinetic energy, is predicted in a physically consistent manner.
Collectively, these findings reveal that \VcPINNs{} successfully learned universal two-phase flow dynamics across diverse experimental conditions.
%%% Outlook: improvement of the proposed approach
While the developed lightweight temporal feature extraction network based on 1D convolutions proved to be effective for velocity and pressure inference, more elaborate sequence models, such as transformers~\cite{Vaswani2017} or state-space models like Mamba~\cite{Gu2024}, could potentially further increase the predictive accuracy.

%%% Outlook: further applications
The developed image- and video-conditioned PINNs seamlessly integrate prior knowledge from numerical simulations and the governing equations of two-phase flows with observations from glare-point shadowgraphy experiments. 
This conditioning with local spatio-temporal features and the physics-informed regularization enables the 3D reconstruction of the spatio-temporal droplet dynamics from a single slice of 2D measurement data. 
The presented approach offers a versatile framework for the post-processing of two-phase flow experiments, as the accurate measurement of the interface position can be easily obtained by the glare-point shadowgraphy technique across various two-phase flows.
Therefore, the application to droplets in fuel cells, sprays, or bubble flows appears straightforward.
By operating directly in image space, the developed PINNs efficiently leverage the high spatial and temporal resolution of the raw image data while inherently avoiding measurement errors associated with conventional techniques such as particle image velocimetry (PIV), which can be challenging for two-phase flows.
The proposed image- and video-conditioned PINNs represent a significant methodological advance, establishing a general paradigm for parameterizing PINNs directly with visual data, with potential applications to various other physical problems beyond fluid mechanics.
While the presented image- and video-conditioned PINNs demonstrate strong potential for reconstructing and analyzing two-phase flow dynamics across diverse applications, further improvements may be achieved by advancing the training methodology itself.
In particular, future work will explore extending \emph{PINNs4Drops} with second-order optimizers~\cite{kiyani2025optimizer}, which can accelerate convergence and enhance the robustness of the learned droplet dynamics.

\section*{Materials and methods}
\label{sec:methods}

In the following, we introduce the materials and methods employed in the proposed \emph{PINNs4Drops} framework, with a more detailed discussion in the Supplementary Materials.

\subsection*{\textbf{Simulation and synthetic training data generation}}

We obtain training and validation data for the phase distribution, velocity, and pressure fields in both phases from the 3D direct numerical simulations conducted by \citet{fink2018} within the framework of the phase-field method~\cite{Cai2015, Jamshidi2019} implemented in the open-source computational fluid dynamics (CFD) software \emph{OpenFOAM}\textsuperscript{\textregistered}~\cite{Weller1998}.
These simulations involved water droplets with an equivalent diameter of $D_0 = 2.1$\,mm impacting at a velocity of $U_0 = 0.62$\,m/s on flat and structured hydrophobic Polydimethylsiloxane (PDMS) substrates.
As shown in Figure~\ref{fig:problem_setup} (B), the initial impingement was followed by a rebound of the droplet~\cite{Rioboo2002}, and subsequent anisotropic wetting of the substrate that resulted in non-axisymmetric droplet deformation.
Based on the simulated interface geometries, we generate synthetic glare-point shadowgraphy images through physics-based ray tracing.
For this purpose, we accurately reproduce the optical setup of the experiments in the rendering environment \emph{Blender}~\cite{Blender} with the \emph{LuxCore}~\cite{LuxCore} package, which enables a physically accurate optical simulation~\cite{Dreisbach2025b}.
Thereby, we generate a labeled dataset for training the networks, comprising synthetic images that exactly correspond to the simulated interface geometries and visually align with the recordings of the experiments.

\subsection*{\textbf{Droplet impingement experiments}}

To facilitate the reconstruction of the 3D droplet dynamics from monocular recordings, we employ an optical measurement technique that embeds additional 3D information about the shape of the gas-liquid interface in the images.
To achieve this, we apply the glare-point shadowgraphy method~\cite{Dreisbach2023}, which extends the canonical shadowgraphy technique by color-coded glare points from additional lateral light sources, to droplet impingement experiments.
A blue LED is used as the backlight for the shadowgraphy setup, which produces an accurate projection of the gas-liquid interface in the image, as shown in Figure~\ref{fig:problem_setup} (A).
Additionally, two lateral red and green LED light sources are positioned at oblique angles relative to the droplet to produce colored glare points on the gas-liquid interface.
Due to the smoothness of the interface, these lateral glare points can be attributed to pure interface reflection~\cite{Hulst1981}.
Additional glare points arise from the two-fold refraction of the backlight as light enters and, subsequently, exits the gas-liquid interface, appearing in the central region of the droplet.
The different colors of the glare points enable the identification of the respective light source.
Given the known geometric configuration of the light propagation, additional 3D information of the gas-liquid is encoded in the position and the shape of the glare points~\cite{Dreisbach2023, Dreisbach2024a, Dreisbach2025a}.
We conduct experiments involving the impingement of droplets on two solid substrates with different wetting properties.
Specifically, we investigate the impingement of water droplets with an equivalent diameter of $D_0=2.2-2.3$\,mm on structured hydrophobic Polydimethylsiloxane (PDMS) and hydrophilic polylactide (PLA) substrates, with impact velocities ranging between $U_0 = 0.43 - 0.88$\,m/s.
Due to the different wettability of the substrates and the variation in the initial kinetic energy of the droplets, the experiments resulted in different droplet dynamics compared to the simulation, particularly different degrees of interface deformation.

\subsection*{\textbf{Volume of Fluid \& phase-field PINNs}}

We model the fluid dynamics during droplet impingement using the governing equations for two-phase incompressible, immiscible flows involving surface tension.
Specifically, we train the developed PINNs on physics-informed losses derived from the single-field two-phase formulation of the Navier-Stokes equations, the continuity equation, and an equation for the interface evolution based either on the Volume of Fluid (VoF)~\cite{HirtNichols1981} or the phase-field (PF) method~\cite{Jacqmin1999}.
The continuity and Navier-Stokes equations are harnessed to learn the dynamics of the two-phase flow, while the interface evolution equation couples the velocity field to the distribution of the two phases and, consequently, the location of the gas-liquid interface~\cite{Tryggvason2011}.
We employ the dimensionless formulation of the Navier-Stokes equations due to previously demonstrated benefits for the accuracy of the predicted flow quantities~\cite{Cai2021b} and for balancing of the different physics-informed loss components~\cite{Cuomo2022} compared to the dimensional formulation.
Volume-of-Fluid-based Physics-Informed Neural Networks (\VoFPINNs{}) represent the interface evolution in two-phase flows using the algebraic VoF method~\cite{HirtNichols1981}, a well-established approach for capturing fluid interfaces.
The phase-field version (\PFPINNs{}) employs the convective Cahn-Hilliard equation~\cite{CahnHilliard1959} to represent interface evolution.

\subsection*{\textbf{Image- and video-conditioned PINNs}}

We propose a novel framework for image- and video-conditioned PINNs (\IcPINNs{} and \VcPINNs{}) that are parameterized by latent spatio-temporal features extracted through Convolutional Neural Networks (CNNs) and Temporal Convolutional Networks (TCNs).
By operating directly in image space, instead of training on measurements of the interface, velocity, or pressure, \IcPINNs{} and \VcPINNs{} inherently avoid potential errors associated with traditional measurement techniques.
Moreover, this approach also offers the opportunity for extracting additional information from the image data, including long-distance correlations and previously unconsidered optical phenomena.
The extracted features from glare-point shadowgraphy images capture the spatio-temporal development of the gas-liquid interface, which provides an accurate basis for velocity and pressure inference.
As illustrated in Figure~\ref{fig:problem_setup} (C), the network architecture of the proposed \VcPINNs{} is comprised of four major components, namely, the feature extraction network (a CNN), the temporal fusion network (a TCN), a multi-layer perceptron (MLP), and the physics-informed network.
To enable the integration of the fixed grid processing inherent in CNNs with the continuous point-wise sampling required for PINNs, we employ two key concepts from implicit neural representation learning~\cite{Saito2019}.
First, pixel-aligned features $\mathcal{I}_i$ are extracted from the input images through a stacked hourglass network~\cite{Newell2016}, which consists of four consecutive convolutional encoder-decoder pairs.
This network progressively extracts global information while conserving local details and provides high-resolution features in the final feature maps that are spatially aligned with the input image.
Second, we apply bilinear interpolation to the final feature maps from the hourglass network, to enable continuous sampling of feature vectors $\mathcal{I}_i(x,y)$ at any point $x,y$ on the image plane.
These operations are performed simultaneously across a sequence of images, with the snapshot considered for reconstruction at $t_n$ being in the middle of the sequence.
The extracted sequence of spatial features is stacked along a new axis, representing physical time, and forwarded to the temporal fusion network, a simple TCN composed of several 1D convolutional layers.
In this network, temporal correlations between the spatial features $\mathcal{I}_i$ are extracted by performing multiple consecutive 1D convolutional operations on the stacked spatial features along the time axis.
In practice, a sequence of $N=5$ images and two 1D convolutional layers is optimal. 
%Furthermore, we employ weight sharing for all parameters in the stacked hourglass network.
The extracted spatio-temporal features $\mathcal{F}_i(x,y,t)$ along with their corresponding dimensionless spatial coordinates $x^*,y^*$ and temporal coordinate $t^*$ are forwarded to the MLP.
To preserve sharp spatial details of the current time step, the pixel-aligned features from the central image $\mathcal{I}_i(x,y)$ are also provided as input to the MLP.
Additionally, the spatial coordinate $z^*$ is given as input to the MLP.
On this basis, the MLP predicts the phase distribution $\phi$, the three components of the dimensionless velocity vector $\vf{u}^*=(u^*,v^*,w^*)^T$, as well as the dimensionless pressure $p^*$ at $x^*,y^*,z^*,t^*$.
The residuals of the single-field two-phase Navier-Stokes, continuity, and interface evolution equations are computed using automatic differentiation applied to the predicted output with respect to the spatio-temporal input coordinates.
The physics-informed losses are defined as the mean squared error (MSE) of the respective residuals, \emph{i.e.}, the continuity equation ($\loss_\text{Conti}$), the advection equation governing interface motion ($\loss_\text{Adv}$) and the momentum equations ($\loss_{\text{NSE},j}$), with $j=(x,y,z)$.
Additionally, data loss terms $\loss_\text{Data}$ for the predicted quantities $\phi,u^*,v^*,w^*,p^*$ are computed as the MSE between the predictions and ground truth obtained by numerical simulations.
The joint neural networks are trained on a composite loss representing the weighted sum of the physics-informed and data loss terms.
Note that the architecture of the \IcPINNs{} is a simplification of \VcPINNs{}, without the temporal fusion network, in which only the spatial features $\mathcal{I}_i$ of the central frame are used to condition the PINNs.
We use layer-wise adaptive activation functions~\cite{Jagtap2020} in the MLP, as recent studies have demonstrated their ability to enhance the convergence rate and accuracy of PINNs applied to two-phase flows~\cite{Buhendwa2021}.
%Furthermore, the output for the phase distribution $\phi$ is confined to prediction within the physical bounds $\alpha \in [0,1]$ or $C \in [-1,1]$ through a sigmoid activation function to avoid unphysical overshoots.
To address potential accuracy and memory issues arising from the fourth-order derivatives in the convective Cahn-Hilliard equation, we employ the Deep Mixed Residual Method (MIM)~\cite{Lyu2022} for the \PFPINNs{}.
%Specifically, the chemical potential $\psi$ is predicted as an auxiliary variable and constrained by an additional identity loss term to bypass the calculation of $\psi$ through AD and reduce the problem to second-order PDEs.
%In the \emph{PF-PINNs}, the capillary width $\epsilon$ is treated as a learnable parameter, allowing the PINNs to determine its optimal value dynamically.

\subsection*{\textbf{Sequential training}}

The training of the \VoFPINNs{} and \PFPINNs{} involves the simultaneous minimization of ten or eleven loss terms, respectively, which results in a complicated optimization process.
Furthermore, accurate prediction of the phase distribution $\phi$ is critical for correctly solving the momentum equation, as the mixture density $\rho=f(\phi)$ significantly impacts the momentum balance~\cite{Buhendwa2021}.
Therefore, accurate convergence of the phase distribution prediction has to be ensured before introducing the physics-informed losses.
We address this through a three-stage sequential training process.
Initially, the data loss terms for $u^*$, $v^*$, $w^*$, $p^*$, and the physics-informed loss terms are gradually weighted during the first $10,000$ iterations, with a factor scaling linearly from zero to one to prioritize the learning of an accurate interface prediction in the early stages of training.
In the second stage, the physics-informed losses are only partially introduced to learn an initial approximation of the two-phase flow dynamics from the simulation data.
After this data-guided initialization, meaningful physics-informed losses can be calculated that provide physical regularization during further optimization of the network in the third training stage.
Here, the weights for the physics-informed losses are increased to fully integrate the governing equations, allowing the network to be trained with a comprehensive consideration of the underlying physics of two-phase flows.

\subsection*{\textbf{Sampling methods}}

Due to steep gradients of the flow fields near the gas-liquid interface, adaptive refinement of the sampling points for loss computation at the interface is essential for the successful application of PINNs to two-phase flow problems~\cite{Buhendwa2021, Chen2024}.
We employ a combination of adaptive and random sampling to accurately capture the physics of the two-phase flow across the entire domain.
Adaptive sampling points are drawn from a normal distribution centered on the interface, with a standard deviation of $5\%$ of the initial droplet diameter $D_0 = 2.1$\,mm.
To further guide the optimization of the PINNs towards regions in the domain that are challenging to optimize, the residual points are additionally weighted by a factor $\lambda_i$ ranging from $0.8$ to $1.2$ based on the relative magnitude of the residual, calculated as
\begin{equation}
    \lambda_i = 0.8 + 0.4 \frac{\left|r_i\right|}{\text{max}_i(\left|r_i\right|)},
\end{equation}
where $r_i$ is the residual at sampling point $i$.

\section*{Acknowledgments}
This work was performed on the HoreKa supercomputer funded by the Ministry of Science, Research and the Arts Baden-Württemberg and by the Federal Ministry of Education and Research.
We gratefully acknowledge the financial support from the Karlsruhe House of Young Scientists (KHYS), which contributed to funding the research visit of Maximilian Dreisbach to the Crunch Group led by Prof. George Em Karniadakis that initiated the collaboration in this work. In addition, research supported as part of the AIM for Composites, an Energy Frontier Research Center funded by the U.S. Department of Energy (DOE), Office of Science, Basic Energy Sciences (BES), under award \#DE-SC0023389 (computational studies, data analysis). Elham Kiyani and George Em Karniadakis acknowledge support by DOE BES grant \#DE-SC0023389 (computational studies, data analysis) and the support from ONR MURI (Machine learning Enabled Two-pHase flow metrologies, models, and Optimized DesignS (METHODS)) \#N000142412545.

\section*{Author contributions}
\noindent
Maximilian Dreisbach: Conceptualization, Writing - Original Draft, Investigation, Experiments, Visualization, Software, Validation\\
Elham Kiyani: Software, Writing - Review \& Editing\\
Jochen Kriegseis: Writing - Review \& Editing, Project administration, Resources\\
George Em Karniadakis: Writing - Review \& Editing, Specialist contact person for physics-informed machine learning\\
Alexander Stroh: Conceptualization, Writing - Review \& Editing, Project administration, Resources\\

\section*{Data and materials availability}

All data needed to evaluate the conclusions in the paper are present in the paper and/or the Supplementary Materials.
Additionally, the data used to generate the results will be made publicly available.
The code for the evaluation and training of the PINNs is publicly available at https://github.com/MaxDreisbach/PINNs4Drops.

%\appendix
%\section{Appendix}

\clearpage
\bibliographystyle{abbrvnat}
\bibliography{cas-refs}

%%%%%%%%%%%%%%%% END OF MAIN TEXT %%%%%%%%%%%%%%%

\clearpage
\newpage

%%%%%%%%%%%%%%%% START OF SUPPLEMENT %%%%%%%%%%%%%%%

% Figures, tables, equations and pages in the supplement are numbered S1, S2 etc.
\renewcommand{\thefigure}{S\arabic{figure}}
\renewcommand{\thetable}{S\arabic{table}}
\renewcommand{\theequation}{S\arabic{equation}}
\renewcommand{\thepage}{S\arabic{page}}
\setcounter{figure}{0}
\setcounter{table}{0}
\setcounter{equation}{0}
\setcounter{page}{1} % not 0 as \newpage already started a supplementary page
% References continue the numbering from the main text.

%%%%%%%%%%%%%%%% SUPPLEMENT TITLE PAGE %%%%%%%%%%%%%%%

\begin{center}
\section*{Supplementary Materials for PINNs4Drops:\\ Video-conditioned physics-informed neural networks \\for two-phase flow reconstruction}

% Author list for the supplement
% Indicate the corresponding authors, but do NOT include institutions here
% It would be nice if the template auto-generated this, but doing so is complicated...
Maximilian Dreisbach$^{\ast}$,
Elham Kiyani,
Jochen Kriegseis,\\
George Em Karniadakis,
Alexander Stroh\\
%First~Author$^{\ast\dagger}$,
%A.~Scientist$^\dagger$,
%Someone~E.~Else\\ % we're not in a \author{} environment this time, so use \\ for a new line
\small$^\ast$Corresponding author. Email: maximilian.dreisbach@kit.edu\\
%\small$^\dagger$These authors contributed equally to this work.
\end{center}

% Fill out the numbers for each type of supplementary material,
% and delete any lines that aren't applicable.
% These are just example numbers that don't match the rest of this template.
\subsubsection*{This PDF file includes:}
\noindent
Supplementary Materials and Methods\\
Supplementary Text\\
Figures S1 to S13\\
Tables S1 to S7\\
%Captions for Movies S1 to S7
%Captions for Data S1 to S2

%\subsubsection*{Other Supplementary Materials for this manuscript:}
%\noindent 
%Movies S1 to S4
%Data S1 to S2

%\newpage

%%%%%%%%%%%%%%%% MATERIALS AND METHODS %%%%%%%%%%%%%%%

%\subsection*{\textbf{Supplementary Materials and Methods}}

\paragraph*{\textbf{Data generation}}
\label{sup:methods:data}

\begin{figure}[h]
    \centering
    \includegraphics[width=1.0\linewidth]{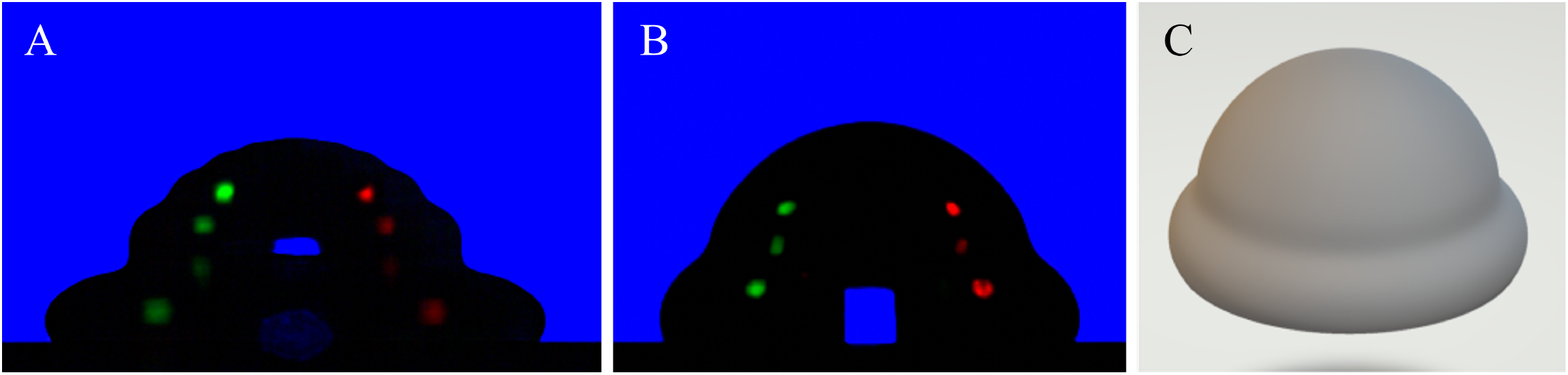}
    \caption{\textbf{Comparison of glare-point shadowgraphy recording and synthetic image.} (\textbf{A}) Recording of an impinging droplet obtained by glare-point shadowgraphy experiments and (\textbf{B}) synthetic image generated through physics-based rendering from (\textbf{C}) a mesh of the gas-liquid interface that was extracted from the numerical simulation.}
    \label{fig:images_exp_synth}
\end{figure}

The glare-point shadowgraphy experiments~\cite{Dreisbach2023} involved the impingement of droplets on two solid substrates with different wetting properties.
Specifically, we investigated the impingement of water droplets with an equivalent diameter of $D_0=2.2-2.3$\,mm on structured Polydimethylsiloxane (PDMS) and polylactide (PLA) substrates.
The surface of the PDMS substrate forms regular square grooves with a height, width, and spacing of $60$\,\textmu m and matches the substrate in the simulation.
This surface structure further enhances the hydrophobicity of the PDMS substrate~\cite{Wenzel1936}, resulting in an equilibrium contact angle of $\theta_{eq,p} = 107^\circ$ in the parallel direction to the grooves and $\theta_{eq,t} = 97^\circ$ in the transversal direction~\cite{Dreisbach2024a}. 
The droplets impacted the PDMS surface at velocities of $U_0 = 0.78-0.88$\,m/s, resulting in the initial spreading of the droplet over the substrate, followed by a pronounced retraction phase that involved the formation of a liquid jet and, finally, the deposition of the droplet~\cite{Rioboo2001}.
The 3D-printed PLA substrate, produced by fused deposition modeling (FDM), features circular arc-shaped ridges with a peak-to-peak spacing of $154$\,\textmu m.
The PLA substrate is hydrophilic, with an equilibrium contact angle of $\theta_{eq,p} = 76^\circ$ in the parallel direction and $\theta_{eq,t} = 63^\circ$ in the transversal direction~\cite{Dreisbach2024a}. 
The water droplets impacted the PLA surface at significantly lower velocities of $U_0 = 0.43-0.44$\,m/s, leading to a deposition of the droplets involving substantially less deformation compared to impingement on the PDMS substrate.
It should be noted that the droplet dynamics in both experiments differed significantly from the results of the numerical simulation, as the initial kinetic energy of the impinging droplets in the experiments ranged from $0.6$ to $2.5 \times E_{\rm{k},0}$ in the simulation.
Furthermore, the different wettability of the substrates influenced the droplet dynamics in these experiments, as the hydrophobic PDMS encouraged the retraction of the droplets, while the hydrophilic PLA encouraged the spreading of the droplets.
The resulting variety in droplet dynamics allows us to evaluate the generalization capability of the developed PINNs.
Experiments on both substrates were conducted at observation angles of $\omega=0^\circ$, $\omega=45^\circ$, and $\omega=90^\circ$ relative to the surface structure.
The impinging droplet is captured by a \emph{Photron Nova R2} fitted with a \emph{Schneider-Kreuznach Apo-Componon} $4.0/60$ enlarging lens, operating at $f=7,500$ frames per second (fps) with a resolution of $1,280\text{\,px} \times 512\text{\,px}$.
The recordings were cut to a resolution of $512\text{\,px} \times 512\text{\,px}$ and resized to match the magnification of the training dataset.
A more in-depth discussion of the glare-point shadowgraphy technique can be found in previous work of the authors~\cite{Dreisbach2023, Dreisbach2025b}.

%Numerical simulation provides suitable ground truth data of the gas-liquid interface, as well as the velocity and pressure fields in both phases, for the supervision of the network optimization through the data loss terms.
We use data from the 3D direct numerical simulations conducted by \citet{fink2018} within the framework of the phase-field method~\cite{Cai2015, Jamshidi2019} implemented in the open-source computational fluid dynamics (CFD) software \emph{OpenFOAM}\textsuperscript{\textregistered}~\cite{Weller1998}, for the training and validation of the developed PINNs.
These simulations featured droplet impingement on flat and structured hydrophobic Polydimethylsiloxane (PDMS) substrates.
The surface structure comprised of repeating regular square grooves with a width, height, and spacing of $60$\,\textmu m and matched the experiments involving the PDMS substrate.
A uniform grid with $18$ million mesh cells was used to simulate a quarter of the domain, exploiting symmetries.
The average numerical time step was of order $0.01\mu$s, and a total of $117$\,ms were simulated.
%- resolution: 18 million mesh cells, $1.9 R_0$ x $1.7 R_0$ x $4.8 R_0$ (quarter of domain exploiting symmetries)
To obtain suitable data for training and validation of the PINNs, the simulated 3D velocity and pressure fields are non-dimensionalized using characteristic quantities (introduced in the following subsection), and gas-liquid interface geometries are extracted from the phase distribution as isosurfaces of the order parameter at $C=0$.
Based on the extracted interface geometries, synthetic glare-point shadowgraphy images are generated through physics-based ray tracing.
For this purpose, the optical setup of the experiments is accurately reproduced in the rendering environment \emph{Blender}~\cite{Blender} with the \emph{LuxCore}~\cite{LuxCore} package, which enables a physically accurate optical simulation~\cite{Dreisbach2025b}.
As illustrated in Figure~\ref{fig:images_exp_synth}, this rendering accurately reproduces the optical phenomena involved in the formation of the glare points and the shadowgraph, and thus ensures a minimal domain gap, required for successful training of the neural network with synthetic data~\cite{Csurka2017,Shrivastava2017}.
The synthetic images were generated by rotating the droplet geometries in the rendering setup in $10^\circ$ increments of the observation angle $\omega$ for a total of $360^\circ$.
Consequently, the dataset consists of $53,244$ synthetic images associated with $1,479$ ground truth droplet shapes, of which $1,015$ stem from the simulation involving the structured substrate and $465$ from the simulation with the flat substrate.
The dataset is split by a ratio of $70\%/10\%/20\%$ into training and separate validation and testing subsets.
The full training dataset comprising the two numerical simulations involving droplet impingement on the flat and structured PDMS substrate is used to train the \IcPINNs{}, while the \VcPINNs{} are trained only on the numerical data of droplet impingement on the structured substrate.

\paragraph*{\textbf{VoF \& phase-field PINNs}}
\label{sup:methods:VoF_PF}

The fluid dynamics of the two-phase flow during droplet impingement are described by the single-field two-phase formulation of the incompressible Navier-Stokes equations, the continuity equation, and an equation for the interface evolution based either on the Volume of Fluid (VoF) or the phase-field (PF) method.
The dimensionless formulation of the Navier-Stokes equations is defined as follows:
\begin{equation}
    \begin{aligned}
    \rho^* \left(\frac{\partial \vf{u}^*}{\partial t^*} + \left(\vf{u}^* \cdot \nabla^* \right) \vf{u}^* \right) =
    - \nabla^* p^* + \nabla^* \cdot \left( \frac{1}{\mathit{Re}} \left(\nabla^*\vf{u}^* + \nabla^*\vf{u}^{*T}\right) \right) \\
    + \frac{1}{\mathit{We}} \frac{\vf{f}_{\sigma}}{\sigma } + \rho^* \frac{1}{\vf{Fr}^2},
    \end{aligned}
    \label{eq:dimensionless_NSE}
\end{equation}
with the following relationships between dimensional and non-dimensional quantities for the velocity $\vf{u} = \vf{u}^* u_R$, the pressure $p = p^* \rho_{\rm{R}} u_{\rm{R}}^2$, the density $\rho = \rho^* \rho_{\rm{R}}$, the spatial coordinates $\vf{x} = \vf{x}^* L_{\rm{R}}$, and the time $t = t^* L_{\rm{R}}/u_{\rm{R}}$, as well as the Reynolds number $\mathit{Re} = \rho_{\rm{R}} u_{\rm{R}} L_{\rm{R}} / \mu$, Weber number $\mathit{We} = \rho_{\rm{R}} u_{\rm{R}}^2 L_{\rm{R}} /\sigma$, and Froude number $\vf{Fr} = u_{\rm{R}} /\sqrt{\vf{g} L_{\rm{R}}}$.
The dynamic viscosity, surface tension coefficient, and gravitational acceleration are indicated by $\mu$, $\sigma$, and $\vf{g}$, respectively.
The reference quantities are the impact velocity $u_{\rm{R}}=U_0$ for the considered case of droplet impingement, the density of the liquid phase $\rho_{\rm{R}}=\rho_{\rm{l}}$, and the reproduction scale $L_{\rm{R}}=r_p$ of the computational domain to the experiments, which ensures the correct scaling of the interface curvature required for the calculation of surface tension.
Both versions of the PINNs employ the continuity equation for incompressible fluids
\begin{equation}
    \nabla \cdot \vf{u}^* = 0.
    \label{eq:fund:conti}
\end{equation}
%%% VOF-PINNs
Volume-of-Fluid-based Physics-Informed Neural Networks (\DFSVOF{}) represent the interface evolution in two-phase flows using the Volume of Fluid (VoF) method, a well-established approach for capturing fluid interfaces.
The transport equation for the volume fraction $\alpha$ is employed to describe the interface evolution, following the formulation of algebraic VoF approaches~\cite{HirtNichols1981}.
\begin{equation}
    \frac{\partial \alpha}{\partial t} + \left(\vf{u}^* \cdot \nabla \right) \alpha = 0.
    \label{eq:algVOF_advection}
\end{equation}
The volume fraction $\alpha$ indicates whether a computational cell is occupied by the liquid ($\alpha=1$), the gaseous phase ($\alpha=0$), or both ($0<\alpha<1$).
The surface tension $f_{\sigma}$ is modeled using the Continuum Surface Force (CSF) model~\cite{Brackbill1992} as a localized body force within the transition region of finite thickness at the interface
\begin{equation}
    \vf{f}_\sigma = -\sigma \kappa \nabla \alpha,
    \label{eq:CSF}
\end{equation}
where $\kappa$ is the cell-averaged curvature of the interface.
The curvature of the interface is approximated by $\kappa = -\nabla \cdot \vf{n}_i$ with the outwards pointing normal vector of the liquid interface $\vf{n}_i$, which is represented by the gradient of the volume fraction $\vf{n}_i = \frac{\nabla \alpha}{|\nabla \alpha|}$.
The mixture density $\rho$ and viscosity $\mu$ are determined by the arithmetic mean of the fluid properties in both phases
\begin{equation}
    \xi = \alpha \xi_{\rm{l}} + (1-\alpha) \xi_{\rm{g}} \quad \text{with} \quad \xi \in \{\rho,\mu\}.
    \label{eq:arithmetic_mixture}
\end{equation}
%%% PF-PINNs
The phase-field version (\PFPINNs{}) employs the convective Cahn-Hilliard equation~\cite{CahnHilliard1959} to represent interface evolution,
\begin{equation}
    \frac{\partial C}{\partial t} + \left(\vf{u}^* \cdot \nabla \right) C = M \nabla^2 \psi
    \label{eq:CahnHilliard}
\end{equation}
with the mobility parameter $M$ that determines the relaxation time of the interface and the conserved order parameter $C$ that represents both phases and takes the value of $C_{\rm{l}} = 1$ in the liquid phase and $C_{\rm{g}} = -1$ in the gaseous phase.
According to the diffuse-interface theory~\cite{VanDerWaals1979}, the phase separation and diffusion in two-phase flows are driven by the chemical potential at the interface $\psi$, which is derived as the variational derivative of the mixing energy with respect to the order parameter $C$
\begin{equation}
    \psi = \frac{\delta F_{mix}}{\delta C} = \frac{\lambda}{\epsilon^2} C (C^2-1) - \lambda \nabla^2 C,
    \label{eq:chem_pot}
\end{equation}
where $\lambda$ represents the magnitude of the mixing energy and $\epsilon$ is the capillary width, which is proportional to the interface thickness.
Equating the surface energy and the mixing energy in the interface region yields~\cite{Yue2004}
\begin{equation}
    \sigma = \frac{2\sqrt{2}}{3} \frac{\lambda}{\epsilon}.
    \label{eq:PF_sigma_lambda}
\end{equation}
As the surface tension coefficient $\sigma$ can be measured by experiments, Equation~\ref{eq:PF_sigma_lambda} can be used to determine the mixing energy $\lambda$.
The value for $\epsilon$, however, needs to be chosen and is typically defined in relation to the characteristic macroscopic length scale of the flow.
In the phase-field version, the continuum surface tension in the potential form~\cite{Jacqmin1999} is employed
\begin{equation}
    \vf{f}_\sigma = \psi \nabla C.
    \label{eq:PF_surface_tension}
\end{equation}
The mixture density $\rho$ and viscosity $\mu$ are determined as
\begin{equation}
    \xi = \frac{1+C}{2} \xi_{\rm{l}} + \frac{1-C}{2} \xi_{\rm{g}} \quad \text{with} \quad \xi \in \{\rho,\mu\}.
    \label{eq:PF_mixture}
\end{equation}
The representation of the gas-liquid interface as a continuous implicit function through a neural network is more akin to diffuse interface methods than sharp interface methods.
Moreover, the physically sound modeling of the interface dynamics in the phase-field approach renders the method promising for accurate predictions of two-phase flows at high density and viscosity ratios, as demonstrated by \citet{Qiu2022}.
Therefore, the convective Cahn-Hilliard equation (Eq.~\eqref{eq:CahnHilliard}) appears as a highly suitable choice for representing the interface evolution. 
However, computing the physics-informed loss derived from the Cahn-Hilliard equation involves fourth-order derivatives, which, coupled with the complexity of the three-dimensional domain, lead to high computational demands.
Furthermore, the repeated calculation of gradients through automatic differentiation to obtain the fourth-order derivatives leads to the accumulation of potential errors.
In contrast, the Volume of Fluid (VoF) approach offers a simplified alternative by representing the interface evolution through the transport equation for the volume fraction, which is purely convective and thus involves only first-order derivatives.
The successful application of PINNs based on the VoF method to inverse two-phase flow problems, as demonstrated by \citet{Buhendwa2021} and \citet{Jalili2024}, highlights the feasibility and efficiency of this approach for tackling such problems.

\begin{figure*}[h]
    \centering
    \includegraphics[width=1.0\textwidth]{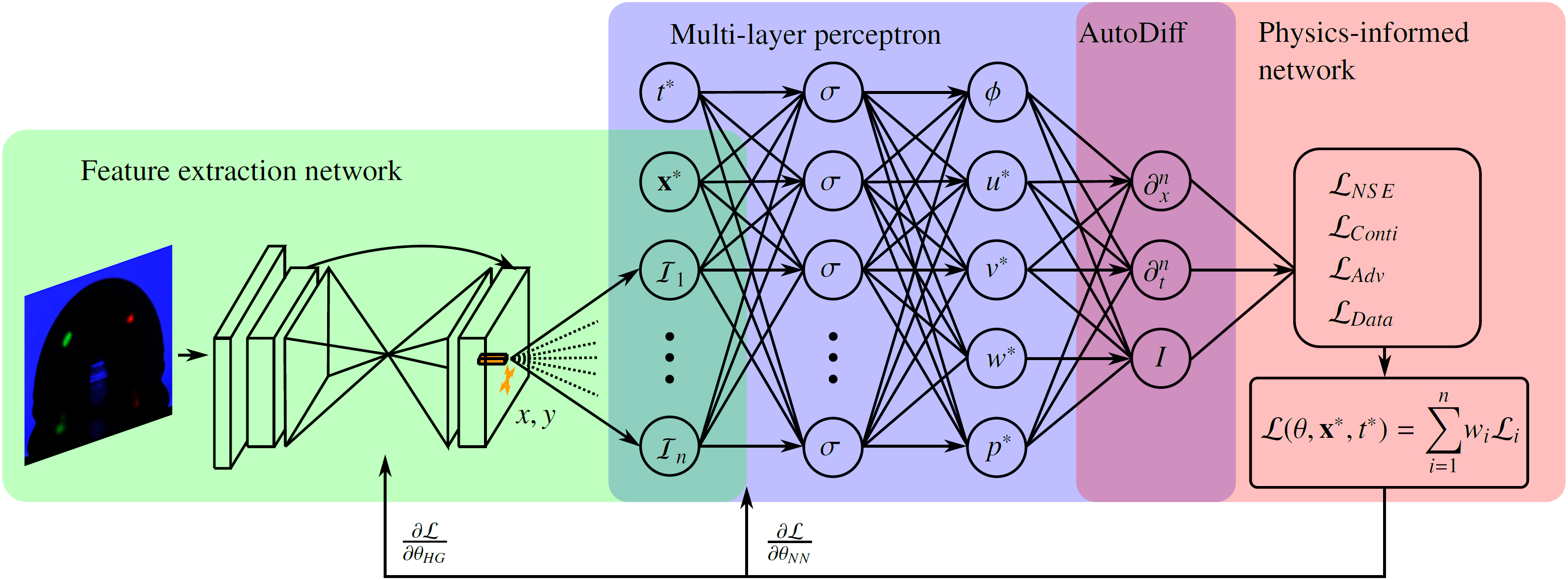}
    \caption{\textbf{Schematics of the image-conditioned PINNs (\IcPINNs{}).} The proposed framework predicts continuous 3D fields of the phase distribution, velocity, and pressure in both phases from 2D snapshots of planar measurement data obtained by glare-point shadowgraphy experiments. The network architecture comprises three main components: a convolutional feature extraction network (CNN), a multi-layer perceptron (MLP), and the physics-informed network. Initially, glare-point shadowgraphy images are processed using a convolutional hourglass network, which extracts pixel-aligned features $\mathcal{I}_i(x,y)$ from the input image at the pixel location $(x, y)$ on the image plane. These extracted features, along with the dimensionless temporal coordinate $t^*$ and the dimensionless spatial coordinates $\vf{x}^*$, are provided as inputs to an MLP. The MLP predicts the phase distribution $\phi$, the three components of the dimensionless velocity vector $\vf{u}^* = (u^*, v^*, w^*)^T$, and the dimensionless pressure $p^*$ at the spatio-temporal coordinates $(x^*, y^*, z^*, t^*)$. The loss function $\loss$ comprises data loss terms $\loss_{\text{Data}}$ for $\phi$, $\vf{u}^*$, and $p^*$, as well as physics-informed loss terms. The physics-informed loss terms enforce the governing equations and are defined as the MSE of the following residuals: $\loss_{\text{Conti}}$, enforcing the continuity equation; $\loss_{\text{Adv}}$, representing the advection equation for phase distribution $\phi$, and $\loss_{\text{NSE},j}$, representing the Navier-Stokes equations for momentum conservation, with $j = (x, y, z)$ indicating the spatial components. The weights $\theta$ of the joint neural networks are updated by minimizing the composite loss $\loss(\theta,\vf{x},t)$ representing the weighted sum of the physics-informed and data loss terms.}
   \label{sec:pinn:IcPINNs_schematics}
\end{figure*}

\paragraph*{\textbf{Image- and video-conditioned PINNs}}
\label{sup:methods:VcPINNs}

Image- and video-conditioned PINNs (\IcPINNs{} and \VcPINNs{}) are parameterized by latent spatio-temporal features extracted through Convolutional Neural Networks (CNNs) and Temporal Convolutional Networks (TCNs).
This parameterization with features extracted from the experimental recordings enables the models to generalize more effectively, allowing them to adapt to different experimental conditions.
Figure~\ref{sec:pinn:IcPINNs_schematics} illustrates the network architecture of the proposed \IcPINNs{}, which consists of three major components, namely, the feature extraction network (a CNN), a multi-layer perceptron (MLP), and the physics-informed network.
The network architecture of \VcPINNs{} is presented in the Manuscript (Figure~\ref{fig:problem_setup}) and is comprised of the same three components as \IcPINNs{}, with an additional temporal fusion network (a TCN) nested between the feature extraction module and the MLP.

The CNN is a stacked hourglass network~\cite{Newell2016} consisting of four consecutive convolutional encoder-decoder pairs (hourglass modules).
In the encoder stage of each hourglass module, successive convolutional layers build a feature pyramid with increasing semantic information, while the spatial resolution of the feature maps is reduced.
In the decoder stage, the spatial resolution is restored through upsampling operations, and semantically rich feature maps are fused with earlier high-resolution feature maps via residual connections~\cite{He2015}.
This process enables the network to integrate information across multiple scales during feature extraction.
Consequently, the hourglass architecture maintains the spatial structure of the input image while enriching the feature maps with contextual information from neighboring pixels.
The repeated down- and upsampling across multiple stacked hourglass modules further facilitates information flow within the feature maps, thereby introducing global context and long-range dependencies between distant regions in the output.

The TCN consists of multiple stacked 1D convolutional layers with a kernel size of $w_{\rm{conv}}=3$.
Different image sequence lengths $N$ and corresponding TCN depths were comparatively evaluated.
Specifically, one convolutional layer was used for a sequence of $N=3$ images, with an additional convolutional layer added for each increase in sequence length ($N=3,5,7$).
Different sequence lengths lead to a trade-off between temporal receptive field and spatial accuracy.
In practice, a sequence of $N=5$ with two convolutional layers yielded the best performance.
Weight sharing is applied across the stacked hourglass network by processing the image sequence as a single batch during feature extraction.
Due to the compact size of the TCN and the parallel processing of the image sequence, the \VcPINNs{} approach adds minimal computational cost compared to \IcPINNs{}.

In the case of \IcPINNs{}, the MLP has $260$ input nodes, with $256$ nodes receiving the spatial features from the hourglass network and four nodes receiving the spatio-temporal coordinates $\vf{x}^*,t^*$.
The MLP implemented in \VcPINNs{} has $516$ input nodes, with $256$ nodes for the spatio-temporal features, $256$ nodes for the spatial features of the central frame, and four nodes for the spatio-temporal coordinates.
The MLP further comprises four hidden layers with $1024, 512, 256, 128$ neurons, respectively, and five output neurons for the prediction of the flow quantities $\phi,u^*,v^*,w^*,p^*$ in case of \VoFPINNs{}, or six output neurons for \PFPINNs{}, when additionally the chemical potential $\psi$ is predicted.
To enable the computation of higher-order derivatives through automatic differentiation, the infinitely differentiable hyperbolic tangent activation function is used as the non-linearity for the hidden layers in the MLP.

Adaptive activation functions~\cite{Jagtap2020} introduce an additional learnable scaling coefficient $n \cdot a$ to the activation function that regulates its slope and thus the sensitivity to its inputs.
The scaling coefficient consists of the fixed scale factor $n$ and the adaptive activation coefficient $a$, which is optimized alongside the parameters of the neural network.
Layer-wise adaptive activation functions have been found to improve the convergence rate and accuracy of PINNs across various types of problems~\cite{Jagtap2020}, including two-phase flows~\cite{Buhendwa2021}.
Therefore, we employ layer-wise adaptive activation functions in the hidden layers of the MLP with a scale factor $n=2$ and an initial value for the adaptive activation coefficient of $a=0.5$.

In the output layer, we use different activation functions for each predicted quantity according to the range of physically possible values.
The sigmoid activation function is used for the output neuron of the phase distribution $\phi$ to confine the prediction within the physical bounds $\alpha \in [0,1]$ or $C \in [-1,1]$ for the volume fraction $\alpha$ in the VoF version and the order parameter $C$ in the phase-field version of the PINNs, respectively.
%The output for the phase distribution $\phi$ is confined to prediction within the physical bounds $\alpha \in [0,1]$ or $C \in [-1,1]$ through a sigmoid activation function to avoid unphysical overshoots.
The prediction of the order parameter $C$ requires an additional scaling of the activation function by a factor of two.
The output neurons for the prediction of the velocity components $u^*,v^*,w^*$ employ linear activation functions, while the output neuron for the prediction of the pressure $p^*$ features an exponential activation function, as suggested by \citet{Buhendwa2021}.
Skip connections~\cite{He2015} are employed at each hidden layer of the MLP to propagate the information of the input feature vector and spatio-temporal coordinates to later layers in the network, which has been shown to improve the accuracy for both data-driven volumetric reconstruction~\cite{Chen2019,Saito2019} and PINNs~\cite{Cheng2021,Wang2024}.

In the \PFPINNs{}, the capillary width $\epsilon$ is treated as a learnable parameter, allowing the PINNs to determine its optimal value dynamically.
To promote the convergence toward a thin interface, we introduce an additional loss term, defined as the Huber loss~\cite{Huber1964} between $\epsilon$ and the value of $\epsilon_{\rm{R}}=2.2 \times 10^{5}$, determined by \citet{fink2018}.
We initialize the capillary width $\epsilon$ with values ranging from $0.01$ to $0.05$ to study the influence of the parameter, and the weight for the additional learnable interface loss is set to $w_\epsilon=100$.
To address potential accuracy and memory issues arising from the fourth-order derivatives in Equation~\ref{eq:CahnHilliard}, we employ the Deep Mixed Residual Method (MIM)~\cite{Lyu2022} for the \PFPINNs{}.
Specifically, for \DFSPFb{}, the chemical potential $\psi$ is directly predicted using the neural network as an additional auxiliary variable, in contrast to the computation of $\psi$ from $C$ in \DFSPFa{}, reducing the highest order of derivatives to the second order.
We introduce an additional loss term derived from the residual of Equation~\ref{eq:chem_pot}, to ensure the consistency between the predicted phase distribution and the distribution of the chemical potential.
This identity loss provides additional supervision for the learning of $\psi$.
The weight for the additional identity loss of the chemical potential is set to $w_\psi=0.1$.

The total number of trainable model parameters for \VcPINNs{} amounted to $16,498,590$ and is distributed in the different modules as follows: $14,420,864$ in the stacked hourglass network, $393.728$ in the TCN, and $1,683,998$ in the MLP.
\IcPINNs{} have $15,612,062$ trainable parameters, with $14,420,864$ in the stacked hourglass network and $1.191.198$ in the MLP.

\paragraph*{\textbf{Sequential training}}
\label{sec:methods:sequential}

%The training of the \VoFPINNs{} and \PFPINNs{} involves the simultaneous minimization of ten or eleven loss terms, respectively, which results in a complicated optimization process.
Previous research has demonstrated the critical role of proper loss weighting for PINNs to achieve the simultaneous convergence of all loss terms~\cite{Buhendwa2021, Cuomo2022, Wang2022a}.
To address this, we use a combination of static loss weights and adaptive loss weighting to balance the data-driven and physics-informed loss terms during training.
For the inverse problem of flow field reconstruction in a flow with natural convection, larger relative weights for the data loss terms compared to the physics-informed losses have been shown to be beneficial~\cite{Cai2021b}.
Furthermore, the accurate prediction of the phase distribution is crucial for correctly solving the momentum equation, as the mixture density $\rho$ significantly affects the momentum balance~\cite{Buhendwa2021}.
Therefore, the optimization process must ensure the convergence of the phase distribution prediction to an accurate level before the physics-informed losses are introduced.
This is addressed through a three-stage sequential training process.
In the first ``warm-up'' stage, the data loss terms for $u^*$, $v^*$, $w^*$, $p^*$, and the physics-informed loss terms are gradually introduced.
%weighted during the first $10,000$ iterations, with a factor scaling linearly from zero to one between training iterations one and $10,000$, to prioritize the learning of an accurate interface prediction in the early stages of training.
In the second ``data-guided'' stage, the weighting coefficients are kept constant at the respective final level reached in the first stage.
Additionally, for \VcPINNs{}, the physics-informed losses are given lower weights relative to the data losses during these first two stages, while for \IcPINNs{}, these weights are set to zero to support the learning of an initial approximation for the velocity and pressure distributions and the accurate interface location from the simulation data.
%After this data-guided initialization of the phase distribution, velocity, and pressure fields, meaningful physics-informed losses can be calculated that provide physical regularization during further optimization of the network.
In the third ``full physics'' stage, the weights for the physics-informed losses are increased to fully integrate the governing equations into the optimization.
%, allowing the network to be trained with a comprehensive consideration of the underlying physics of two-phase flows.
For \VcPINNs{}, the third stage starts after a fixed number of training iterations.
For \IcPINNs{}, the physics-informed losses are introduced dynamically by a threshold of the data loss term for the phase distribution $\mathcal{L}_{\phi,T}$, to further emphasize the learning of an accurate interface.
Specifically, the physics-informed losses were only considered in the total loss whenever $\loss_\phi < \loss_{\phi,T}$, to dynamically introduce the physics of the flow, starting with samples for which the phase distribution was already learned sufficiently well, while retaining purely data-driven training for samples with high uncertainty of the interface location.
A threshold value $\loss_{\phi,T}=0.03$ was determined to provide a sufficient convergence of the phase distribution prediction to obtain adequate physics-informed losses.
We determined optimal weights for achieving the simultaneous convergence of all loss terms experimentally for both \VcPINNs{} and \IcPINNs{}, which are detailed in Tables~\ref{tab:l_w_VcPINNs} and~\ref{tab:l_w_IcPINNs}, respectively.

\begin{table}[ht]
    \centering
    \caption{\textbf{Global static weights for \VcPINNs{} during the three stages of sequential training.} In the first ``warm-up'' stage, the optimization is focused on learning the phase distribution $\phi$ from the simulation data. In the second ``data-guided'' stage, \VcPINNs{} learn an initial approximation for the velocity and pressure distributions from the simulation data. In the third ``full physics'' stage, \VcPINNs{} are trained on the governing equations of the flow by increasing the weights of the physics-informed losses. The identity loss of the chemical potential $\psi$ is only calculated for \PFvcPINNs{}.}
	\begin{tabular}{l|lll}		%\toprule
		stage & loss type & loss term & global weight ($w_i$) \\
        \hline \rule{0pt}{1.0\normalbaselineskip}%
        1 & data & $\phi$ & $1$ \\
        1 & data & $u^*,v^*,w^*,p^*$ & $0.1-100$ \\
        1 & equations & momentum $x,y$ & $10^{-6}-10^{-3}$ \\
        1 & equations & momentum $z$ & $10^{-7}-10^{-4}$ \\
        1 & equations & continuity & $10^{-6}-10^{-3}$ \\
        1 & equations & interface & $10^{-6}-10^{-3}$ \\
        1 & identity & $\psi$ & $10^{-4} - 0.1$ \\
        \hline \rule{0pt}{1.0\normalbaselineskip}%
        2 & data & $\phi$ & $1$ \\
        2 & data & $u^*,v^*,w^*,p^*$ & $100$ \\
        2 & equations & momentum $x,y$ & $10^{-3}$ \\
        2 & equations & momentum $z$ & $10^{-4}$ \\
        2 & equations & continuity & $10^{-3}$ \\
        2 & equations & interface & $10^{-3}$ \\
        2 & identity & $\psi$ & $0.1$ \\
        \hline \rule{0pt}{1.0\normalbaselineskip}%
        3 & data & $\phi$ & $1$ \\
        3 & data & $u^*,v^*,w^*,p^*$ & $100$ \\
        3 & equations & momentum $x,y$ & $0.1$ \\
        3 & equations & momentum $z$ & $0.1$ \\
        3 & equations & continuity & $0.1$ \\
        3 & equations & interface & $0.1$ \\
        3 & identity & $\psi$ & $0.1$ \\
	\end{tabular} 
	\label{tab:l_w_VcPINNs}
\end{table}

\begin{table}[ht]
    \centering
    \caption{\textbf{Global static weights for \IcPINNs{} during the three stages of sequential training.} Same as Table~\ref{tab:l_w_VcPINNs}, but for \IcPINNs{}.}
	\begin{tabular}{l|lll}		%\toprule
		stage & loss type & loss term & global weight ($w_i$) \\
        \hline \rule{0pt}{1.0\normalbaselineskip}%
        1 & data & $\phi$ & $1$ \\
        1 & data & $u^*,v^*,w^*,p^*$ & $0.01-10$ \\
        1 & equations & momentum $x$ & $0$ \\
        1 & equations & momentum $y$ & $0$ \\
        1 & equations & momentum $z$ & $0$ \\
        1 & equations & continuity & $0$ \\
        1 & equations & interface & $0$ \\
        1 & identity & $\psi$ & $0$ \\
        \hline \rule{0pt}{1.0\normalbaselineskip}%
        2 & data & $\phi$ & $1$ \\
        2 & data & $u^*,v^*,w^*,p^*$ & $10$ \\
        2 & equations & momentum $x$ & $0$ \\
        2 & equations & momentum $y$ & $0$ \\
        2 & equations & momentum $z$ & $0$ \\
        2 & equations & continuity & $0$ \\
        2 & equations & interface & $0$ \\
        2 & identity & $\psi$ & $0$ \\
        \hline \rule{0pt}{1.0\normalbaselineskip}%
        3 & data & $\phi$ & $1$ \\
        3 & data & $u^*,v^*,w^*,p^*$ & $10$ \\
        3 & equations & momentum $x$ & $0.01$ \\
        3 & equations & momentum $y$ & $0.01$ \\
        3 & equations & momentum $z$ & $0.01$ \\
        3 & equations & continuity & $0.01$ \\
        3 & equations & interface & $0.01$ \\
        3 & identity & $\psi$ & $0.1$ \\
	\end{tabular} 
	\label{tab:l_w_IcPINNs}
\end{table}

In addition to using static loss weights, we employ the adaptive loss weighting scheme \emph{SoftAdapt}~\cite{Heydari2019} to dynamically balance the loss terms during training.
This simple loss weighting scheme is based on the relative convergence rate of the different loss terms, with inversely proportional loss weights determined dynamically.
\emph{SoftAdapt} significantly improves the training dynamics and enhances the accuracy of the trained model, all while incurring minimal computational costs.
We apply both static and adaptive loss weights in a multiplicative manner.

We jointly train the coupled neural networks on the labeled synthetic dataset by supervised learning with the RMSProp optimization algorithm~\cite{Tieleman2012}, which is an extension of stochastic gradient descent with momentum (SGDM)~\cite{Qian1999}. 
In comparison to the Adam optimizer~\cite{Kingma2015}, which is commonly employed for the optimization of PINNs, the RMSProp optimization algorithm facilitates better network convergence of the proposed PINNs and higher reconstruction accuracy.
To reach a similar number of training iterations for \VcPINNs{} and \IcPINNs{} trained on different partitions of the dataset, \IcPINNs{} were trained for eight epochs on the $37,296 $ training samples of the complete training dataset, and \VcPINNs{} were trained for twelve epochs on the $25,560$ training samples of the partial training dataset representing only the numerical data of droplet impingement on the structured substrate.
We set the initial learning rate to $1 \times 10^{-4}$, and employ learning rate decay.
Specifically, the learning rate of \IcPINNs{} is reduced twice during the training, by a factor of ten at the start of epochs six and eight, and the learning rate of \VcPINNs{} is reduced at the start of epochs eight and eleven.
We employ training data augmentation by random scaling and translation.
A discussion of the training dynamics for \VcPINNs{} and \IcPINNs{} can be found in the Supplementary Text.
As a technical note, the training duration for both \IcPINNs{} and \VcPINNs{} amounted on average to $156$\,h on a single Nvidia RTX A6000 graphics processing unit.
% similar time required for both \IcPINNs{} and \VcPINNs{}, as \VcPINNs{} had a lower number of sampling points, due to memory constraints 
On the same hardware, the PINNs require $9.8$\,s on average per time step for the volumetric reconstruction at an output resolution of $256^3$ grid nodes.
%The inference time is required proportionally for the following processes on average, i.e. $8.2$\,s for the prediction by the neural network, $0.2$\,s for marching cubes, and $1.4$\,s for data handling.

\paragraph*{\textbf{Sampling methods}}
\label{sec:methods:sampling}

Previous studies have demonstrated that an adaptive refinement of the sampling points for the loss computation at the gas-liquid interface is essential for the successful application of PINNs to two-phase flow problems \cite{Buhendwa2021, Chen2024}.
This necessity arises from the steep gradients of the solution near the interface \cite{Mao2020,Lu2021}, which require a higher sampling density to accurately capture the physics in these regions.
To consistently learn the underlying physics of the two-phase flow across the entire domain, we employ a combination of adaptive and random sampling, with different ratios for the individual predicted fields.
For the phase distribution, a $16:1$ ratio of adaptive to random sampling points is used, as this field exhibits a sharp variation at the interface and a uniform distribution in the bulk phases.
The dense adaptive sampling near the interface facilitates the learning of an accurately localized prediction for the gas-liquid interface, while sparse random sampling across the rest of the domain is required to prevent overfitting.
For \VcPINNs{}, velocity and pressure data points are sampled at a $1:1$ ratio, as these fields change more smoothly across the entire domain.
Residual points are also sampled at a $1:1$ ratio to additionally guide the learning of the interface dynamics through a tight coupling of the velocity and phase distribution at the interface.
For \IcPINNs{}, this coupling is further enforced by a higher ratio of $16:1$ for all three sets of sampling points.

Following the adaptive surface sampling method introduced by \citet{Saito2019}, sampling points are randomly drawn from a normal distribution centered on the interface with a standard deviation $\sigma=3.9\%$ of the domain size.
This corresponds to an adaptive sampling at a thickness of $0.107$\,mm near the interface in physical space or $5\%$ of the initial droplet diameter $D_0 = 2.1$\,mm, respectively.
The data loss terms and the physics-informed loss terms are calculated on different sets of sampling points to effectively exploit the continuous nature of PINNs, which allows for the computation of residuals at any sampling point in the spatio-temporal domain.
Thereby, prior knowledge from the numerical simulation is incorporated into the PINNs through supervised learning, while further residual sampling points at different locations in the domain provide supplementary supervision by the physics-informed losses, which ensure that the solution adheres to governing equations.
To ensure sufficient coverage of the spatio-temporal domain, $n_\phi=2,500/5,000$ (\VcPINNs{} / \IcPINNs{}) data points for the phase distribution, $n_{u,v,w,p}=2,500/5,000$ data points for the velocity and pressure, and $n_{phys}=6,500/10,000$ residual points are sampled at each time step.
Moreover, both data and residual points are resampled at every epoch to provide additional coverage of the spatio-temporal domain during the training.

Motivated by the residual-based attention scheme~\cite{Anagnostopoulos2024} and adaptive sampling methods based on the residuals of the PDEs~\cite{Chen2024, Mao2020}, we employ a residual-based weighting of the residual points to further guide the optimization of the PINNs towards regions in the domain that are challenging to optimize.
Such adaptive sampling methods have been shown to improve the prediction of dynamic interfaces \cite{Chen2024} by focusing the optimization of the PINNs on parts of the domain where the governing equations are not fulfilled.
%In the proposed approach, the residual points are additionally weighted by a factor $\lambda_k$ ranging from $0.8$ to $1.2$ based on the relative magnitude of the residual, calculated as
%\begin{equation}
%    \lambda_i = 0.8 + 0.4 \frac{\left|r_i\right|}{\text{max}_i(\left|r_i\right|)},
%\end{equation}
%where $r_i$ is the residual at sampling point $i$.

\paragraph*{\textbf{Evaluation metrics}}
\label{sec:methods:metrics}

%We evaluate the performance of the PINNs considering the reconstructed three-dimensional interface geometries, the velocity and pressure fields, as well as the availability of ground truth data.
We use the following metrics for the evaluation:

\begin{itemize}
    \item The three-dimensional intersection over union 
    \begin{equation}
        \rm{3D{\texttt{-}}IOU} = \frac{R \cap GT}{R \cup GT}\label{eq:3D-IOU}
    \end{equation}
    is calculated as the fraction of the intersection volume between the reconstructed interface $\rm{R}$ and ground truth $\rm{GT}$ and the union volume of $\rm{R}$ and $\rm{GT}$.
    The 3D-IOU provides a measure for the spatial volumetric accuracy of the interface reconstruction in 3D space.
    \item The bias error of the reconstructed volume $\delta_{\rm{V}}$ is calculated by the absolute deviation of the arithmetic mean 
    \begin{equation}
        \overline{V} = \frac{1}{n}\sum_{i=1}^{n} V_{\rm{R},i} \label{eq:Vmean}
    \end{equation}
     of the reconstructed volumes $V_{\rm{R},i}$ from the ground truth volume $V_{\rm{GT}}$, and given relative to the ground truth volume \cite{Bendat2010} 
    \begin{equation}
        \delta_{\rm{V}} = \left|\frac{V_{\rm{GT}} - \overline{V}}{V_{\rm{GT}}}\right|.\label{eq:deltaV}
    \end{equation}
    \item The measured uncertainty of the reconstructed volume $\sigma_{\rm{V}}$ is calculated by the standard deviation of the deviation between the reconstructed volume and the ground truth volume, and given relative to the ground truth volume \cite{Bendat2010}  
    \begin{equation}
        \sigma_{\rm{V}} = \frac{1}{V_{\rm{GT}}} \sqrt{ \frac{1}{n-1}\sum_{i=1}^{n} (V_{\rm{R},i} - \overline{V})^2}.\label{eq:sigmaV}
    \end{equation}
    \item The mean absolute error (MAE) of the quantities $q \in \{u,v,w,p\}$ is calculated as the absolute deviation of the predicted quantities $q_{\rm{pred}}$ from the ground truth values $q_{\rm{GT}}$, averaged over all sampling points $i=0\ldots n$ with coordinates $\vf{x}_i$ in the spatio-temporal domain for one reconstructed time step.
    \begin{equation}
        \text{MAE}_q = \frac{1}{n} \sum_{i=1}^{n} \left| q_{\rm{GT}}(\vf{x}_i) - q_{\rm{pred}}(\vf{x}_i) \right|,
    \end{equation}
    \item The root mean squared error (RMSE) is calculated as
    \begin{equation}
        \text{RMSE}_q = \sqrt{\frac{1}{n} \sum_{i=1}^{n} \left( q_{\rm{GT}}(\vf{x}_i) - q_{\rm{pred}}(\vf{x}_i) \right)^2}.
    \end{equation}
    \item The relative $\rm{L}^1$ and $\rm{L}^2$ error norms are calculated as
    \begin{equation}
        RL^1_q =\frac{\sum_{i=1}^{n}\left|q_{\rm{GT}}(\vf{x}_i) - q_{\rm{pred}}(\vf{x}_i)\right|}{\sum_{i=1}^{n}\left|q_{\rm{GT}}(\vf{x}_i)\right|}, \\
    \end{equation}
    and
    \begin{equation}
        RL^2_q=\frac{\sqrt{\sum_{i=1}^n\left(q_{\rm{GT}}(\vf{x}_i) - q_{\rm{pred}}(\vf{x}_i)\right)^2}}{\sqrt{\sum_{i=1}^n q_{\rm{GT}}(\vf{x}_i)^2}}.
    \end{equation}
\end{itemize}

Before impingement, the shape of the droplets is approximately spheroidal with axisymmetry around the vertical axis, which allows for an accurate estimation of the droplet volume.
For that purpose, the shadowgraph contour is fitted by an ellipse with the horizontal $D_{\rm{h}}$ and the vertical semi-axis $D_{\rm{v}}$, and the volume is then calculated as \mbox{$V_{\rm{GT}} = \frac{\pi}{6} D_{\rm{h}}^2 D_{\rm{v}}$}.
The volume-equivalent spherical droplet diameter is calculated as $D_0 = (D_{\rm{h}}^2 D_{\rm{v}})^\frac{1}{3}$.
The impact velocity $U_0$ was determined by tracking the vertical displacement $\Delta s$ of the fitted ellipse center over $n_{\rm{f}}$ frames and tracking the time difference $\Delta t=n_{\rm{f}}/f$ between the initial and final frame, yielding $U_0 = \Delta s/\Delta t=f\Delta s/n_{\rm{f}}$.
The uncertainty of the volume measurement amounts to $\sigma_{\rm{V,GT}} = 0.06\%$ of $V_{\rm{GT}}$, and the uncertainty of the velocity measurement is estimated to be below $\sigma_u = 0.006$\,m/s.
After impingement, the horizontal locations of the left and right contact lines $CL_{\rm{r/l}}$ are determined from the shadowgraph contour, and the in-plane spreading diameter $\xi_{\rm{in}}$ is calculated as $\xi_{\rm{in}}=\frac{D_0}{D_{\rm{in}}} = \frac{D_0}{CL_{\rm{r}} - CL_{\rm{l}}}$.

Three energy contributions are quantified for the droplet impingement process based on the inferred flow quantities and reconstructed interface geometries:
\begin{align}
    \text{surface energy } & E_{\rm{s}}=\sigma A_{\rm{LG}}+\left(\sigma_{\rm{LS}}-\sigma_{\rm{SG}}\right) A_{\rm{LS}}, \\
    \text{kinetic energy } & E_{\rm{k}}=\int_{\Omega} \rho(\mathbf{u} \cdot \mathbf{u}) \rm{d} V, \\
    \text{gravitational energy } & E_{\rm{g}}=\int_{\Omega} \rho|\mathbf{g}| h \rm{~d} V,
\end{align}
within the domain of the liquid phase $\Omega$ and calculated based on the area of the gas-liquid $A_{\rm{LG}}$ and liquid-solid interfaces $A_{\rm{LS}}$, as well as the interfacial energies at the gas-liquid interface $\sigma$, the liquid-solid interface $\sigma_{\rm{LS}}$, and the solid-gas interface $\sigma_{\rm{SG}}$.
The total energy $E_{\rm{t}}$ is calculated as the sum of $E_{\rm{s}}$, $E_{\rm{k}}$, and $E_{\rm{g}}$.
The accurate measurement of the impact velocity and droplet volume enable a precise estimation of the initial energy contributions $E_{\rm{q}, 0}$ with $q \in \{\rm{k},\rm{s},\rm{g}\}$ and the initial total energy $E_{\rm{t}, 0}$ in the experiments before droplet impingement:
\begin{multline}
E_{\rm{t}, 0}=E_{\rm{s}, 0}+E_{\rm{k}, 0}+E_{\rm{g}, 0}= \\ 
\sigma \pi D_0^2+\frac{1}{2} \rho U_0^2\left(\frac{1}{6} \pi D_0^3\right)+\rho|\rm{g}| \frac{D_0}{2}\left(\frac{1}{6} \pi D_0^3\right).
\end{multline}
%- assuming spherical droplet shape
Using the principle of energy conservation, the viscous energy dissipation $E_{\rm{d}}$ is additionally calculated as 

\begin{equation}
E_{\rm{d}} = E_{\rm{t},0} - E_{\rm{k}} - E_{\rm{s}} - E_{\rm{g}}.
\end{equation}

\subsection*{\textbf{Supplementary Text}}
\label{sup:results}

\paragraph*{\textbf{Training dynamics of \VcPINNs{} optimized for velocity and pressure inference}}
\label{sup:results:VcPINNs_dynamics}

\begin{figure}[h]
    \centering
    \includegraphics[width=1.0\linewidth]{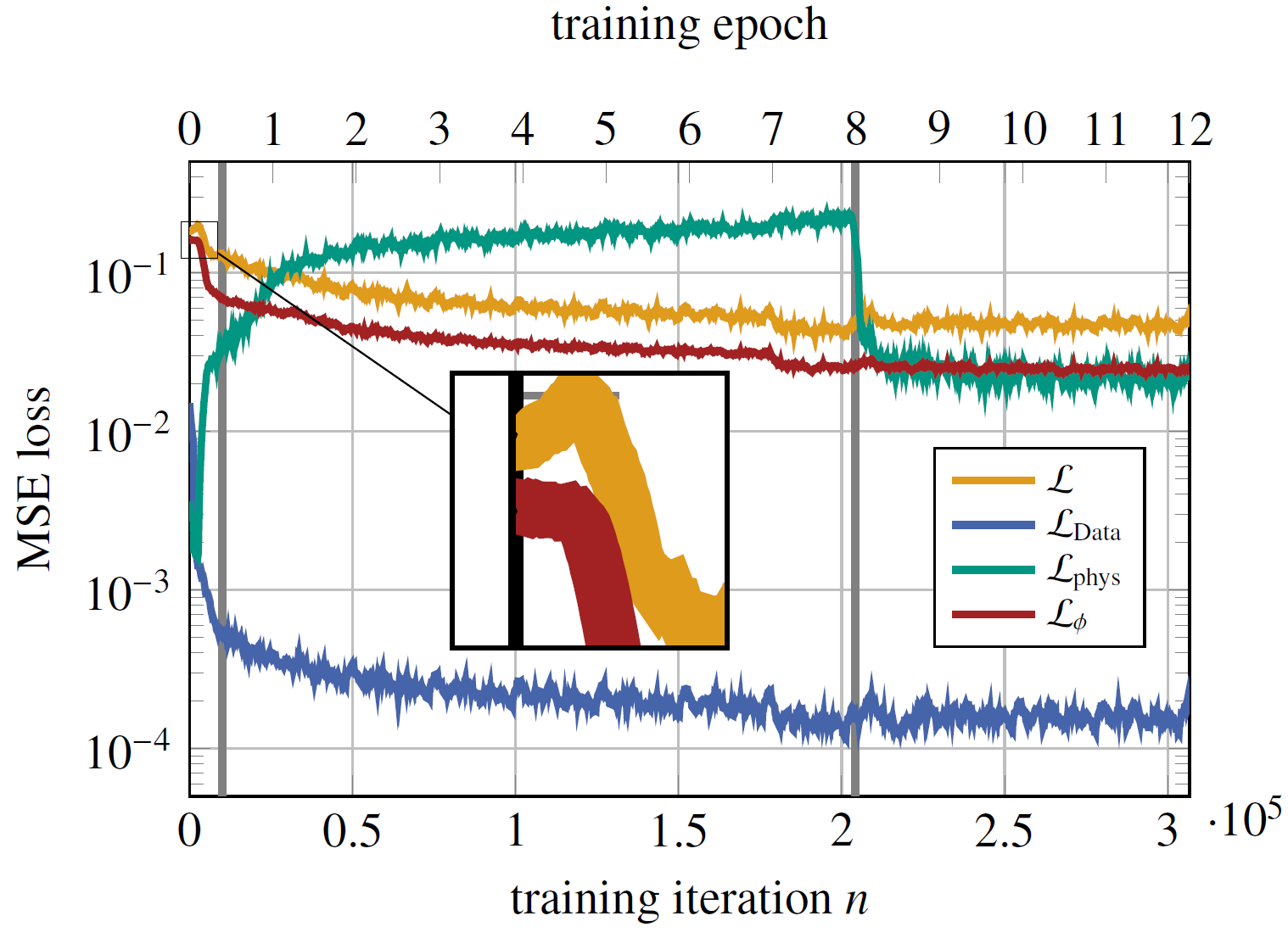}
    \caption{\textbf{Evolution of the loss terms during training of \PFvcPINNs{} optimized for velocity and pressure inference.} The weighted total loss $\loss$, the summed data loss terms (excluding the interface loss) $\loss_\text{Data}$, the summed physics-informed loss terms $\loss_\text{phys}$, and the interface loss term $\loss_\phi$ are indicated by the colored lines. The gray horizontal lines indicate the three stages of sequential training. In particular, the first gray line indicates the end of the ``warm-up'' period for the data and physics-informed loss terms, and the second gray line indicates the increase of the weights for the physics-informed losses to incorporate the full physics. The magnified view highlights the early plateau of $\loss_\phi$, followed by its rapid decrease during the ``warm-up'' period.}
    \label{fig:VcPINNs_train}
\end{figure}

Figure~\ref{fig:VcPINNs_train} shows the evolution of the weighted total loss $\loss$, the unweighted summed data loss terms
\begin{equation}
    \loss_\text{Data} = \loss_u + \loss_v + \loss_v + \loss_p~,
\end{equation}
the unweighted summed physics-informed loss terms 
\begin{equation}
    \loss_\text{phys} = \loss_\text{Conti} + \loss_\text{Adv} + \loss_\text{NSE,x} + \loss_\text{NSE,y} + \loss_\text{NSE,z}~,
\end{equation}
and the data loss for the phase distribution $\loss_\phi$ during the training of the \PFvcPINNs{}, optimized for velocity and pressure inference using the sampling and sequential training schemes detailed in the Materials and Methods.
The \VoFvcPINNs{} show similar dynamics of losses during training.
Moreover, \IcPINNs{} trained under the same conditions have similar training dynamics to the results presented for \VcPINNs{}, demonstrating that the difference in the network architectures of \IcPINNs{} and \VcPINNs{} has a negligible influence on the training dynamics. 
As can be seen in Figure~\ref{fig:VcPINNs_train}, the development of the different loss contributions follows the three stages of the sequential training scheme.
During the initial phase of the training, $\loss_\phi$ only decays slowly, which is followed by a sudden drop that can be attributed to the learning of the gas-liquid interface.
A further drop can be observed at the start of epoch eight, at which the first learning rate decay occurs.
The first drop of $\loss_\phi$ is correlated with a sudden increase in $\loss_\text{phys}$.
Similar training dynamics were already observed by \citet{Buhendwa2021}, who reported that the sudden learning of the interface location leads to a rapid increase in the magnitude of the gradients in the phase distribution at the interface and, in turn, an increase in the magnitude of the surface force term in the momentum equation.
Furthermore, \citet{Buhendwa2021} found that the sudden increase of the physics-informed losses destabilizes the optimization and can cause complete divergence.
The reason is that $\loss_\text{phys}$ offer only limited guidance when the interface is not accurately learned and may even mislead the optimization of the neural network.
This is particularly problematic for the considered water-air flow, where an erroneous prediction of the occupancy field introduces large errors in the momentum equation due to the high fluid density ratio.
In the proposed \VcPINNs{} and \IcPINNs{}, this issue is mitigated by ensuring that the prediction of the gas-liquid interface and the flow topology is sufficiently converged before introducing $\loss_\text{phys}$ to a significant degree.
This is successfully achieved by the sequential training scheme, as indicated by the monotonic decrease of all loss terms.
In particular, the gradual weighting of $\loss_\text{Data}$ and $\loss_\text{phys}$ during the first $10,000$ iterations accelerated the learning of the phase distribution significantly, which was reflected by a pronounced reduction in the early plateau of $\loss_\phi$, as can be seen in the inset in Figure~\ref{fig:VcPINNs_train}.
The data loss terms for the velocity components $u,v,w$ and $p$ are decreasing rapidly during the first stage of training as well, which indicates that the learning of the flow field was not impeded by the early focus of the optimization on the phase distribution.
During the second training stage (see Figure~\ref{fig:VcPINNs_train}, between the first and second gray vertical line), in which the weights of $\loss_\text{phys}$ are still kept relatively low, $\loss_\phi$ and $\loss_\text{Data}$ keep decreasing, while $\loss_\text{phys}$ are slowly increasing.
This indicates that during the second stage, the influence of $\loss_\text{phys}$ is low compared to the data losses.
However, the modest introduction of $\loss_\text{phys}$ during the second stage reduces the potential spike in the total loss at the beginning of the third stage, resulting from the suddenly raised weights of $\loss_\text{phys}$.
As indicated by the second gray line in Figure~\ref{fig:VcPINNs_train}, the higher weights for $\loss_\text{phys}$ in the third stage lead to a sudden decrease of all physics-informed loss terms and a simultaneous but smaller jump in the total loss.
For lower weights of $\loss_\text{phys}$ or purely data-driven training in the second stage, the spike in the total loss leads to unstable training dynamics due to initially high $\loss_\text{phys}$.
Conversely, higher weights during the second stage impaired the convergence of $\loss_\text{Data}$, indicating an optimum at moderately low weights for $\loss_\text{phys}$ during the first and second stages of training.
Consequently, the sequential training scheme, consisting of a data-driven initialization and gradual introduction of $\loss_\text{phys}$, ensures the simultaneous convergence of all loss terms.
Furthermore, the difference in the magnitude of $\loss_\text{Data}$ in comparison to $\loss_\phi$ and $\loss_\text{phys}$ highlights the necessity of loss weighting.
\VoFvcPINNs{} reached a lower final phase distribution loss of $\loss_\phi = 0.021$ compared to \PFvcPINNs{} with $\loss_\phi = 0.0241$, which is also reflected in a higher reconstruction accuracy for the gas-liquid interface from experimental data (see Table~\ref{tab:uncertainty_bias_IcPINNs}).
Conversely, \PFvcPINNs{} reached lower physics-informed losses of $\loss_\text{phys} = 0.0235$ compared to $\loss_\text{phys} = 0.0539$ for \VoFvcPINNs{}, which is likewise reflected in the lower residuals of the governing equations evaluated on the reconstructed fields (see Table~\ref{tab:residuals}).

\paragraph*{\textbf{Training dynamics of \IcPINNs{} optimized for interface reconstruction}}
\label{sup:results:IcPINNs_dynamics}

\begin{figure}[h]
    \centering
    \includegraphics[width=1.0\linewidth]{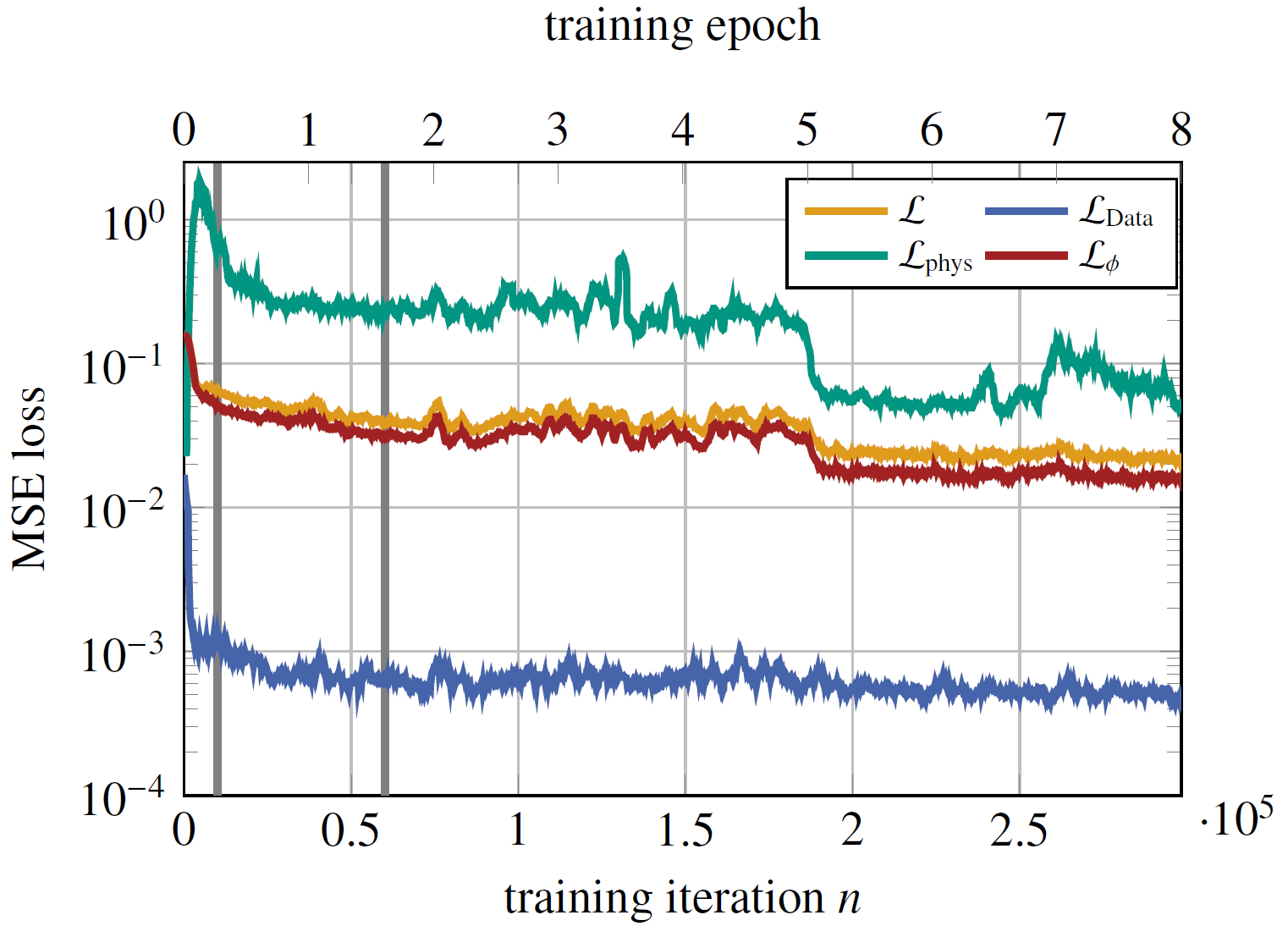}
    \caption{\textbf{Evolution of the loss terms during training of \VoFicPINNs{} optimized for interface reconstruction.} The weighted total loss $\loss$, the summed data loss terms (excluding the interface loss) $\loss_\text{Data}$, the summed physics-informed loss terms $\loss_\text{phys}$, and the interface loss term $\loss_\phi$ are indicated by the colored lines. The gray horizontal lines indicate the three stages of sequential training. In particular, the first gray line indicates the end of the ``warm-up'' period for the data loss terms, and the second gray line indicates the crossing of the moving average of $\loss_\phi$ below the threshold value $\loss_{\phi,T}=0.03$ that determines the consideration of the physics-informed losses.}
    \label{fig:IcPINNs_train}
\end{figure}

Figure~\ref{fig:IcPINNs_train} shows the evolution of the weighted total loss $\loss$, and the unweighted loss terms $\loss_\text{Data}$, $\loss_\text{phys}$, and $\loss_\phi$ during the training of the \VoFicPINNs{}, optimized for interface reconstruction with the sampling and sequential training schemes detailed in the Materials and Methods.
Both phase-field variants of \IcPINNs{} -- \DFSPFa{} and \DFSPFb{} -- show similar dynamics of losses during training.
As can be seen, the dynamics of $\loss_\text{Data}$ and $\loss_\phi$ behave similarly to the \VcPINNs{}, optimized for velocity and pressure inference, but distinct differences in the dynamics of $\loss_\text{phys}$ occur.
As the \IcPINNs{} are optimized for interface reconstruction, more focus was placed on the learning of the phase distribution.
This was achieved by purely data-driven training until the loss for the phase distribution reaches the threshold value $\loss_{\phi,T}=0.03$.
%Particularly, $\loss_\text{phys}$ were only considered in the total loss whenever $\loss_\phi < \loss_{\phi,T}$, to dynamically introduce the physics of the flow, starting with samples for which the phase distribution was already learned sufficiently well, while retaining purely data-driven training for samples with high uncertainty of the interface location.
This dynamical introduction of $\loss_\text{phys}$ effectively started briefly after the ``warm-up'' period for $\loss_\text{Data}$ and $\loss_\text{phys}$, indicated by the first gray line in Figure~\ref{fig:IcPINNs_train} and applied to the majority of the weight updates around the time at which the moving average of $\loss_\phi$ crossed $\loss_{\phi,T}$, which is indicated by the second gray line.
Consequently, $\loss_\text{phys}$ are already declining in this second stage of sequential training of \IcPINNs{} optimized for interface reconstruction, in contrast to \VcPINNs{} optimized for velocity and pressure inference.
Moreover, the \IcPINNs{} converge to lower values of $\loss_\phi$ but remain at higher values of $\loss_\text{Data}$ in comparison to \VcPINNs{}, which likely results from the focus of the optimization on learning an accurate interface location, encouraged by lower static weights for $\loss_\text{Data}$ and a more concentrated distribution of sampling points at the interface.
Note that \IcPINNs{} and \VcPINNs{} were trained using different sampling schemes, which likely leads to differences in the absolute magnitudes of the loss terms.

%\paragraph{Influence of Mixed residual precision}
Both phase-field versions of the \IcPINNs{} converged to similarly low losses in comparison to the VoF version, except for $\loss_\phi$.
Specifically, \DFSPFa{} with an initial value of $\epsilon_0=0.01$, remained at a considerably higher $\loss_\phi$ in comparison to \DFSVOF{} with around $3.9 \times \loss_{\phi,\text{VoF}}$, while \DFSPFb{} with $\epsilon_0=0.01$ reached $1.7 \times \loss_{\phi,\text{VoF}}$.
These results suggest that the proposed VoF approach is more appropriate for training PINNs aimed at the accurate reconstruction of the gas-liquid interface in the considered two-phase droplet flows.
Furthermore, the lower $\loss_\phi$ achieved by \DFSPFb{} trained with Mixed residual precision \cite{Lyu2022} indicates that the application of MIM by the separate prediction of the chemical potential is beneficial for learning the phase distribution in phase-field PINNs.
The cause for the better performance of \DFSPFb{} might be an improvement of the training dynamics by avoiding the fourth-order derivative in the Cahn-Hilliard equation.
The learnable interface thickness of \DFSPFb{} remains close to the initial values of $\epsilon_0=0.01$ and $\epsilon_0=0.05$ throughout the training and only marginally decreases from $\epsilon=0.01$ to $\epsilon=0.0093$ for \DFSPFa{} towards the end of the training.
It was found that a higher weighting of the identity loss for the chemical potential term leads to lower values of $\epsilon$; however, at the cost of significantly reduced convergence for the other loss terms, which resulted in a degraded accuracy of the interface reconstruction.
Similarly, an increased initial value of $\epsilon_0$ from $0.01$ to $0.05$ led to a significantly improved convergence of $\loss_\phi$ for \DFSPFb{}, reaching similarly low values of $\loss_\phi$ at the end of the training in comparison to \DFSVOF{}.
Consequently, a more diffuse interface of the phase-field PINNs was found to be beneficial for their optimization.

\paragraph*{\textbf{Validation of interface reconstruction on synthetic data}}
\label{sup:results:interface_val}

\begin{figure}[h]
    \centering
    \includegraphics[width=1.0\linewidth]{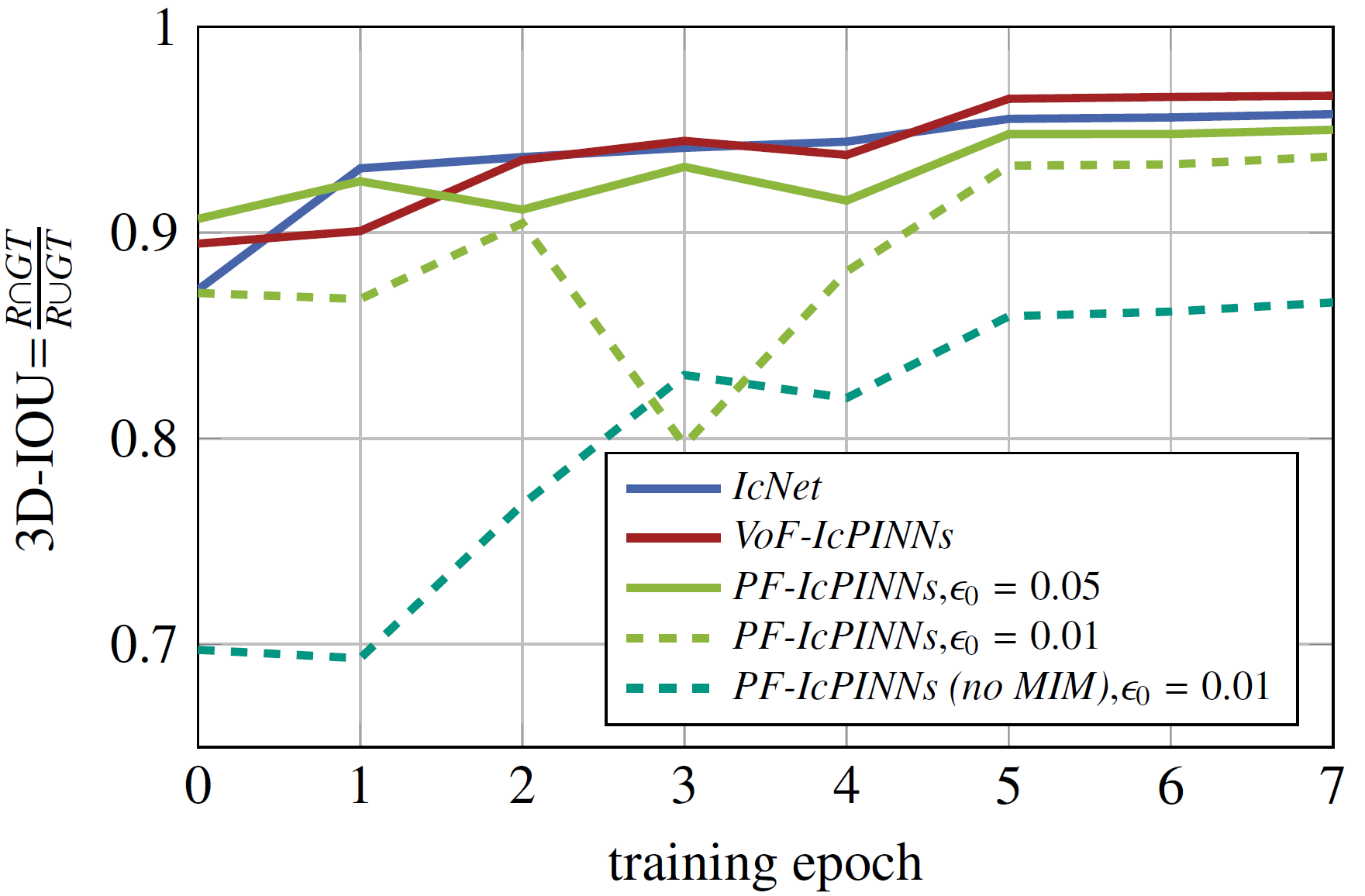}
    \caption{\textbf{Volumetric interface reconstruction accuracy of all \IcPINNs{} variants during training.} We measure the volumetric accuracy by the average 3D-IOU on the validation dataset at the end of each training epoch and compare the \IcPINNs{} variants with the data-driven baseline \DFSC{}. The comparison includes \DFSVOF{} and phase-field PINNs with (\DFSPFb{}) and without mixed residual precision (\DFSPFa{}), as well as different initial values of the capillary width $\epsilon_0$.}
    \label{fig:PINNs:3D-IOU_full}
\end{figure}

Figure~\ref{fig:PINNs:3D-IOU_full} shows the evolution of the 3D-IOU on the synthetic validation data during the training of the \IcPINNs{} in comparison to the purely data-driven \DFSC{} network. 
As can be seen, the different training dynamics between the VoF variant and the two phase-field versions of the \IcPINNs{} are reflected in the reconstruction accuracy of the gas-liquid interface.
The accuracy of \DFSPFb{} with $\epsilon_0=0.01$ sharply drops between training epochs two and three, correlating with the convergence of $\loss_\phi$ below $\loss_{\phi,T}=0.03$, which marks the point in training at which the physics-informed losses are applied to the majority of weight updates.
A potential cause might be the onset of the identity loss for the chemical potential that couples the phase distribution to the hidden field chemical potential, which at that point in the training is still in the condition of random initialization.
Consequently, the randomness of distribution and magnitude of the chemical potential might introduce an erroneous objective for the phase distribution through the identity loss, as long as the prediction for the chemical potential is not yet sufficiently converged.
An increase of the capillary width from $\epsilon_0=0.01$ to $\epsilon_0=0.05$ helps to mitigate this issue, leading to a substantial improvement in reconstruction accuracy, as shown in Figure~\ref{fig:PINNs:3D-IOU_full}.
By means of an ablation study for the residual-based weighting of the sampling points for the calculation of the physics-informed losses and the layer-wise adaptive activation functions, we found that both measures only marginally improved the accuracy of the interface prediction.
Consequently, the gains in reconstruction accuracy can predominantly be attributed to the introduction of physics-informed losses to the reconstruction framework.

\paragraph*{\textbf{Interface reconstruction from experimental data}}
\label{sup:results:interface_exp}

\begin{table}[ht]
    \centering
    \caption{\textbf{Uncertainty and bias error of the droplet volume reconstructed from experiments by \IcPINNs{}.} Shown are the uncertainty~$\sigma_{\rm{V}}$ and bias error~$\delta_{\rm{V}}$ of the reconstructed integral volume for \DFSC{}, \DFSVOF{}, and \DFSPFb{} with an initial capillary width of~$\epsilon_0 = 0.05$, expressed as percentages of the ground-truth volume for droplet impingement experiments on PLA and PDMS substrates observed at different viewing angles.}
	\begin{tabular}{l|rrrrrrrrr}%\toprule
		case & \multicolumn{2}{c}{\DFSC{}} & \multicolumn{2}{c}{\DFSVOF{}} & \multicolumn{2}{c}{\PFicPINNs{}} \\ 
        $\sigma_{\rm{V}}$, $\delta_{\rm{V}}$ (in \%) & $\sigma_{\rm{V}}$ & $\delta_{\rm{V}}$ & $\sigma_{\rm{V}}$ & $\delta_{\rm{V}}$ & $\sigma_{\rm{V}}$ & $\delta_{\rm{V}}$ \\%
        \hline \rule{0pt}{1.0\normalbaselineskip}%
        PLA $0^\circ$   & 8.2 & 9.1 & 1.7 & 2.9 & 2.5 & 4.7 \\
        PLA $45^\circ$  & 3.5 & 0.7 & 1.3 & 2.6 & 1.4 & 1.8 \\
        PLA $90^\circ$  & 8.2 & 5.4 & 1.8 & 0.7 & 2.3 & 0.3 \\
        PDMS $0^\circ$  & 11.6 & 4.3 & 2.2 & 1.1 & 2.8 & 2.8 \\
        PDMS $45^\circ$ & 2.9 & 5.1 & 1.3 & 2.1 & 1.4 & 3.8 \\
        PDMS $90^\circ$ & 3.9 & 6.1 & 1.6 & 0.1 & 1.9 & 0.9 \\
        average         & 6.5 & 5.1 & 1.6 & 1.6 & 2.0 & 2.4 \\
	\end{tabular} 
	\label{tab:uncertainty_bias_IcPINNs}
\end{table}

The images recorded in experiments involving the impingement of droplets on the structured PLA and PDMS substrates at different observation angles with respect to the orientation of the surface structure are reconstructed by \DFSVOF{} and \DFSPFb{} with an initial capillary width of $\epsilon_0=0.05$.
The uncertainty, as well as bias errors of the integral reconstructed volume, are compared to the results by \DFSC{} and detailed in Table~\ref{tab:uncertainty_bias_IcPINNs}.
The \IcPINNs{}, optimized for interface reconstruction, consistently reduce the uncertainty and bias errors of the reconstruction across droplet impingement experiments with substantially different gas-liquid interface dynamics, resulting from the varying initial kinematic conditions and different wettability of the substrates.
Particularly, \DFSVOF{} demonstrate a significant improvement in the reconstruction accuracy over \DFSC{}.

\begin{table}[ht]
    \centering
    \caption{\textbf{Uncertainty and bias error of the droplet volume reconstructed from experiments by \VcPINNs{}.} Shown are the uncertainty~$\sigma_{\rm{V}}$ and bias error~$\delta_{\rm{V}}$ of the reconstructed integral volume for \RefvcPINNs{}, \VoFvcPINNs{}, and \PFvcPINNs{}, expressed as percentages of the ground-truth volume for droplet impingement experiments on PLA and PDMS substrates observed at different viewing angles.}
	\begin{tabular}{l|rrrrrrrrr}		%\toprule
		case & \multicolumn{2}{c}{\RefvcPINNs{}} & \multicolumn{2}{c}{\VoFvcPINNs{}} & \multicolumn{2}{c}{\PFvcPINNs{}} \\ 
        $\sigma_{\rm{V}}$, $\delta_{\rm{V}}$ (in \%) & $\sigma_{\rm{V}}$ & $\delta_{\rm{V}}$ & $\sigma_{\rm{V}}$ & $\delta_{\rm{V}}$ & $\sigma_{\rm{V}}$ & $\delta_{\rm{V}}$ \\%
        \hline \rule{0pt}{1.0\normalbaselineskip}%
        PLA $0^\circ$   & 7.3 & 4.8 & 7.0 & 1.5 & 9.0 & 5.0 \\
        PLA $45^\circ$  & 3.3 & 17.6 & 3.6 & 16.1 & 3.3 & 17.7 \\
        PLA $90^\circ$  & 7.7 & 11.4 & 7.7 & 9.8 & 9.4 & 11.6 \\
        PDMS $0^\circ$  & 13.4 & 9.1 & 13.5 & 7.0 & 13.2 & 8.4 \\
        PDMS $45^\circ$ & 3.5 & 12.2 & 4.7 & 13.5 & 4.4 & 15.1 \\
        PDMS $90^\circ$ & 6.4 & 10.1 & 6.2 & 8.1 & 6.3 & 8.9 \\
        average         & 6.9 & 10.9 & 7.1 & 9.3 & 7.6 & 11.1 \\
	\end{tabular} 
	\label{tab:uncertainty_bias_VcPINNs}
\end{table}

Table~\ref{tab:uncertainty_bias_VcPINNs} compares the reconstruction accuracy of \VcPINNs{}.
As can be seen, all \VcPINNs{} models, optimized for velocity and pressure inference, yield a lower interface reconstruction accuracy compared to \IcPINNs{}, which were optimized for interface reconstruction.
For comparison, we train \IcPINNs{} under the same conditions as \VcPINNs{} with the methods described in the Materials and Methods, specifically on the same smaller dataset of one simulation case, more distributed sampling points for the data and physics-informed losses, and the same sequential training scheme used for \VcPINNs{}.
These \IcPINNs{}, optimized for velocity and pressure inference, achieved comparable uncertainty and bias errors to their respective \VcPINNs{} counterparts.
This version of \VoFicPINNs{} reaches an average uncertainty of $\sigma_{\rm{V}}=7.1\%$ and a bias error of $\delta_{\rm{V}}=8.9\%$, and \PFicPINNs{} reach an average uncertainty of $\sigma_{\rm{V}}=6.9\%$ and a bias error of $\delta_{\rm{V}}=9.8\%$.
Likewise, similar results were obtained for the data-driven reference model \ReficPINNs{} with $\sigma_{\rm{V}}=6.8\%$ and $\delta_{\rm{V}}=8.5\%$.
These observations suggest that the reduction in the reconstruction accuracy of \VcPINNs{} resulted from the employed sampling and weighting schemes, rather than the network architecture.
This indicates a trade-off between accurate interface reconstruction and consistent velocity and pressure inference across the entire domain.
Notably, \VcPINNs{} trained with the optimization strategies tailored for velocity and pressure inference demonstrate significantly more consistent results than \IcPINNs{} trained under identical conditions.
This highlights the advantage of the conditioning with spatio-temporal features in \VcPINNs{}, which outperforms the purely spatial conditioning of \IcPINNs{} in terms of velocity and pressure inference, despite both models achieving similar accuracy in interface reconstruction.
Moreover, these results suggest that the interface reconstruction accuracy of \VcPINNs{} can be improved further by more optimized sampling and weighting schemes.
However, if the primary objective is interface reconstruction alone, \IcPINNs{} might be the preferable model due to their lower complexity and memory demands.
In fact, \IcPINNs{} yield a marginally higher interface reconstruction accuracy in comparison to \VcPINNs{} under the same training conditions, which might be related to the purely spatial features used in \IcPINNs{} that are more spatially accurate than the spatio-temporal features used in \VcPINNs{}.
Although \VcPINNs{} also incorporate purely spatial features, the network must learn to balance and integrate both spatial and spatio-temporal inputs, adding complexity to the optimization process.

\begin{figure}[h!]
    \centering
    \includegraphics[width=\linewidth]{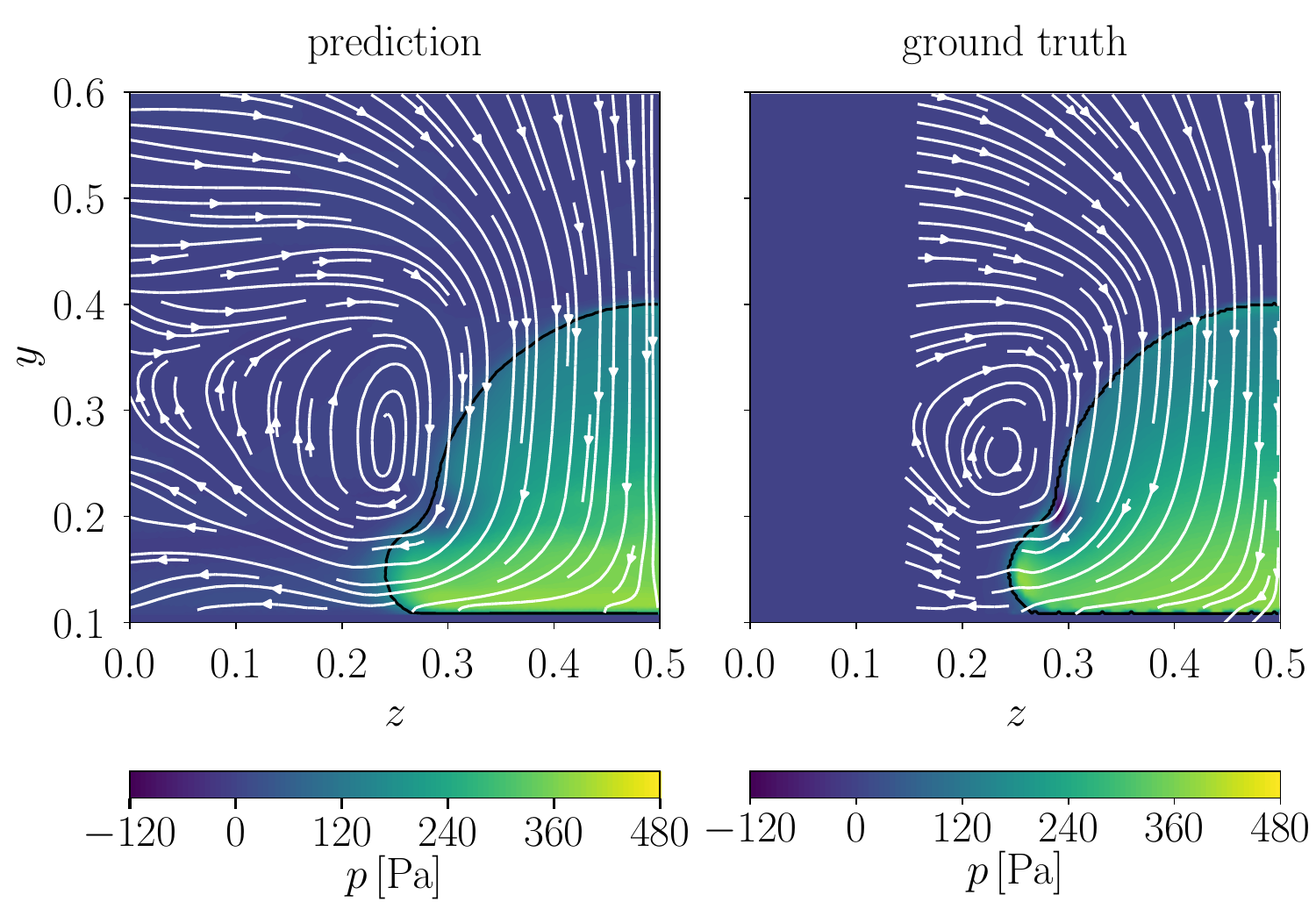}
    \caption{\textbf{Out-of-plane velocity and pressure inference of \VoFvcPINNs{} for one snapshot of the validation dataset.} Inferred velocity field visualized by streamlines and the corresponding pressure contours along the center plane of the droplet in the out-of-plane direction at $t = 1.05$ ms after impingement (left) in comparison to the ground truth simulation data (right). The contour of the gas-liquid interface is indicated by the black solid line. The synthetic input image rendered from a simulated droplet during impingement is shown in the Manuscript Figure~\ref{fig:PINNs:compound_z}. \VoFvcPINNs{} complete the flow field in a physically consistent way beyond the boundary of the training data domain.}
    \label{fig:PINNs:compound_x}
\end{figure}

\begin{figure}[h!]
    \centering
    \includegraphics[width=\linewidth]{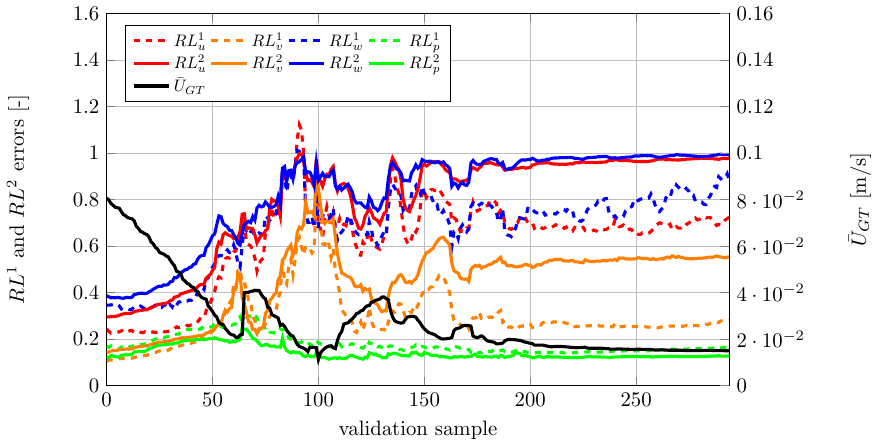}
    \caption{\textbf{Temporal evolution of the $RL^1$ and $RL^2$ errors for velocity and pressure predicted by \VoFvcPINNs{} in the entire validation domain.} Dashed colored lines indicate the relative $L^1$ errors and solid colored lines the relative $L^2$ errors of the inferred fields of the velocity components $u$, $v$, and $w$ and the pressure $p$, evaluated in the entire domain for all snapshots of the validation dataset. The averaged ground truth velocity magnitude of the flow in the entire domain is indicated by the black solid line.}
    \label{fig:PINNs:err_rel_complete}
\end{figure}

\begin{figure}[h!]
    \centering
    \includegraphics[width=\linewidth]{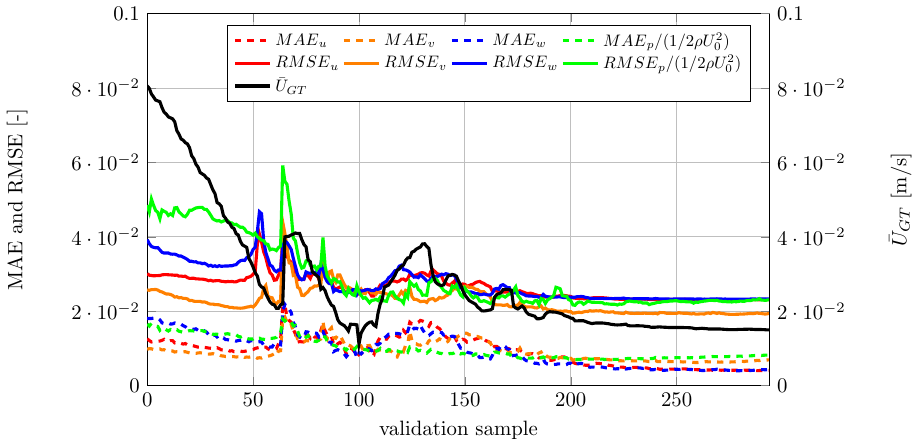}
    \caption{\textbf{Temporal evolution of the MAE and RMSE for velocity and pressure predicted by \VoFvcPINNs{} in the entire validation domain.} Dashed colored lines indicate the MAE and solid colored lines the RMSE of the inferred fields of the velocity components $u$, $v$, and $w$ and the pressure $p$, evaluated in the entire domain for all snapshots of the validation dataset. The averaged ground truth velocity magnitude of the flow in the entire domain is indicated by the black solid line. The errors of the pressure prediction are scaled by the dynamic pressure for better visibility.}
    \label{fig:PINNs:err_abs_complete}
\end{figure}

\begin{figure}[h!]
    \centering
    \includegraphics[width=\linewidth]{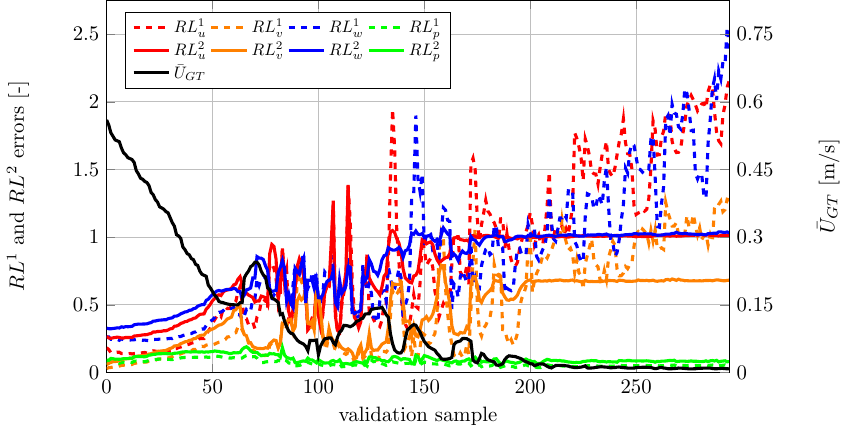}
    \caption{\textbf{Temporal evolution of the $RL^1$ and $RL^2$ errors for velocity and pressure predicted by \VoFvcPINNs{} in the internal validation domain.} Dashed colored lines indicate the relative $L^1$ errors and solid colored lines the relative $L^2$ errors of the inferred fields of the velocity components $u$, $v$, and $w$ and the pressure $p$, evaluated only in the liquid domain for all snapshots of the validation dataset. The averaged ground truth velocity magnitude of the internal flow in the droplet is indicated by the black solid line.}
    \label{fig:PINNs:err_rel_drop}
\end{figure}

\begin{figure}[h!]
    \centering
    \includegraphics[width=\linewidth]{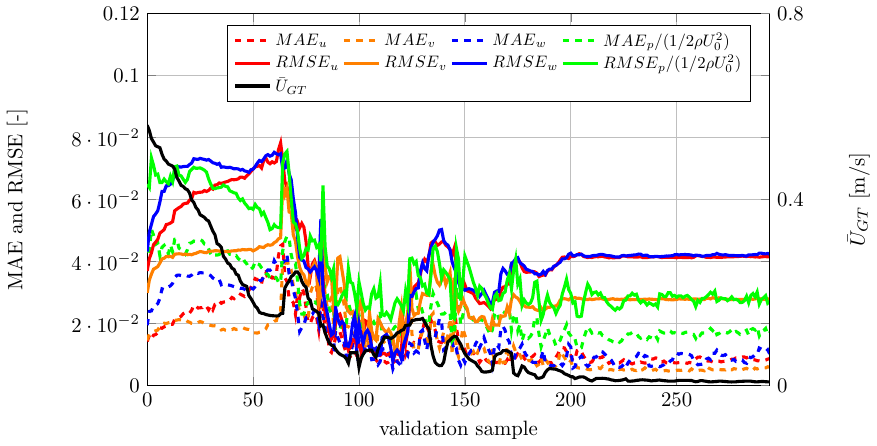}
    \caption{\textbf{Temporal evolution of the MAE and RMSE for velocity and pressure predicted by \VoFvcPINNs{} in the internal validation domain.} Dashed colored lines indicate the MAE and solid colored lines the RMSE of the inferred fields of the velocity components $u$, $v$, and $w$ and the pressure $p$, evaluated only in the liquid domain for all snapshots of the validation dataset. The averaged ground truth velocity magnitude of the internal flow in the droplet is indicated by the black solid line. The errors of the pressure prediction are scaled by the dynamic pressure for better visibility.}
    \label{fig:PINNs:err_abs_drop}
\end{figure}

\begin{table}[h!]
    \centering
    \caption{\textbf{Error metrics for the internal flow predicted by \VoFvcPINNs{}.} MAE, RMSE, and relative $L^1$ and $L^2$ errors of the predicted flow quantities $u$, $v$, $w$, and $p$ evaluated only in the liquid phase.}
    \begin{tabular}{l|rrrrrrrr}%\toprule
    quantity & MAE [m/s] & RMSE [m/s] & $RL^1$ [$\%$] & $RL^2$ [$\%$]\\%
    \hline \rule{0pt}{1.0\normalbaselineskip}%
    $u$ & 0.0147 & 0.0442 & 36.5 & 48.4 \\
    $v$ & 0.0134 & 0.0324 & 16.6 & 20.7 \\
    $w$ & 0.0164 & 0.0466 & 44.1 & 54.4 \\
    $p$ & 9.584  & 15.945 & 6.8  & 10.7 \\
    average & -  & -      & 26.0 & 33.6 \\
    \end{tabular} 
    \label{tab:VoFVcPINNs_err}
\end{table}

\begin{table}[h!]
    \centering
    \caption{\textbf{Error metrics for the internal flow predicted by \PFvcPINNs{}.} MAE, RMSE, and relative $L^1$ and $L^2$ errors of the predicted flow quantities $u,v,w$, and $p$ evaluated only in the liquid phase.}
    \begin{tabular}{l|rrrrrrrr}%\toprule
    quantity & MAE [m/s] & RMSE [m/s] & $RL^1$ [$\%$] & $RL^2$ [$\%$]\\%
    \hline \rule{0pt}{1.0\normalbaselineskip}%
    $u$ & 0.0143 & 0.0442 & 35.6 & 48.4 \\
    $v$ & 0.0133 & 0.0318 & 16.4 & 20.3 \\
    $w$ & 0.0165 & 0.0473 & 44.2 & 55.3 \\
    $p$   & 6.823 & 12.623 & 4.8 & 8.5 \\
    average & -   & -     & 25.3 & 33.1 \\
    \end{tabular} 
    \label{tab:PFVcPINNs_err}
\end{table}

\paragraph*{\textbf{Velocity and pressure inference on validation data}}
\label{sup:results:velocity}

Figure~\ref{fig:PINNs:compound_x} shows the out-of-plane pressure distribution and visualized streamlines of the velocity field in the center plane of the droplet predicted by \VoFvcPINNs{} (left) in comparison to the ground truth (right) for one snapshot of the synthetic validation dataset displayed in the Manuscript Figure~\ref{fig:PINNs:compound_z}.
The relative $L^1$, $L^2$ errors and the absolute MAE, RMSE errors of the inferred velocity and pressure fields were evaluated across the entire spatio-temporal domain at sampling points on a uniform grid with $512^3$ grid nodes and are subsequently averaged over the subset of the validation data that features droplet impingement on the structured surface.
Additionally, the errors were evaluated only in the liquid domain, \emph{i.e.}, only for the internal flow in the droplet.
The temporal evolution of the relative $L^1$, $L^2$ errors for the velocity and pressure fields predicted by \VoFvcPINNs{} evaluated for the entire flow field is plotted in Figure~\ref{fig:PINNs:err_rel_complete}, while Figure~\ref{fig:PINNs:err_abs_complete} shows the development of the MAE and RMSE errors.
Figures~\ref{fig:PINNs:err_rel_drop} and~\ref{fig:PINNs:err_abs_drop} show the relative $L^1$, $L^2$ errors and the MAE and RMSE errors, of the internal flow inferred by \VoFvcPINNs{}.
Tables~\ref{tab:VoFVcPINNs_err} and~\ref{tab:PFVcPINNs_err} detail the MAE and RMSE, as well as the relative $\rm{L}^1$ and $\rm{L}^2$ errors for the prediction of the velocity and pressure distributions of the internal flow for \VoFvcPINNs{} and \PFvcPINNs{}, respectively.

\begin{figure*}[h]
    \centering
    \includegraphics[width=0.85\textwidth]{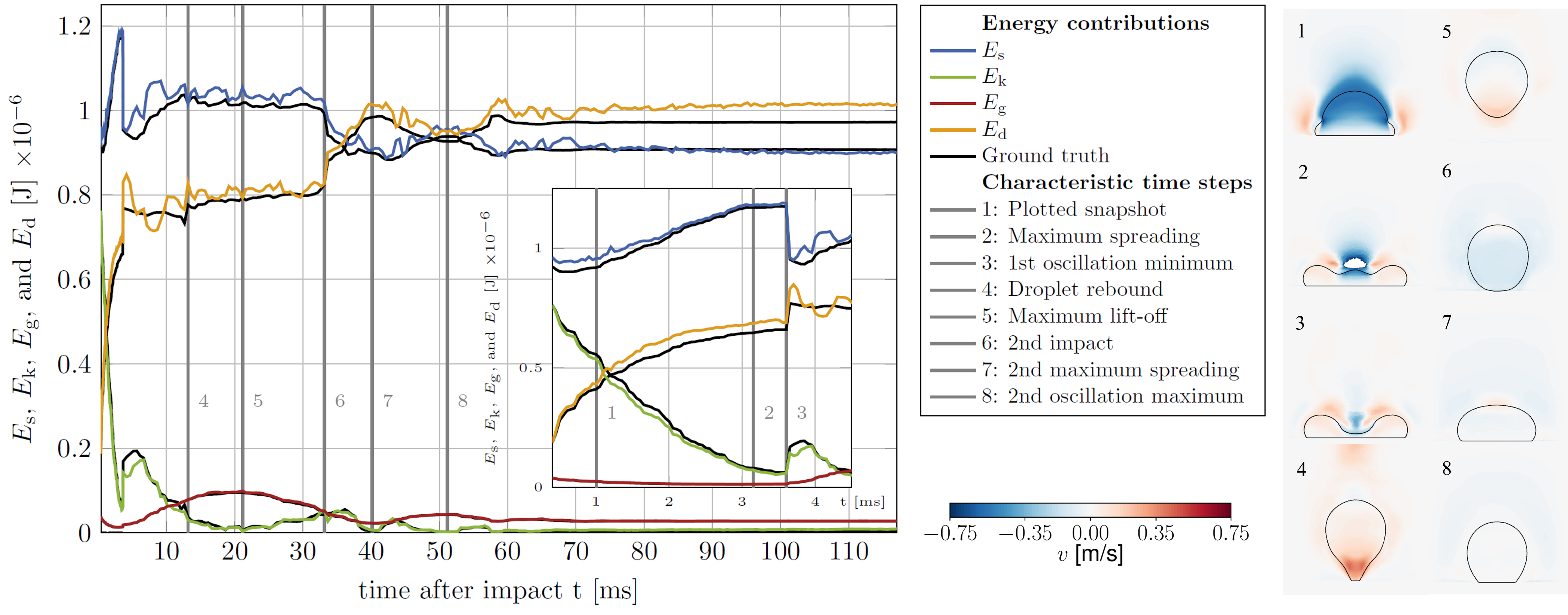}
    \caption{\textbf{Temporal development of the energy contributions predicted by \VoFvcPINNs{} for validation data.} The colored lines represent the kinetic energy of the droplet $E_{\rm{k}}$, the cumulative surface energy $E_{\rm{s}}$ of the gas-liquid and the liquid-solid interfaces, and the potential energy $E_{\rm{g}}$, evaluated for the subset of the validation data featuring droplet impingement on the structured PDMS substrate. The inset shows a magnified view of the early time steps. The details (1) to (8) display the in-plane prediction of the vertical velocity component $v$ at characteristic time steps, indicated by the gray vertical lines in the main plot. The alternating sign of the vertical velocity component reflects the oscillatory nature of the droplet dynamics after impingement.}
    \label{fig:energy_validation}
\end{figure*}

\paragraph*{\textbf{Temporal evolution of the energy contributions for validation data}}
\label{sup:results:energy}

Figure~\ref{fig:energy_validation} shows the temporal development of the individual energy contributions for predictions obtained with \VoFvcPINNs{} in comparison to the ground truth DNS validation data.
The energy contributions were evaluated across the entire spatio-temporal domain at sampling points on a uniform grid with $256^3$ grid nodes and averaged across the spatial domain.
As can be seen, the potential energy $E_{\rm{g}}$ constitutes the smallest contribution throughout the process.
After the initial impact, both $E_{\rm{k}}$ and $E_{\rm{g}}$ decrease, while their energy is partly converted into surface energy $E_{\rm{s}}$ and partly dissipated (increasing $E_{\rm{d}}$).
The kinetic energy reaches a minimum shortly after maximum droplet spreading (2), at which point $E_{\rm{s}}$ is maximal.
Just afterwards, a rapid flow reversal occurs related to the start of the retraction phase (3) and characterized by a sudden increase in $E_{\rm{k}}$, which corresponds to a sharp decrease in $E_{\rm{s}}$ and a jump in $E_{\rm{d}}$ caused by the sudden change of the flow quantities.
During the formation of a liquid jet, $E_{\rm{g}}$ increases again, while $E_{\rm{k}}$ decreases, and eventually the droplet rebounds from the substrate (4).
As the droplet lifts off, $E_{\rm{g}}$ further increases and reaches its maximum, while $E_{\rm{k}}$ becomes minimal due to the second reversal in flow direction (5).
Consecutively, the droplet then falls back down, converting $E_{\rm{g}}$ into $E_{\rm{k}}$, and the droplet impacts the surface for a second time (6).
This is followed by a damped oscillatory motion during which the droplet remains attached to the surface.

\begin{table}[ht]
    \centering
    \caption{\textbf{Errors of the energy contributions predicted for validation data.} Relative $L^1$ and $L^2$ errors of the surface energy $E_{\rm{s}}$, kinetic energy $E_{\rm{k}}$, gravitational energy $E_{\rm{g}}$, and viscous energy dissipation $E_{\rm{d}}$ for the prediction of \RefvcPINNs{}, \VoFvcPINNs{}, and \PFvcPINNs{} on the validation dataset featuring droplet impingement on the structured PDMS substrate.}
	\begin{tabular}{l|rrrrrrrrr}		%\toprule
		model & \multicolumn{2}{c}{\RefvcPINNs{}} & \multicolumn{2}{c}{\VoFvcPINNs{}} & \multicolumn{2}{c}{\PFvcPINNs{}} \\ 
  $RL^1$/$RL^2$ [$\%$] & $RL^1$ & $RL^2$ & $RL^1$ & $RL^2$ & $RL^1$ & $RL^2$ \\
  \hline \rule{0pt}{1.0\normalbaselineskip}%
            $E_{\rm{s}}$ & 1.8 & 2.0 & 1.8 & 2.3 & 1.6 & 2.0 \\
            $E_{\rm{k}}$ & 6.7 & 3.8 & 9.2 & 6.6 & 12.5 & 9.5 \\
            $E_{\rm{g}}$ & 2.9 & 3.1 & 2.7 & 3.0 & 1.2 & 1.7 \\
            $E_{\rm{d}}$ & 5.6 & 6.0 & 4.8 & 5.0 & 2.7 & 3.0 \\
	\end{tabular} 
	\label{tab:energy_val}
\end{table}

The comparison between the energy contributions inferred by \VoFvcPINNs{} and those obtained by DNS reveals a good agreement over the entire temporal development.
The integral energy contributions closely follow the ground truth from DNS, with only minor uniform deviations and no significant outliers.
The \PFvcPINNs{} and the baseline model \RefvcPINNs{} exhibit similar temporal trends, with only minor differences.
Table~\ref{tab:energy_val} details the relative $L^1$ and $L^2$ errors of the energy contributions evaluated across the validation dataset for the prediction by \VoFvcPINNs{}, \PFvcPINNs{}, and \RefvcPINNs{}.
As can be seen in Figure~\ref{fig:energy_validation}, the potential energy $E_{\rm{g}}$ is predicted consistently well by \VoFvcPINNs{}, which is reflected in the low relative $L^2$ error of $RL^2_{E_{\rm{g}}}=3.0\%$.
\PFvcPINNs{} achieve the most accurate results with $RL^2_{E_{\rm{g}}}=1.7\%$.
The surface energy $E_{\rm{s}}$ is initially overestimated, but a good agreement for the long-term prediction is achieved, with $RL^2_{E_{\rm{s}}}=2.3\%$ for \VoFvcPINNs{} and $RL^2_{E_{\rm{s}}}=2.0\%$ for \PFvcPINNs{}.
Both \VoFvcPINNs{} and \PFvcPINNs{} underestimate $E_{\rm{k}}$ during the early time steps, most notably \PFvcPINNs{}, whereas at later times all models tend to overestimate $E_{\rm{k}}$ marginally.
Despite these trends, the overall agreement remains high, with $RL^2_{E_{\rm{k}}}=6.6\%$ reached by \VoFvcPINNs{} and $RL^2_{E_{\rm{k}}}=9.5\%$ by \PFvcPINNs{}.
As a cumulative quantity, $E_{\rm{d}}$ should ideally increase monotonically and thus serves as an indicator for energy conservation of the predictions.
Since $E_{\rm{d}}$ is derived from the other energy contributions, errors in those terms can partially compensate, affecting the overall accuracy of $E_{\rm{d}}$.
The approximate monotonic increase of $E_{\rm{d}}$ is captured well by all models, consistent with minor nonphysical decreases observed in the simulation during the second oscillation period (see Figure~\ref{fig:energy_validation} in the time between (3) and (4)). 

\begin{table}[h]
    \centering
    \caption{\textbf{Governing equation residuals of the predictions for experimental data.} Mean absolute errors (MAE) of the residuals for the dimensionless continuity equation, interface evolution equation, and the Navier-Stokes momentum equations in $x$, $y$, and $z$ for the prediction by \VoFvcPINNs{}, \PFvcPINNs{}, and \RefvcPINNs{}, averaged over all experimental test cases featuring droplet impingement on structured PDMS and PLA substrates.}
    \begin{tabular}{l|lll}
        equation & \RefvcPINNs{} & \VoFvcPINNs{} & \PFvcPINNs{} \\ 
        \hline \rule{0pt}{1.0\normalbaselineskip}%
        Continuity       & $6.679\times 10^{-2}$ & $1.706\times 10^{-2}$ & $1.213\times 10^{-2}$ \\
        Interface        & $7.704\times 10^{-5}$ & $7.285\times 10^{-5}$ & $6.678\times 10^{-4}$ \\
        Momentum $x$     & $5.087\times 10^{-3}$ & $1.529\times 10^{-3}$ & $9.140\times 10^{-4}$ \\
        Momentum $y$     & $6.006\times 10^{-3}$ & $3.022\times 10^{-3}$ & $2.664\times 10^{-3}$ \\
        Momentum $z$     & $4.205\times 10^{-2}$ & $1.732\times 10^{-2}$ & $1.098\times 10^{-2}$ \\
    \end{tabular}
    \label{tab:residuals}
\end{table}

\begin{figure}[h]
    \centering
    \includegraphics[width=1.0\linewidth]{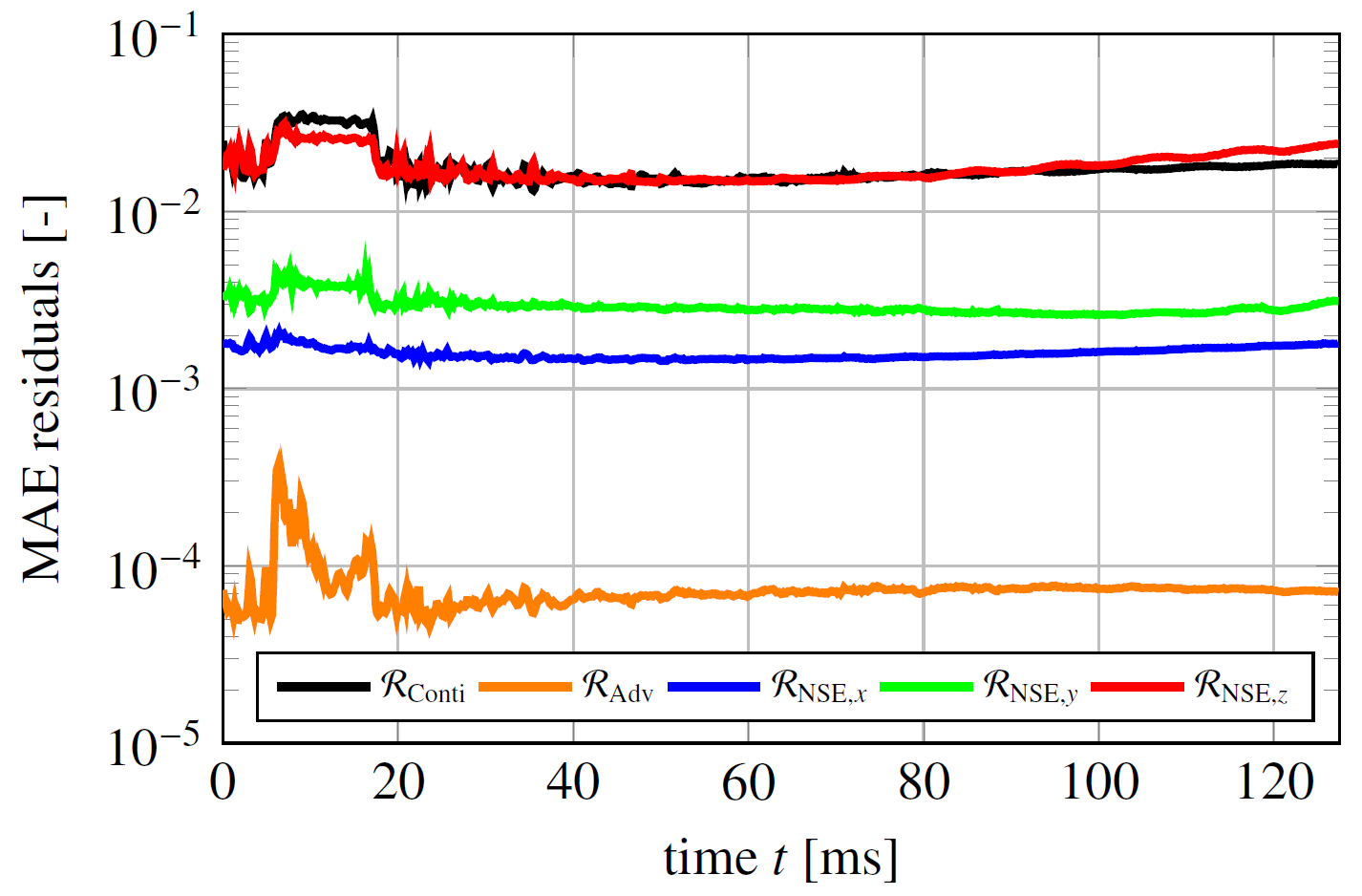}
    \caption{\textbf{Temporal evolution of the governing equation residuals predicted by \VoFvcPINNs{} for experimental data.} Mean absolute errors (MAE) of the residuals for the dimensionless continuity equation $\mathcal{R}_{\text{Conti}}$, interface evolution equation $\mathcal{R}_{\text{Adv}}$, and the Navier-Stokes momentum equations $\mathcal{R}_{\text{NSE},j}$ with $j = (x, y, z)$, for one droplet impingement experiment involving the structured PDMS substrate.}
    \label{fig:VoFvcPINNs_PDMS90_residuals}
\end{figure}

\begin{figure}[h]
    \centering
    \includegraphics[width=1.0\linewidth]{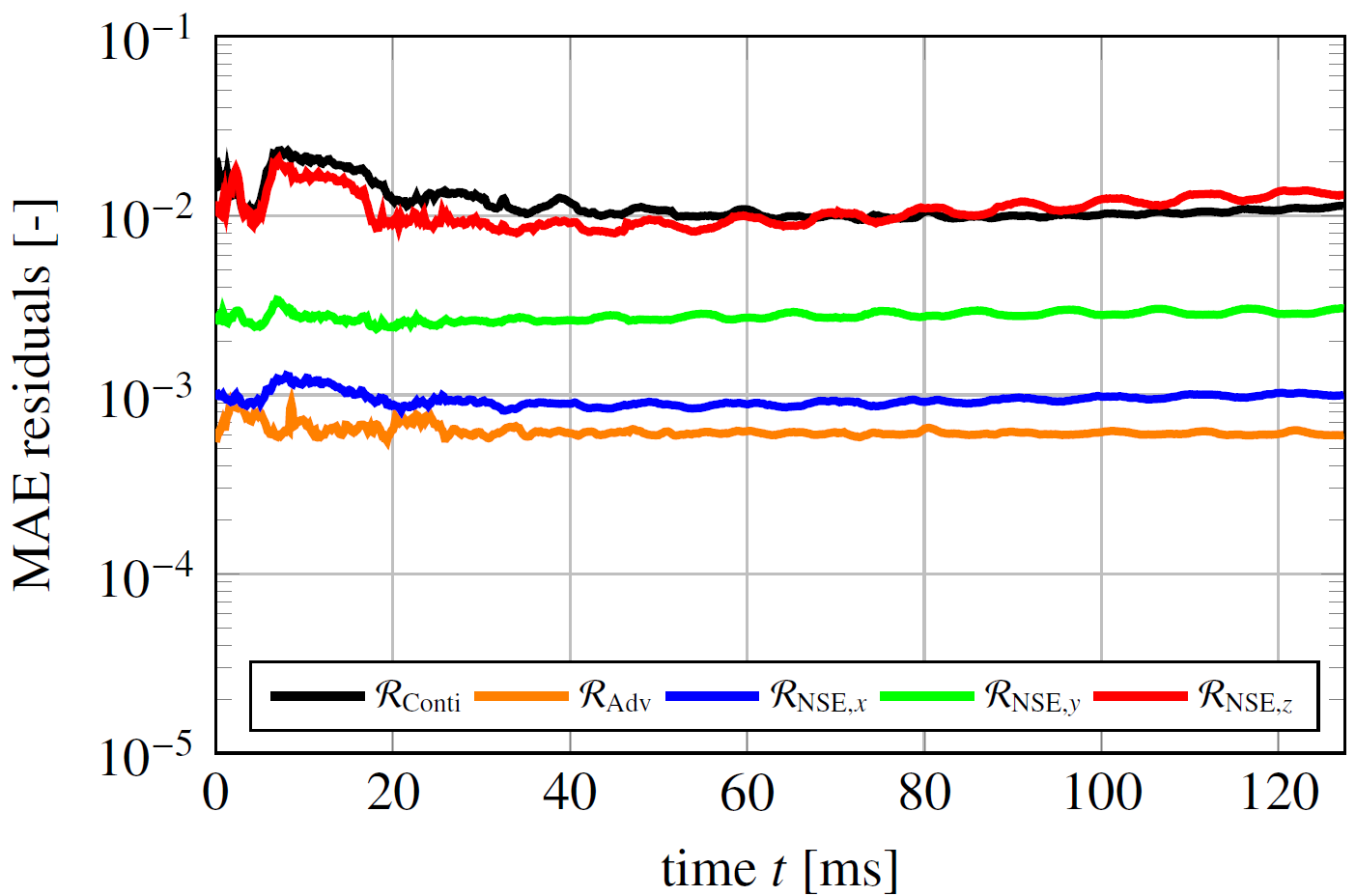}
    \caption{\textbf{Temporal evolution of the governing equation residuals predicted by \PFvcPINNs{} for experimental data.} Same as Figure~\ref{fig:VoFvcPINNs_PDMS90_residuals}, but for \PFvcPINNs{}.}
    \label{fig:PFvcPINNs_PDMS90_residuals}
\end{figure}

\paragraph*{\textbf{Residuals of the governing equations}}
\label{sup:results:residuals}

Table~\ref{tab:residuals} details the mean absolute error (MAE) of the governing equation residuals for the reconstruction of experimental data by \VoFvcPINNs{}, \PFvcPINNs{}, and \RefvcPINNs{}.
The residuals were evaluated across the entire spatio-temporal domain at sampling points on a uniform grid with $32^3$ grid nodes and averaged across the domain.
Across the different experiments, consistently similar magnitudes of residuals were obtained with a standard deviation of less than $12.5\%$ of the mean over all experiments presented in Table~\ref{tab:residuals}.
These results further underline that the prediction of the flow field by \VcPINNs{} generalizes well to different droplet dynamics.
% physics-informed loss graphs in section x are computed on adaptive sampling points with higher sampling density at the interface, thus different magnitudes.
Figures~\ref{fig:VoFvcPINNs_PDMS90_residuals} and~\ref{fig:PFvcPINNs_PDMS90_residuals} additionally show the temporal evolution of the residuals for the reconstruction of the same single droplet impingement experiment reconstructed by \VoFvcPINNs{} and \PFvcPINNs{}, respectively.
For the reconstruction of the other experiments, similar results were obtained by both \VoFvcPINNs{} and \PFvcPINNs{}.
As can be seen, the distribution of all residual terms is uniform in time, except for a peak between $5$\,ms and $20$\,ms.
This peak correlates with the time period in which the droplet dynamics in the experiment differ most from the simulation.
In the experiments involving the PDMS substrate, a liquid column forms during the retraction phase \cite{Rioboo2001}.
%, while in the experiments involving the PLA substrate, the droplet undergoes less severe deformation during the retraction phase.
Conversely, in the simulation, the droplet rebounds from the surface, leading to significantly different droplet dynamics in comparison to the experiment.
This discrepancy between the test data and the training data might be a cause for the higher residuals during that time frame.
Furthermore, during this time, the interface shape as well as the flow field change rapidly, leading to high velocity gradients, which might additionally contribute to higher residuals.
Nonetheless, the overall uniformity of the distribution of the residuals in time suggests that the physics of the two-phase flow were learned consistently throughout the entire droplet impingement process.
The comparison of Figures~\ref{fig:VoFvcPINNs_PDMS90_residuals} and~\ref{fig:PFvcPINNs_PDMS90_residuals} reveals consistently lower residuals of \PFvcPINNs{} in comparison to \VoFvcPINNs{} for all governing equations except the interface evolution equation.
These results corroborate the earlier observation that the Cahn-Hilliard equation (Eq.~\ref{eq:CahnHilliard}) employed in \PFPINNs{} is more difficult to optimize than the advection equation (Eq.~\ref{eq:algVOF_advection}) in the algebraic VoF formulation used to train \VoFPINNs{}.

\end{document}